\documentclass{aa}  
\usepackage[flushleft]{threeparttable}
\usepackage{multirow}
\usepackage{float}
\usepackage{graphics}
\usepackage{graphicx}
\usepackage{amssymb, amsmath}
\usepackage{tikz}
\usepackage{booktabs}
\usepackage{txfonts}

\usetikzlibrary{graphs, positioning, quotes, shapes.geometric}
\usepackage[colorlinks=true,linkcolor=blue,urlcolor=blue,citecolor=blue]{hyperref}

\newcommand{\figref}[1]{\hyperref[#1]{\textcolor{red}{\autoref{#1}}}}

\usepackage{amstext}
\usepackage[T1]{fontenc}
\usepackage{mathrsfs}
\usepackage[normalem]{ulem}
\usepackage{natbib,twoopt}
\makeatletter
\def\@to{to}
\def\as     {\ifmmode {\rlap.}$\,$''$\,$\! \else ${\rlap.}$\,$''$\,$\!$\fi}
     \def\decsec  {\ifmmode {\rlap.}$\,$^{\rm s}$\,$\! \else ${\rlap.}$\,$^{\rm s}$\,$\!$\fi}\def\decs  {\ifmmode {\rlap.}$\,$^{\rm s}$\,$\! \else ${\rlap.}$\,$^{\rm s}$\,$\!$\fi}
\makeatother

\newcommand{\rah}{$^{\mbox{\scriptsize h}}$}
\newcommand{\ram}{$^{\mbox{\scriptsize m}}$}
\newcommand{\ras}{$^{\mbox{\scriptsize s}}$}
\newcommand{\decd}{$^{\circ}$}
\newcommand{\decm}{$'$}

\begin{document}


\title{Massive clumps in W43-main: Structure formation in an extensively shocked molecular cloud}

 \author{Y. Lin
          \inst{1,2} \and F. Wyrowski \inst{2} \and H. B. Liu \inst{3} \and Y. Gong \inst{2} \and O. Sipil{\"a} \inst{1}  \and A. Izquierdo \inst{4,5} \and T. Csengeri \inst{6} \and A. Ginsburg \inst{7} \and G. X. Li \inst{8} \and S. Spezzano \inst{1} \and J. E. Pineda \inst{1} \and S. Leurini \inst{9} \and P. Caselli \inst{1} \and K. M. Menten \inst{2}}

   \institute{Max-Planck-Institut f{\"u}r Extraterrestrische Physik, Giessenbachstr. 1, D-85748 Garching bei M{\"u}nchen\\
   \email{ylin@mpe.mpg.de}
      \and
   Max Planck Institute for Radio Astronomy, Auf dem H\"{u}gel 69, 53121 Bonn
   \and
             Department of Physics, National Sun Yat-Sen University, No. 70, Lien-Hai Road, Kaohsiung City 80424, Taiwan, R.O.C.
             \and 
               European Southern Observatory, Karl-Schwarzschild-Str. 2, 85748 Garching bei München, Germany
               \and 
               Leiden Observatory, Leiden University, P.O. Box 9513, 2300 RA Leiden, the Netherlands
         \and
        OASU/LAB-UMR5804, CNRS, Universit\'e Bordeaux, all\'ee Geoffroy Saint-Hilaire, 33615 Pessac, France
          \and 
        Department of Astronomy, University of Florida, PO Box 112055, USA
        \and 
         South-Western Institute for Astronomy Research, Yunnan University, Chenggong District, Kunming 650091, P. R. China
         \and 
        INAF – Osservatorio Astronomico di Cagliari, Via della Scienza 5, I-09047 Selargius (CA), Italy}
        
\abstract
{}
{W43-main is a massive molecular complex located at the interaction of the Scutum arm and the Galactic bar undergoing starburst activities. We aim to investigate the gas dynamics, in particular, the prevailing shock signatures from the cloud to clump scale and assess the impact of shocks on the formation of dense gas and early-stage cores in the OB cluster formation process.}
{We have carried out NOEMA and IRAM-30m observations at 3\,mm towards five molecular gas clumps in W43 main which are located inside large-scale interacting gas components. We use CH$_{3}$CCH and H$_{2}$CS lines to trace the extended gas temperature and CH$_{3}$OH lines to probe the volume density of the dense gas components ($\gtrsim$10$^{5}$ cm$^{-3}$). We adopt multiple tracers sensitive to different gas density regimes to reflect the global gas motions. The density enhancements constrained by CH$_{3}$OH and a population of NH$_{2}$D cores are correlated (in spatial and velocity domain) with SiO emission, which is a prominent indicator of shock processing in molecular clouds. } 
{The emission of SiO (2-1) is extensive across the region ($\sim$4\,pc) and is contained within a low-velocity regime, hinting at a large-scale origin of the shocks. The position-velocity maps of multiple tracers show systematic spatio-kinematic offsets supporting the cloud-cloud collision/merging scenario. We identify an additional extended velocity component in CCH emission, which coincides with one of the velocity components of the larger scale $^{13}$CO (2-1) emission, likely representing an outer, less dense gas layer in the cloud merging process. We find that the ``V''-shaped, asymmetric SiO wings are tightly correlated with localised gas density enhancements, which is direct evidence of dense gas formation and accumulation in shocks. The formed dense gas may facilitate the accretion of the embedded, massive pre-stellar and protostellar cores. We resolve two categories of NH$_{2}$D cores: ones exhibiting only subsonic to transonic velocity dispersion, and the others with an additional supersonic velocity dispersion. The centroid velocities of the latter cores are correlated with the shock front seen by SiO. The kinematics of the $\sim$0.1 pc NH$_{2}$D cores are heavily imprinted by shock activities, and may represent a population of early-stage cores forming around the shock interface.} 
{}
\maketitle

\section{Introduction}\label{sec:intro}
Massive stars profoundly influence the chemical and kinetic evolution of galaxies throughout their lifetimes, by intense feedback effects, such as ionisation, stellar winds, and fierce death as exploding supernovae, yet our understanding of their formation process remains primitive (\citealt{ZY07}, \citealt{Motte18}). Compared to low-mass stars, OB stars originate from more massive and denser molecular environments, characterised by intense cloud-scale dynamics at $\gtrsim$1-10 pc (\citealt{VS19}, \citealt{Padoan20}, \citealt{Kumar20}). Understanding the formation and evolution of extreme star-forming clouds, which are in the high-mass end of the cloud mass spectrum and/or located in highly pressurised, turbulent environments, is of paramount importance for unravelling the origin of massive stars and clusters. 

Supersonic shocks are common features associated with OB cluster forming molecular clouds, the origin of which can be stellar winds, expanding H\textsc{ii} regions (\citealt{Hill78}, \citealt{Bertoldi89}), cloud-cloud collisions (\citealt{Fukui21}), as well as accretion/infall flows (e.g. \citealt{Hennebelle13}).
In particular, cloud-cloud collision has been invoked as a possible mechanism of forming the dense, highly turbulent massive molecular clumps (for a recent review see \citealt{Fukui21}). 
From a theoretical perspective, the shock-compressed gas layer is density-enhanced and prone to fragmentation (\citealt{Whitworth94a, Whitworth94b}, \citealt{Wu17}, \citealt{Balfour17}), triggering the formation of gravitationally unstable cores and clumps at the collision interface (\citealt{HO92}, \citealt{Ana10}, \citealt{Takahira14, Takahira18}, \citealt{Mocz18}, \citealt{Sakre21}, \citealt{Cosentino22}). Shock compression can also cause convergent flows that further focus on the gas material towards the forming cores and clumps therein (\citealt{IF13}, \citealt{Inoue18}). 
The collision velocity, initial density structures and the kinematic state of the pre-collision clouds can be critical factors in setting the morphology, star formation efficiency (SFE) and protostellar mass distribution (\citealt{Balfour15, Balfour17}, \citealt{Liow20}) of the compressed gas layer. 


The W43 molecular cloud (of mass $M\,\sim\,$$1\times10^{6}$ $M_{\odot}$ and bolometric luminosity, $L_{\mathrm{bol}}$$\,\sim\,$10$^{7}$ $L_{\odot}$, $d$$\sim$5.5 kpc) is among the most massive star-forming complexes in the Galaxy (\citealt{Bally10}, \citealt{Lin16}, \citealt{Motte22a}). It is a cloud hosting several star-burst clusters (\citealt{Blum99}, \citealt{Motte03}), and therefore may be subject to ionisation induced shocks (``radiation driven implosion'', \citealt{Bertoldi89}, \citealt{Bisbas11}). Located at the the intersection of the (near end) Galactic bar and the Scutum arm at a distance of 5.49~kpc (\citealt{Zhang14}), the cloud is enduring violent interactions of gas streams. There is wide-spread low-velocity ($\lesssim$20 km s$^{-1}$) SiO emission discovered toward W43, interpreted as signatures of shocks from large-scale colliding flows (\citealt{NL11}). At even larger spatial scales, there are $\sim$ 200 pc HI filaments showing velocity gradients feeding W43 (\citealt{Motte14}), indicating that the turbulent convergent flows are continuous out to atomic gas. These facts suggest a highly dynamical process of OB cluster formation inside the cloud: W43 is an ideal site for understanding the impact of extensive shocks on the formation of OB clusters.   

Molecular line and dust continuum surveys towards W43 at large scale (several tens of pc) reveal that there is a large amount of dense gas (\citealt{Motte03}, \citealt{Bally10}, \citealt{NL11}, \citealt{Carlhoff}). The cloud is composed of two sub-clouds: W43-main and W43-south. W43-main is more massive and has a prominent ``Z''-shaped filamentary morphology (\citealt{Motte18}). 
$^{13}$CO (3-2) observations reveal two velocity components towards W43-main, 82 km s$^{-1}$ and 94 km s$^{-1}$, intersecting at the southern ridge of the ``Z'' cloud, which indicate cloud-cloud collisions (\citealt{Kohno21}).

There are a number of massive star-forming clumps inside W43-main (\citealt{Motte03}, \citealt{Carlhoff}, \citealt{Lin16}); clump MM2 and MM3 are the second and third most massive clumps in this cloud, after clump MM1, which all have masses of $\gtrsim$1000 $M_{\odot}$ and are actively forming stars (\citealt{Motte22a}, \citealt{Cortes19}, \citealt{Nony23}), representing potential OB cluster progenitor. In fact, considering the number of embedding clumps, among the seven OB cluster forming molecular clouds sampled in \citet{Lin16}, W43-main and W43-south have richer fragmentation than the other clouds, which may be a consequence of shock compression and subsequent self-gravitational fragmentation. 

In order to investigate the cloud-clump connections of gas dynamics, the physical properties of the extended gas structures within massive clumps, of $\sim$0.1 pc scale, are of primary interest.  
We conducted 3\,mm wide-band observations with NOEMA and the IRAM 30m telescope towards five selected clumps located adjacently in the southern ridge of W43-main, where previous large-scale CO (2-1) observations reveal extensive overlap of gas components of different velocities (Fig. \ref{fig:w43_npts}, \citealt{Carlhoff}). These target clumps were identified from a 10 $''$ hydrogen column density map of W43-main, derived by iterative Spectral Energy Distribution fitting based on image combination technique utilising continuum from ground-based bolometers and space telescopes (\citealt{Lin16}, Figure \ref{fig:w43_npts}). They span a wide mass range of 200-4000 $M_{\odot}$, while showing similar luminosity-to-mass ratio ($\sim$20, Table \ref{tab:source_info}) which is indicative of similar evolutionary phase (\citealt{Molinari10}) and may be formed coevally.    

The paper is organised as follows: Sect.~\ref{sec:obs_reduc} describes the observations and data reduction procedure. In Sect.~\ref{sec:3mm_cont} we report the gas mass derived from 3\,mm dust continuum and derive the dynamical timescale of the UCH\textsc{ii} region in clump MM3. 
 In Sect. ~\ref{sec:temp_dens} we constrain the temperature and density structure of the target clumps through several thermometers and CH$_{3}$OH lines. In Sect.~\ref{sec:vo_vw}, we present emission features of SiO and investigate their relation with dense gas formation. 
In Sect.~\ref{sec:veloshift} we reveal systematic velocity offsets of molecules tracing different gas densities. In Sect.~\ref{sec:cch}-Sect.~\ref{sec:nh2d} we describe the hyperfine fitting of CCH and NH$_{2}$D lines and report the discovery of a population of cold dense cores traced by compact NH$_{2}$D emission, and analyze the chemical evolution of NH$_{2}$D with the aid of chemical models. In Sect.~\ref{sec:diss} we discuss the shock dynamics, and put a special focus on the relation of the NH$_{2}$D cores with shocked gas and their multi-scale fragmentation properties.

\section{Observations and data reduction} \label{sec:obs_reduc}

The NOEMA pointed observations at 3 mm were taken between August to September 2019 with the D-array configuration in track-sharing mode (Project ID: S19AJ, PI: Y. Lin). 
Nine antennas were used during the observations, covering a baseline range of 32-176 m.
On-source observations were conducted in 20-30 minute intervals, which are interluded with observations on 1851$+$0035 as phase and amplitude calibrator. The bandpass calibrators are 3C273, 3C454.3 and 3C345, and flux calibrator MWC349. The wide-band correlator PolyFix was used, which covers an instantaneous bandwidth of 31 GHz separated into eight subbands, with a fixed spectral resolution of 2 MHz ($\sim$ 6.2 km s$^{-1}$ at 96 GHz). Multiple high-resolution spectral windows with 62.5 kHz ($\sim$ 0.2 km s$^{-1}$) channel spacing were placed to cover the molecular lines of interest. At a frequency of 96 GHz, the achieved angular resolution is 4.4$''$ and each pointing has a primary beam size (FWHM) of 53$''$.

We use the \textsc{CLIC} and \textsc{MAPPING} modules in the \textsc{GILDAS} software package \footnote{http://www.iram.fr/IRAMFR/GILDAS/} for calibration and imaging. The channels with line emission were identified by visual inspection with the \textsc{IMAGER} module and subtracted from the visibilities of low-resolution windows in the full frequency range to form the continuum. Similarly, to extract line emission, the continuum level was fitted and removed from each of the high-resolution backend chunks. We adopted natural weighting and cleaned the continuum and line cubes using the Hogbom algorithm (\citealt{hogbom}).

As a short-spacing complement to add extended molecular line emission, IRAM 30m observations towards these clumps were taken during November 2019 to June 2020 with the EMIR receiver. A region of 1.7$'$ by 1.7$'$ was mapped around each source with the on-the-fly observing mode and then the five small datacubes were combined together to form a large datacube. Focus was checked on Saturn every 4 hours and the pointing was determined every 1-1.5 hours on 1749+096 or 1741-038. We follow the standard data reduction procedure with the {\textsc{CLASS}} module in {\textsc{GILDAS}}. The main beam efficiency ($\eta_{\mbox{\scriptsize{mb}}}$\footnote{https://publicwiki.iram.es/Iram30mEfficiencies}) was applied for conversion to main-beam temperatures ($T_{\mathrm{mb}}$). 

\begin{figure*}
\begin{tabular}{p{0.33\linewidth}p{0.33\linewidth}p{0.33\linewidth}}
\hspace{-0.2cm}\includegraphics[scale=0.26]{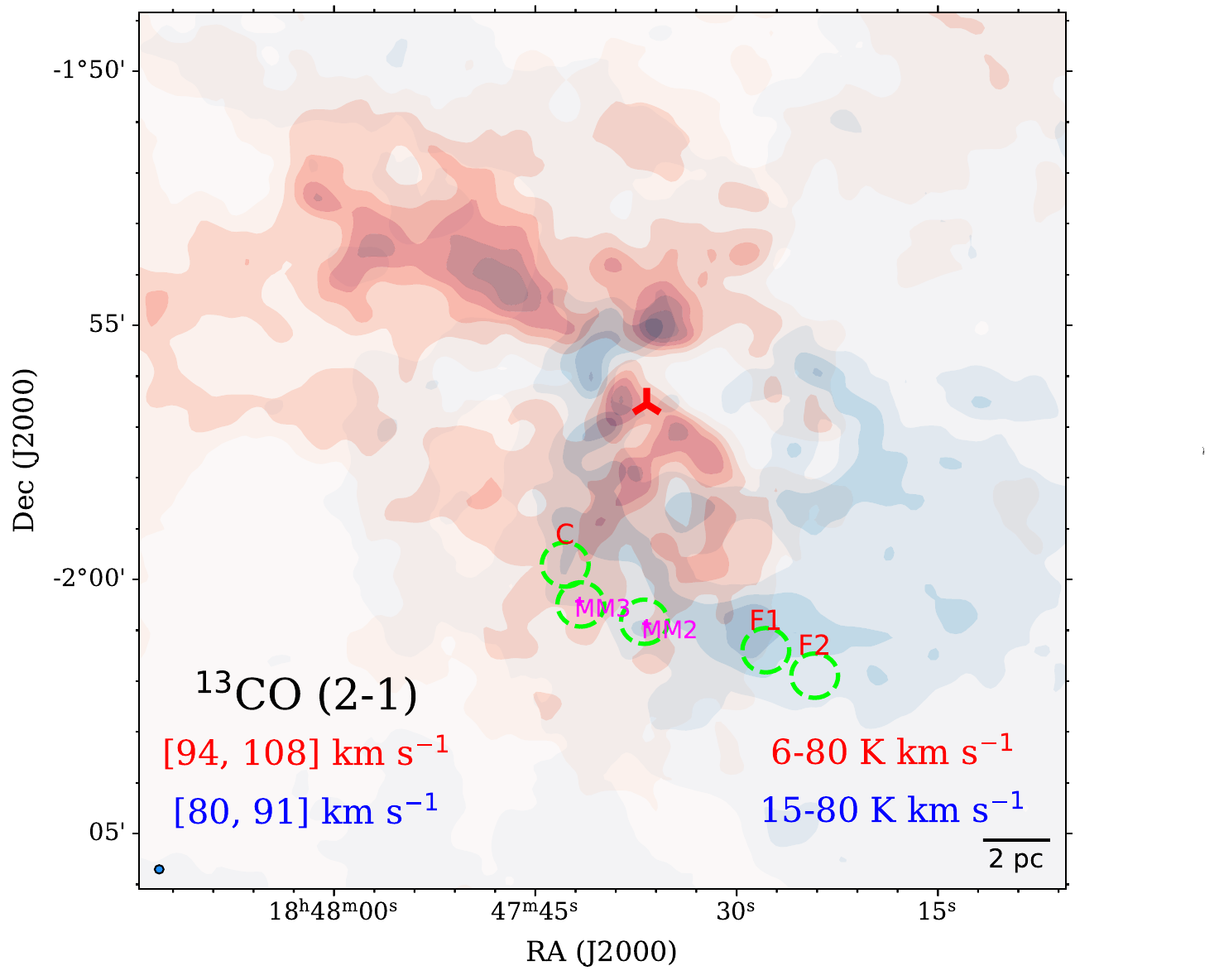}&\hspace{-0.5cm}\includegraphics[scale=0.23]{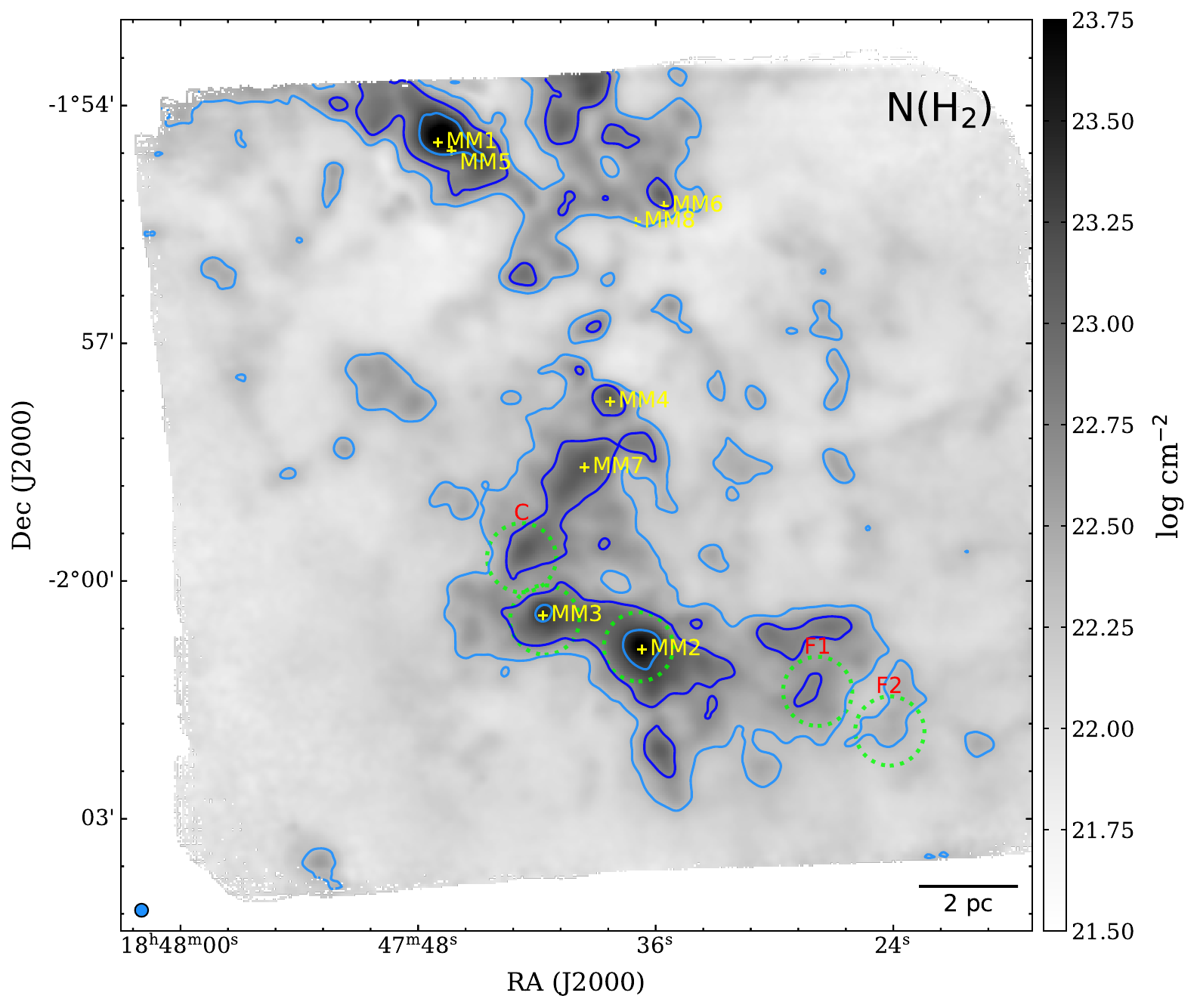}&\hspace{-0.5cm}\includegraphics[scale=0.23]{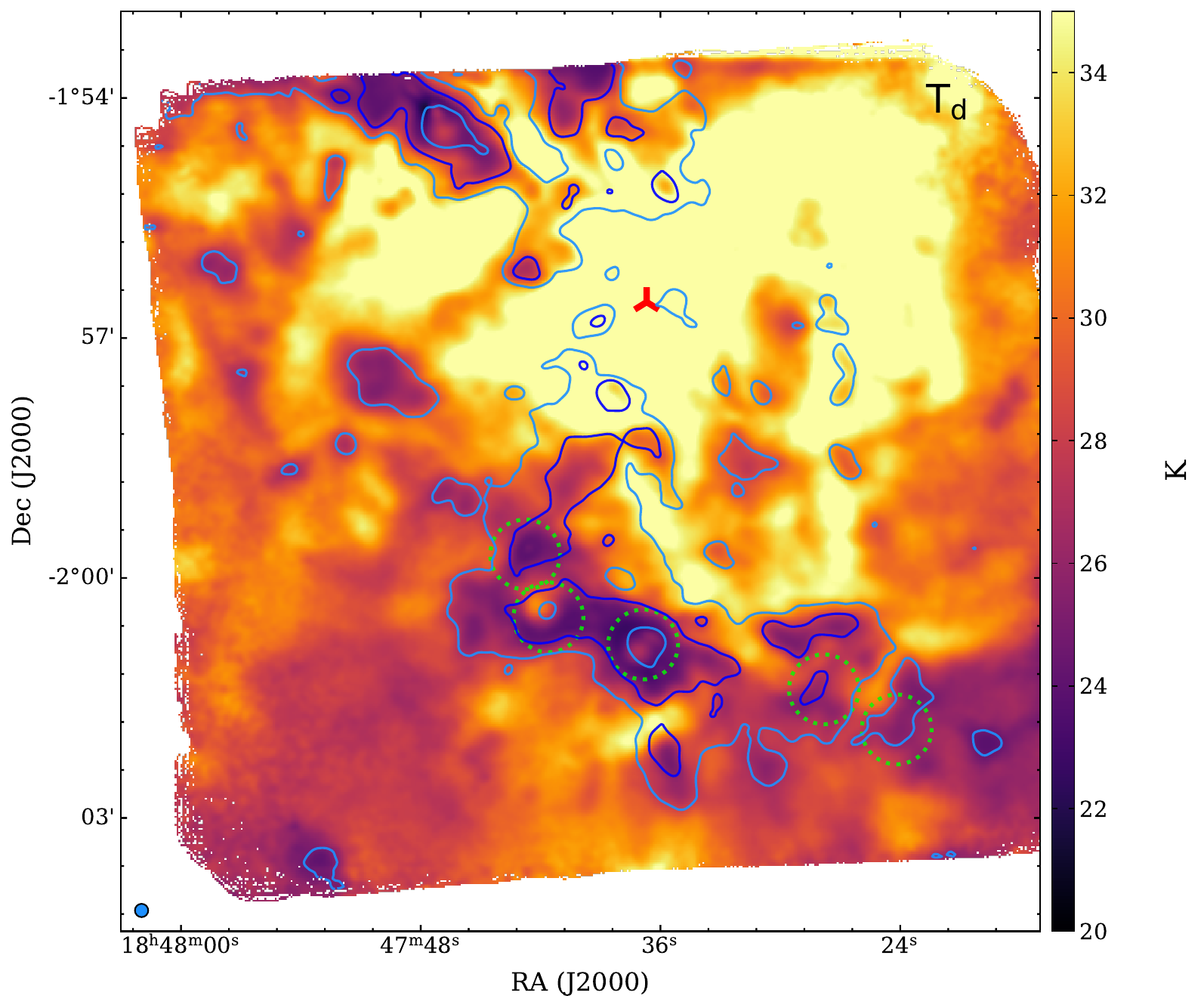}\\
\end{tabular}
\caption{{\emph{Left panel}}: Target clumps overlaid on $^{13}$CO (2-1) (obtained by IRAM 30m as in \citealt{Carlhoff}) integrated intensity maps (red and blue contours are integrated intensities over the respective velocity ranges shown in the figure); {\emph{Middle and right panels: }}Molecular hydrogen column density N(H$_{2}$) and dust temperature map $T_{\mathrm d}$ (>35~K temperatures are truncated for better contrast of the cold regions) of the W43-main molecular complex (10$''$ resolution, \citealt{Lin16}). The three contour levels indicate column densities of 2.2$\times$10$^{22}$, 5$\times$10$^{22}$, and 2$\times$10$^{23}$ cm$^{-3}$ ($\sim$1 g cm$^{-2}$), respectively. The pointed observations towards the 5 selected clumps are shown in green circles of primary beam size (full-width half-maximum) at 96 GHz. In the {\emph Middle panel}, the eight most massive clumps identified in \citet{Motte03} are marked with yellow labels. Other target clumps are marked with red labels, following the names of C, F1 and F2. In the {\emph{Right}} panel, the red three-branched triangle marks the position of the W-R/OB cluster (\citealt{Blum99}).}
\label{fig:w43_npts}
\end{figure*}

We use the short-spacing data from the 30m telescope to generate pseudo-visibilities (task {\textsc{uvshort}}) which were added to the interferometry data. We image the three pointings of clump MM2, MM3 and C, and the two pointings of clump F1 and F2 together as two mosaic fields. We note that we do not utilise Nyquist sampling in the observations (e.g., as shown in Fig. \ref{fig:w43_npts}), so the resultant mosaic fields have a spatially varying noise level. Joint imaging was then performed to obtain two set of combined spectral cubes; again we adopted the Hogbom algorithm to obtain the clean images without providing any masks for the deconvolution process to search for the clean components; plane-specific mask is usually not necessary when short-spacing information is available. In the final step, primary beam correction is applied. The synthesized beam of the final datacube is 4.7$''$ at 96 GHz and varies with frequency in the wide-band dataset. The achieved typical noise level (5$\sigma$) is $\sim$0.05 K. 
Molecular lines of particular interest are listed in Table \ref{tab:3mm_lines_info}, which are all covered by the high-resolution chunks. For all these lines, we use the combined spectral cubes for the analysis. 

\begin{table*}[]
    \centering
     \begin{threeparttable}

    \caption{Properties of the five target clumps in W43-main.}
    \label{tab:source_info}
    \begin{tabular}{cccccccc}
    \toprule

        Source & R.A. & Decl. & Gas Mass$^{a}$ & Luminosity$^{a}$ & $L/M$ &Category&$V_{\mathrm{LSR}}$$^{b}$\\
                &(J2000)&(J2000)& (10$^{2}$ M$_{\odot}$)&(10$^{3}$ L$_{\odot}$)&(L$_{\odot}$/M$_{\odot}$)& & km s$^{-1}$\\
                \midrule
         MM2&18\rah47\ram36\ras.86&-02\decd00\decm49\decs8&38&58&15& 70 $\mu$m bright&91.3\\
         MM3&18\rah47\ram41\ras.61&-02\decd00\decm29\decs5&19&51&26&UCH\textsc{ii}&93.9\\
         F1&18\rah47\ram42\ras.77&-01\decd59\decm42\decs4&3.0&6.6&22&IR dark&89.5\\
         F2&18\rah47\ram27\ras82&-02\decd01\decm23\decs5&2.3&43&19&IR dark&85.5\\
         C&18\rah47\ram42\ras77&-01\decd59\decm42\decs4&7.6&12&16&IR dark&91.7\\
            \bottomrule
     \end{tabular}
        \begin{tablenotes}
      \small
      \item $^{a}$: Gas mass and luminosity are calculated above column density thresholds N$_{\mathrm{thres}}$: for MM2, MM3 and C, $N_{\mathrm{thres}}$ = 5$\times$10$^{22}$ cm$^{-2}$, for F1, $N_{\mathrm{thres}}$ = 4.5$\times$10$^{22}$ cm$^{-2}$, for F2, $N_{\mathrm{thres}}$ = 2.2$\times$10$^{22}$ cm$^{-2}$; the thresholds are chosen based on visual inspection of the N(H$_{2}$) and bolometric luminosity maps (derived from integrating SED profile in a pixel-by-pixel basis) from \citealt{Lin16}.
     \item $^{b}$: The system velocities ($V_{\mathrm{LSR}}$) are with reference to \citet{Urquhart18} based on finding the identical or nearest clumps. 
    \end{tablenotes}
  \end{threeparttable}
\end{table*}

\section{Results}
\subsection{The 3 mm continuum}\label{sec:3mm_cont}
\begin{figure*}
\begin{tabular}{p{0.45\linewidth}p{0.45\linewidth}}
   \hspace{-0.5cm}\includegraphics[scale=0.35]{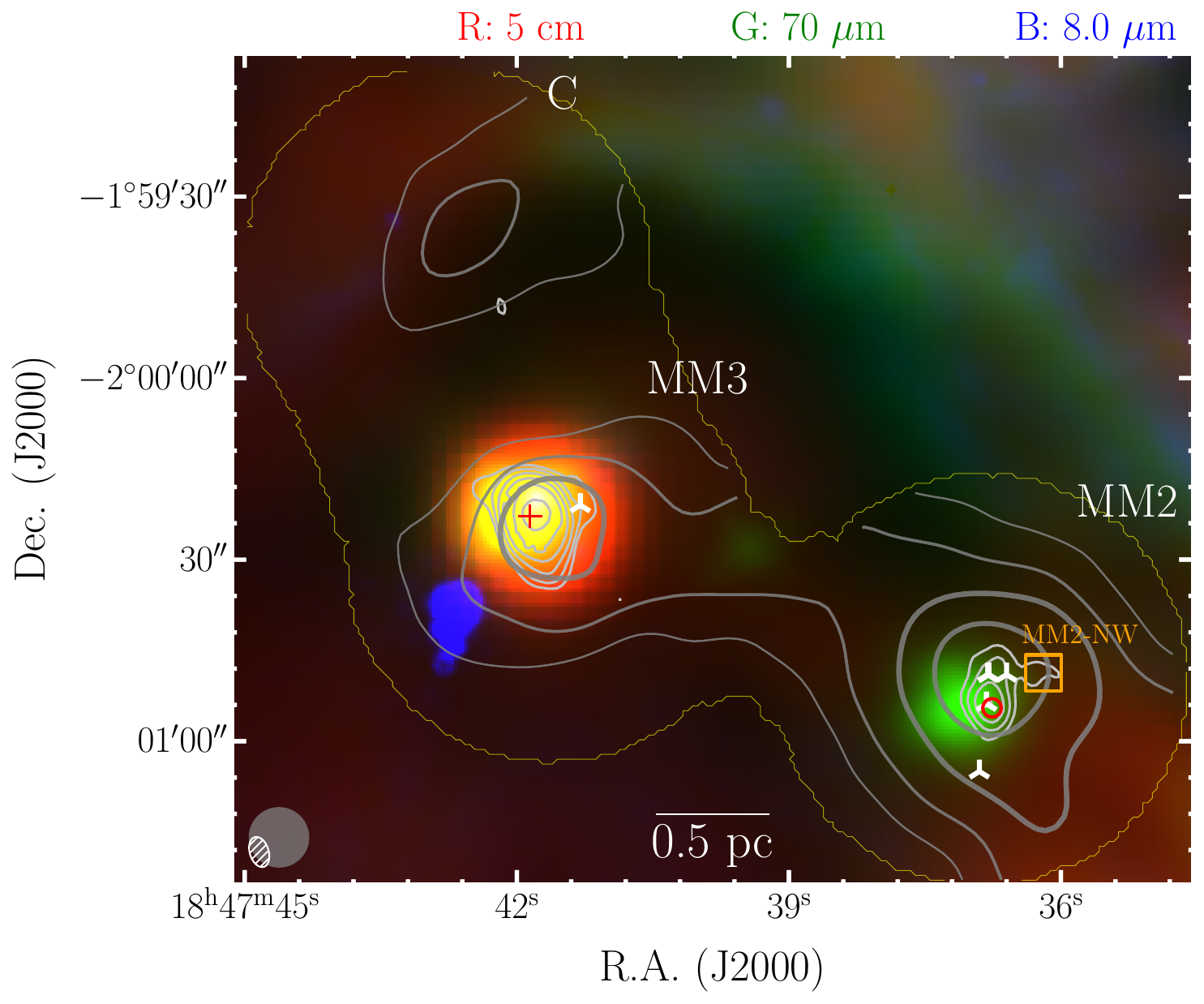}&\hspace{0.3cm}\includegraphics[scale=0.35]{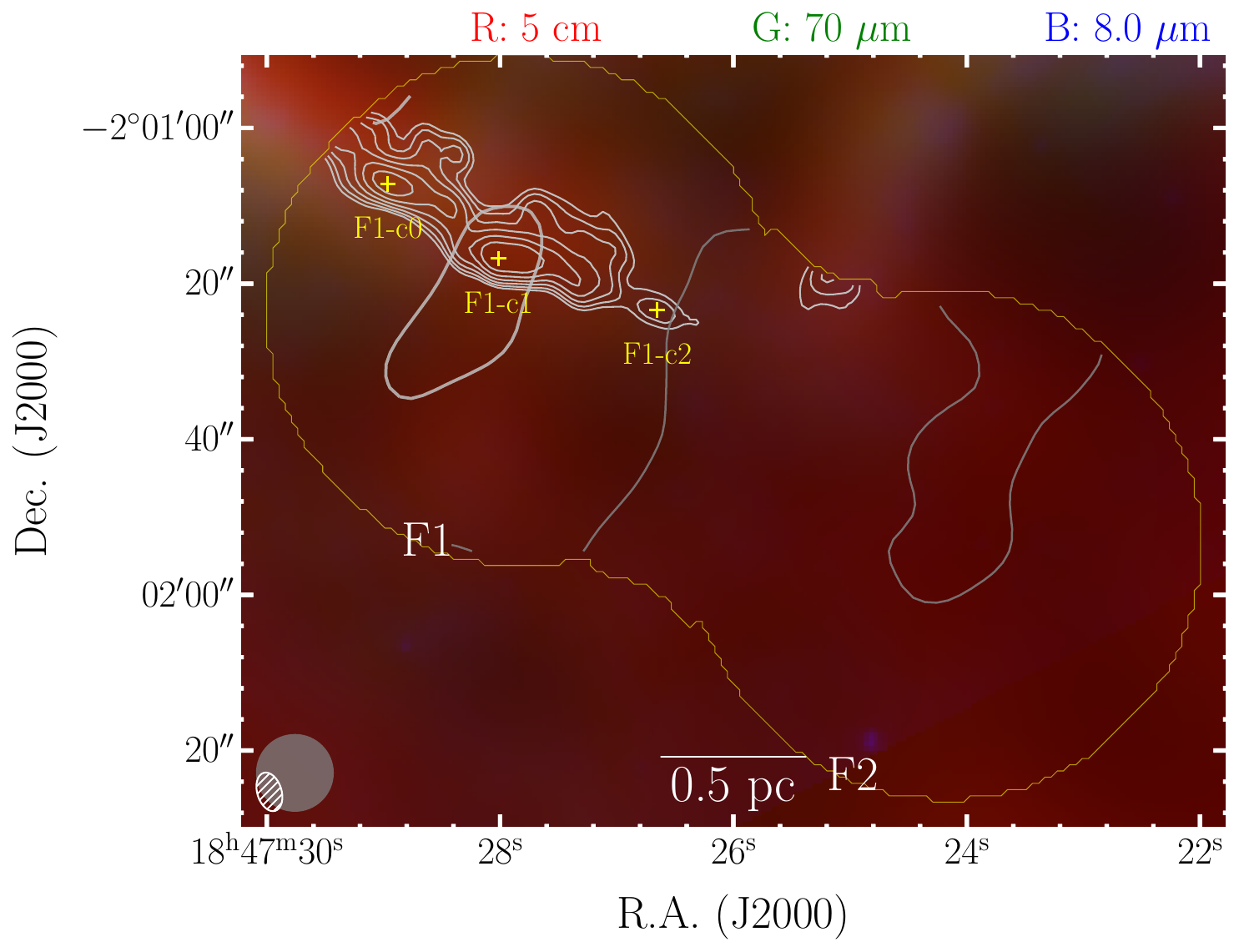}\\
    \end{tabular}
    \caption{Three-color images of the clumps (red: GLOSTAR 5 cm radio continuum; green: {\emph Herschel} Hi-GAL PACS 70 $\mu$m; blue: {\emph Spitzer} 8.0 $\mu$m), overlaid with contour levels of NOEMA 3 mm continuum emission (white contours) and $N(\mathrm{H_{\mathrm 2}})$ (gray contours). 
    The beams for the 3 mm continuum and the $N(\mathrm{H_{\mathrm 2}})$ map (10$''$) are shown in bottom left corner. The outer yellow solid lines indicate the NOEMA field-of-view of the observations. 
    {\emph Left}: Target clump C, MM2 and MM3. The UCH\textsc{ii} region (red cross) and OH maser (red circle), methanol masers (white three-branched triangles), and the northwestern substructure in clump MM2 (orange rectangle) are marked.  Contour levels of 3 mm continuum are logarithmic-spaced from 2$\sigma_{\mathrm{v}}$ ($\sigma_{\mathrm{v}}$ = 1.6 mJy~beam$^{-1}$) to 225.0 mJy~beam$^{-1}$ with 5 intervals. Contour levels of the $N(\mathrm{H_{\mathrm 2}})$ map have values of log$_{10}$($N(\mathrm{H_{\mathrm 2}})$) = [22.75,  23.0, 23.25, 23.5] cm$^{-2}$. 
    {\emph Right}: Target clump F1 and F2.  Contour levels of 3 mm continuum are logarithmic-spaced from 2$\sigma_{\mathrm{v}}$ ($\sigma_{\mathrm{v}}$ = 0.4 mJy~beam$^{-1}$) to 3.3 mJy~beam$^{-1}$ with 5 intervals. The three compact sources are marked with yellow pluses. Contour levels of the $N(\mathrm{H_{\mathrm 2}})$ map have values of 10$^{[22.4,  22.7]}$ cm$^{-2}$.}
    \label{fig:rgb_cont}
\end{figure*}

The 3 mm continuum maps of the five target clumps are shown in Figure \ref{fig:rgb_cont}, overlaid on RGB maps of 8\,$\mu$m (1.9$''$, \citealt{Carey09}), 70\,$\mu$m (6$''$, \citealt{Molinari10}) and 5\,cm continuum (18$''$, GLOSTAR survey, \citealt{AB21}). Clumps MM2 and MM3 are resolved with one dominant compact source in the center. These two sources are slightly elongated along the north-south direction. For clump MM2, there is an adjacent substructure to the northwest of the central compact source, marked as MM2-NW in the figure. The location of this substructure coincides well with a series of cores revealed by ALMA observations at 1.3 mm at 0.01-pc angular resolution (\citealt{Cortes19}, \citealt{Pouteau22}). For clump C and F2, there is no robust compact source detection towards the phase center. Clump F1 exhibits a filamentary structure of $\sim$0.2 pc in width, extending up to $\sim$1.2 pc in length. We marked the three compact sources along the filament of F1 as F1-c0, F1-c1 and F1-c2 (Figures \ref{fig:rgb_cont}). However, these emission structures in F1 were not fully covered by our observations, truncating at the edge of the primary beam. 


Clump MM3 is associated with an ultra-compact H\textsc{ii} (UCH\textsc{ii}) region, and it is the brightest UCH\textsc{ii} in W43 (\citealt{Bally10}). Assuming optically thin free-free emission with a spectral index $\alpha$ = 0.1, the intensity varies with frequency as $\nu^{-0.1}$. The free-free emission contribution to the 96 GHz band (3 mm) can be approximated by re-scaling the VLA X-band (8.46 GHz) continuum map \footnote{Obtained from NRAO archive: http://www.aoc.nrao.edu/~vlbacald/ArchIndex.shtml} with (96/8.46)$^{-0.1}$. With preprocessing of the X-band image including convolution and regridding, the free-free contribution (integrated total flux, $S_{\mathrm{int\,free-free}}$ = 0.24 Jy) to the 3 mm continuum was subtracted in a pixel by pixel manner; the percentage of free-free emission to the total flux at 96 GHz is $\sim$34$\%$.  

The integrated fluxes at 3 mm, peak intensity and source size of the three clumps (MM2, MM3 and F1) estimated from 3 mm continuum are listed in Table \ref{tab:3mm_cont}. Assuming an average dust temperature of 30 K for MM2 and F1, and 100 K for MM3 (determined by the gas temperature in Sect.~\ref{sec:temp_dens}), and $\kappa_{\mathrm{100\,GHz}}$ = 0.245 cm$^{-2}$ g$^{-1}$, a gas-to-dust ratio of 100, we derived core masses from 3 mm continuum, which are listed in Table \ref{tab:3mm_cont}. 

We additionally made an estimate on the dynamical age ($t_{\mathrm{dyn}}$) of the UCH\textsc{ii} region embedded in clump MM3. We first estimated the number of Lyman continuum photons, $N_{\mathrm{uv}}$, with the equation (\citealt{Mezger67}, \citealt{Rubin68}):
\begin{equation}
    N_{\mathrm{uv}} = 7.5\times10^{46}\,\left(\frac{S_{\nu}}{\mathrm{Jy}}\right)\left(\frac{D}{\mathrm{kpc}}\right)\left(\frac{T_{\mathrm e}}{10^{4}\,\mathrm K}\right)^{-0.45}\left(\frac{\nu}{\mathrm{GHz}}\right)^{0.1},\label{eq:nuv}
\end{equation}
in which $S_{\nu}$ is the integrated flux, and $D$ the distance, $T_{\mathrm e}$ the electron temperature, and $\nu$ the frequency.
$S_{\nu}$ is obtained by summing up regions of flux densities $>$5 sigma with the VLA 8.46 GHz image, of $\sim$0.29 Jy. 
We assume $T_{\mathrm e}$ as a typical value of 7300 K (\citealt{Tremblin14}) considering the galacto-centric
distance of W43, which is 4.5 kpc (\citealt{Zhang14}). 
The $N_{\mathrm{uv}}$ is calculated to be 6.6$\times$10$^{47}$ s$^{-1}$. The spectral type of the powering source based on $N_{\mathrm{uv}}$ is likely O9.5-B0 ZAMs star (\citealt{Panagia}). 

With $N_{\mathrm{uv}}$, the dynamical timescale of the H\textsc{ii} region, t$_{\mathrm{dyn}}$, can be computed following (\citealt{Spitzer78}, \citealt{Dyson97}), 
\begin{equation}
    t_{dyn} = \left(\frac{4R_{s}}{7c_{s}}\right)\left[\left(\frac{R_{H\textsc{ii}}}{R_{s}}\right)^{7/4}-1\right],\label{eq:tdyn_hii}
\end{equation}
where $c_{s}$ is the isothermal sound speed in the ionized gas, $\sim$10-11 km s$^{-1}$ (\citealt{Bisbas09}). $R_{\mathrm{H\textsc{ii}}}$ is the radius of the H\textsc{ii} region, and $R_{\mathrm s}$ is the radius of the Str\"{o}mgen sphere. Based on $\pi$$R_{\mathrm{H\textsc{ii}}}^{2}$ = $A$, in which $A$ is the $>$5 $\sigma$ area from the 8.46 GHz image, $R_{\mathrm{H\textsc{ii}}}$ is estimated to be 0.075 pc. $R_{s}$ is calculated from $R_{\mathrm s}$ = $\left(\frac{3N_{\mathrm{uv}}}{4\pi n_{0}^{2}\alpha_{B}}\right)^{1/3}$ in which $\alpha_{\mathrm B}$ is the radiative recombination coefficient; with $T_{e}$ = 7300 K, $\alpha_{\mathrm B}$ = 3.3$\,\times\,$10$^{-13}$ cm$^{3}$ s$^{-1}$ (\citealt{Spitzer78}). For $n_{0}$ which represents initial gas density, we use a typical value of 10$^{5}$ cm$^{-3}$ for UCH\textsc{ii} region, which is close to the average gas density estimated from the 3 mm continuum (of 10$^{5.5}$ cm$^{-3}$, using a gas mass of 900 $M_{\odot}$ and a radius of 0.20 pc with spherical assumption). 
$t_{\mathrm{dyn}}$ is estimated to be $\sim$0.016 Myr, which is at the lower range of the typical lifetime for UC H\textsc{ii} regions (of 0.01-0.1 Myr, \citealt{Churchwell02}, \citealt{Mclow07}).

\begin{table}[]
\begin{threeparttable}
 \caption{Line parameters for the 3 mm molecular lines used in the analysis.}
    \label{tab:3mm_lines_info}
    \begin{tabular}{ccccc}
    \toprule
    Transition & Rest frequency & $E_{\mbox{\scriptsize{up}}}$ & $n_{\mbox{\scriptsize{crit}}}$$^{b}$ \\
               &(GHz)&(K)&(cm$^{-3}$)\\
                \midrule
         
         HC$_{3}$N (9-8)&81.881&19.6&5.6$\times$10$^{4}$ $^{c}$\\
         CS (2-1)&97.980&7.1&8.9$\times$10$^{4}$\\
         H$^{13}$CO$^+$ (1-0)&86.754&4.2&3.5$\times$10$^{4}$\\
         CH$_{3}$OH 2$_{-1,1}$-1$_{-1,0}$ $E$&96.755&28.0&2.2$\times$10$^{5}$\\
         CH$_{3}$CCH 5$_{0}$-4$_{0}$&85.457&12.3&2.8$\times$10$^{3}$\\
         CH$_{3}$CCH 6$_{3}$-5$_{3}$&102.530&82.0&4.0$\times$10$^{3}$\\
         CCH 1$_{2,2}$-0$_{1,1}$ &87.407&4.2&9.3$\times$10$^{3}$\\
         SO 3$_{2}$-2$_{1}$&99.299&9.2&4.3$\times$10$^{4}$\\
         SiO (2-1)&86.846&6.3&4.8$\times$10$^{4}$ $^{b}$\\
        H$_{2}$CS 3$_{0,3}$-2$_{0,2}$&103.040&9.9&3.0$\times$10$^{4}$ $^{b}$\\
        H$_{2}$CS 3$_{2,1}$-2$_{2,0}$&103.051&62.6&1.1$\times$10$^{4}$ $^{b}$\\
        NH$_{2}$D 1$_{1,1}$-1$_{0,1}$&85.926&20.7&4.3$\times$10$^{4}$ $^{b}$\\
            \bottomrule
     \end{tabular}
        \begin{tablenotes}
      \item Note: a: Line frequencies are taken from Cologne Database for Molecular Spectroscopy (CDMS) database (\citealt{Endres16}). 
     \item b: The critical density $n_{\mathrm{crit}}$ is calculated assuming a gas kinetic temperature of 30 K with no optical depth correction.
      \item c: The references of collisional rates for HC$_{3}$N is \citealt{Faure16}, for SiO is \citealt{Balanca18}, for H$_{2}$CS is \citealt{Wiesenfeld13}, and for NH$_{2}$D \citealt{Daniel14}. 
\end{tablenotes}
  \end{threeparttable}
\end{table}

\begin{table*}
\centering
     \begin{threeparttable}
     \caption{Source properties from 3 mm continuum emission.}
    \label{tab:3mm_cont}
    \begin{tabular}{cccccccc}
    \toprule
      Source & R.A.$^{a}$&Decl.$^{a}$&$F_{\mathrm{int}}^{b}$& $I_{\mathrm{peak}}^{b}$ & Size$^{c}$&$\sigma_{\mathrm{v}}$& Gas Mass \\
                &(J2000)&(J2000)&(Jy)&(mJy arcsec$^{-2}$)& ($''$)&(mJy~beam$^{-1}$)&($M_{\odot}$)\\
                \midrule
         MM2&18\rah47\ram36\ras.80&-02\decd00\decm54\decs0&0.051&1.90&4.28&1.6&200\\
         MM3$^{d}$&18\rah47\ram41\ras.79&-02\decd00\decm21\decs7&0.463&7.40&7.75&1.6&900\\
         F1&18\rah47\ram27\ras.84&-02\decd01\decm17\decs0&0.029&0.24&7.93&0.4&350\\
            \bottomrule
     \end{tabular}
    \begin{tablenotes}
      \item a: The coordinates correspond to the position of the peak intensity.
      \item b: The integrated intensity (above 5$\sigma$ emission level) and peak intensity.
      \item       c: Radius R of the emission area A of $>$5$\sigma$, A = $\pi$R$^{2}$.
      \item       d: After subtraction of free-free emission.
    \end{tablenotes}
  \end{threeparttable}
\end{table*}

\begin{table*}[]
\footnotesize{\hspace{-.75cm}
     \begin{threeparttable}
    \caption{Information of lines used as thermometers.}
    \label{tab:ther_dens_info}
  \begin{tabular}{lllcccccccccc}
    \hline
\multirow{3}{*}{Molecule}&\multirow{3}{*}{Transition} & \multirow{3}{*}{E$_{\mbox{\scriptsize{up}}}$}&\multicolumn{2}{c}{C}&\multicolumn{2}{c}{MM3}&\multicolumn{2}{c}{MM2}&\multicolumn{2}{c}{F1}&\multicolumn{2}{c}{F2}\\
        &&&$T_{\mbox{\scriptsize{mb}}}$&$\Delta V$&$T_{\mbox{\scriptsize{mb}}}$&$\Delta V$&$T_{\mbox{\scriptsize{mb}}}$&$\Delta V$&$T_{\mbox{\scriptsize{mb}}}$&$\Delta V$&$T_{\mbox{\scriptsize{mb}}}$&$\Delta V$\\
        &&(K)&(K)&(km s$^{-1}$)&(K)&(km s$^{-1}$)&(K)&(km s$^{-1}$)&(K)&(km s$^{-1}$)&(K)&(km s$^{-1}$)\\
                \hline
         \multirow{2}{*}{CH$_{3}$CCH}&5(0)-4(0)&12.3&0.75&5.18, 2.36&1.80&1.88, 4.29&2.4&5.20&0.38&2.34, 1.9&0.20&3.5\\
         &6(3)-5(3)&81.5&0.1&&0.65&&1.3&&$-$&&$-$&\\
         \hline
        \multirow{2}{*}{H$_{2}$CS}&3(1,3)-2(1,2)&22.9&0.65,0.62&3.41,3.20&3.0&2.33,5.19&5.50&4.20,11.70&0.20,0.12&2.18,5.69&0.10&2.0\\
        &3(0,3)-2(0.2)$^{a}$&9.89&0.60,0.40&&1.78&&3.80&&0.15,0.08&&0.08\\
        &3(2,2)-2(2,1)$^{a}$&62.6&&&&&&&&&\\
        &3(2,1)-2(2,0)&62.6&$<$0.05&&0.20&&1.0&&$<$0.05&&$<$0.05\\
        \hline
       \multirow{4}{*}{CH$_{2}$CHCN$^{b}$}&9(3,7)-8(3,6)&39.9&&&&&0.80&5.89&&\\
       &9(6,3)-8(6,2)&98.2&&&&&0.75&&&&\\
       &9(3,6)-8(3,5)&39.9&&&&&0.80&&&&\\
       &9(7,2)-8(7,1)&126.2&&&&&0.30&&&&\\
       \hline
            \hline
     \end{tabular}
        \begin{tablenotes}
      \small
      \item{Note: The peak intensity ($T_{\mathrm{mb}}$) and corresponding line-width ($\Delta V$ = 2.355$\sigma_{\mathrm{v}}$) are shown for each source. Whenever there are two velocity components, both line-widths are listed while the peak intensity is contributed from both components.}
      \item{For CH$_{3}$CCH only two representative transitions are shown whilst for both (5-4) and (6-5) line series all the k ladders are used in the fits, whenever robustly detected.}
      \item{a. The two lines blend with each other. The listed line parameters are contributed from both lines.}
      \item{b. Vinyl Cyanide (CH$_{2}$CHCN) is only detected towards the center of clump MM2.}
      \end{tablenotes}
      \label{tab:line_peak}
\end{threeparttable}
}
\end{table*}

\subsection{Gas temperature and density distribution from LTE and non-LTE modeling}\label{sec:temp_dens}

To characterise the physical properties of the dense gas inside the clumps, we first estimate the rotational temperature, $T_{\mathrm{rot}}$ and hydrogen volume density $n(\mathrm{H_{\mathrm2}})$. For this purpose, we adopt the CH$_{3}$CCH (6-5) and (5-4) ladders as thermometers and CH$_{3}$OH (2-1) $A/E$ lines as densitometers. 

CH$_{3}$CCH is widely distributed in lukewarm gas of relatively evolved massive star-forming clumps (\citealt{Molinari16b}, \citealt{Giannetti17}) and is an ideal tracer of the bulk gas structures ($\gtrsim$0.1-1\,pc, \citealt{Lin22a}). 
Whenever detected, we also utilised H$_{2}$CS (3-2) and CH$_{2}$CHCN (9-8) lines which have multiple transitions to derive $T_{\mathrm{rot}}$. Compared to the emission of H$_{2}$CS and CH$_{3}$CCH lines, emission of CH$_{2}$CHCN lines is confined within 0.1 pc in the central region of MM2. 
The line parameters of these transitions and the properties of the observed lines at the position of peak emission of the five clumps are listed in Table \ref{tab:ther_dens_info}.

Line series of CH$_{3}$OH have very different critical densities of different $k$ components, hence they compose an ideal densitometer of relatively high dynamical range, especially for the dense gas ($\gtrsim$10$^{5}$~cm$^{-3}$, \citealt{Leurini04, Leurini07}, \citealt{Lin22a}).

\subsubsection{Deriving the gas rotational temperature from thermometer lines}\label{sec:xclass}
\begin{figure*}
\begin{tabular}{p{0.3\linewidth}p{0.3\linewidth}p{0.3\linewidth}}
 \includegraphics[scale=0.35]{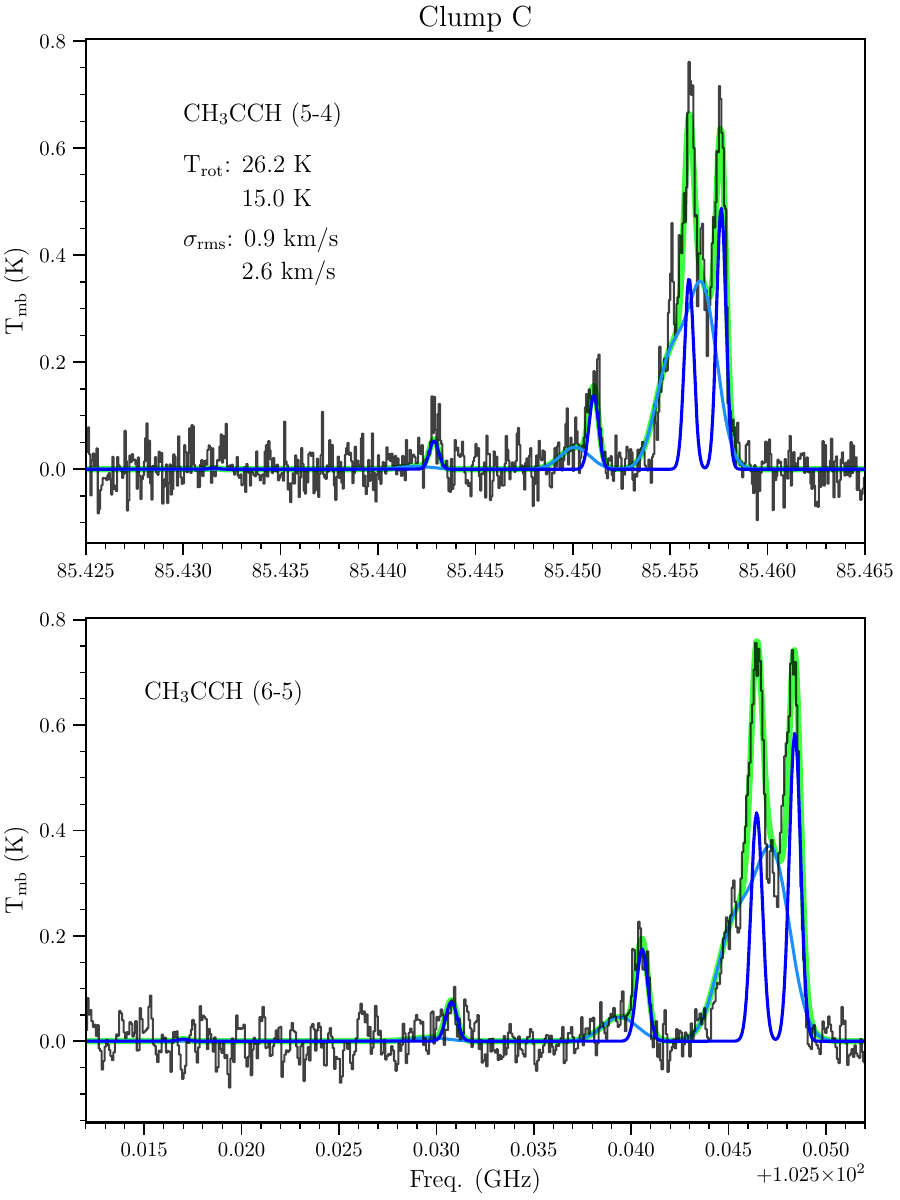}&\hspace{-0.cm}\includegraphics[scale=0.35]{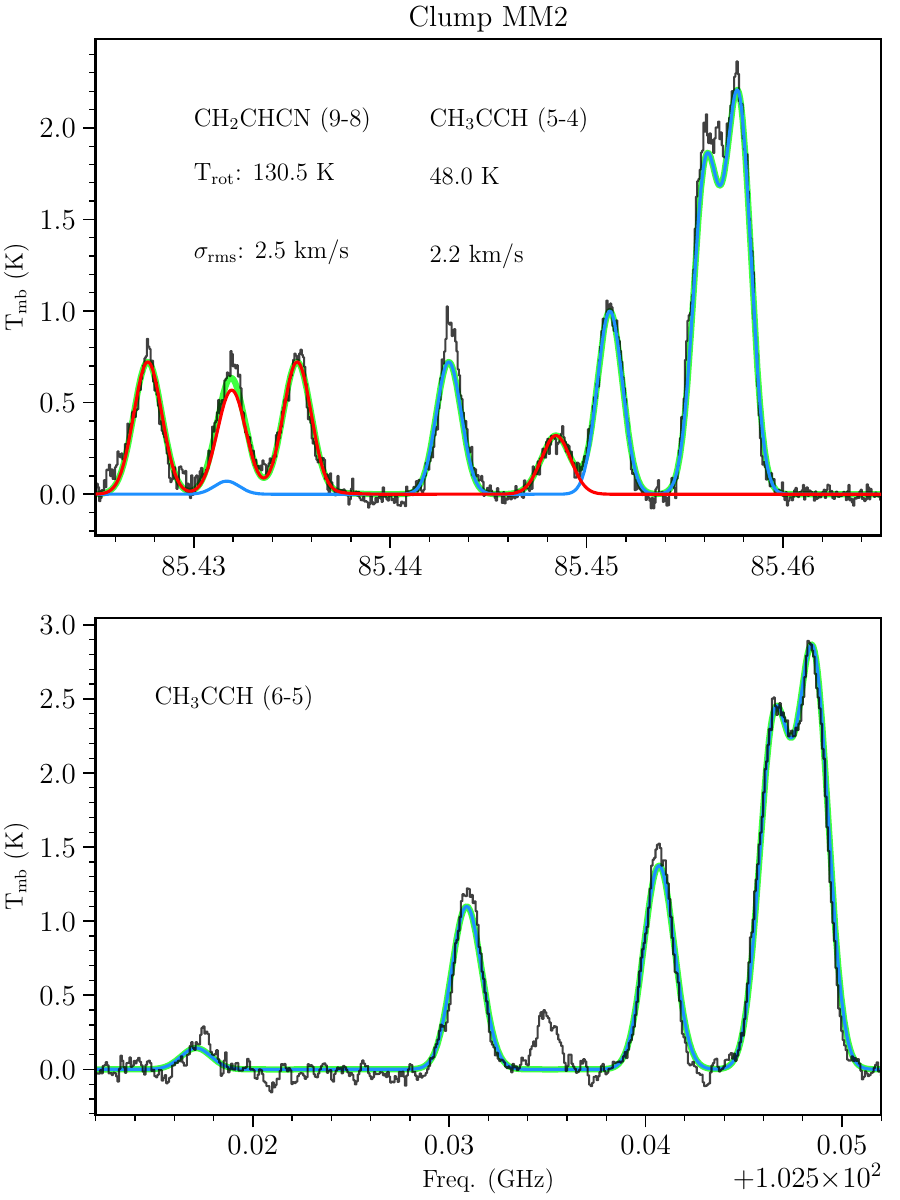}&
 \includegraphics[scale=0.35]{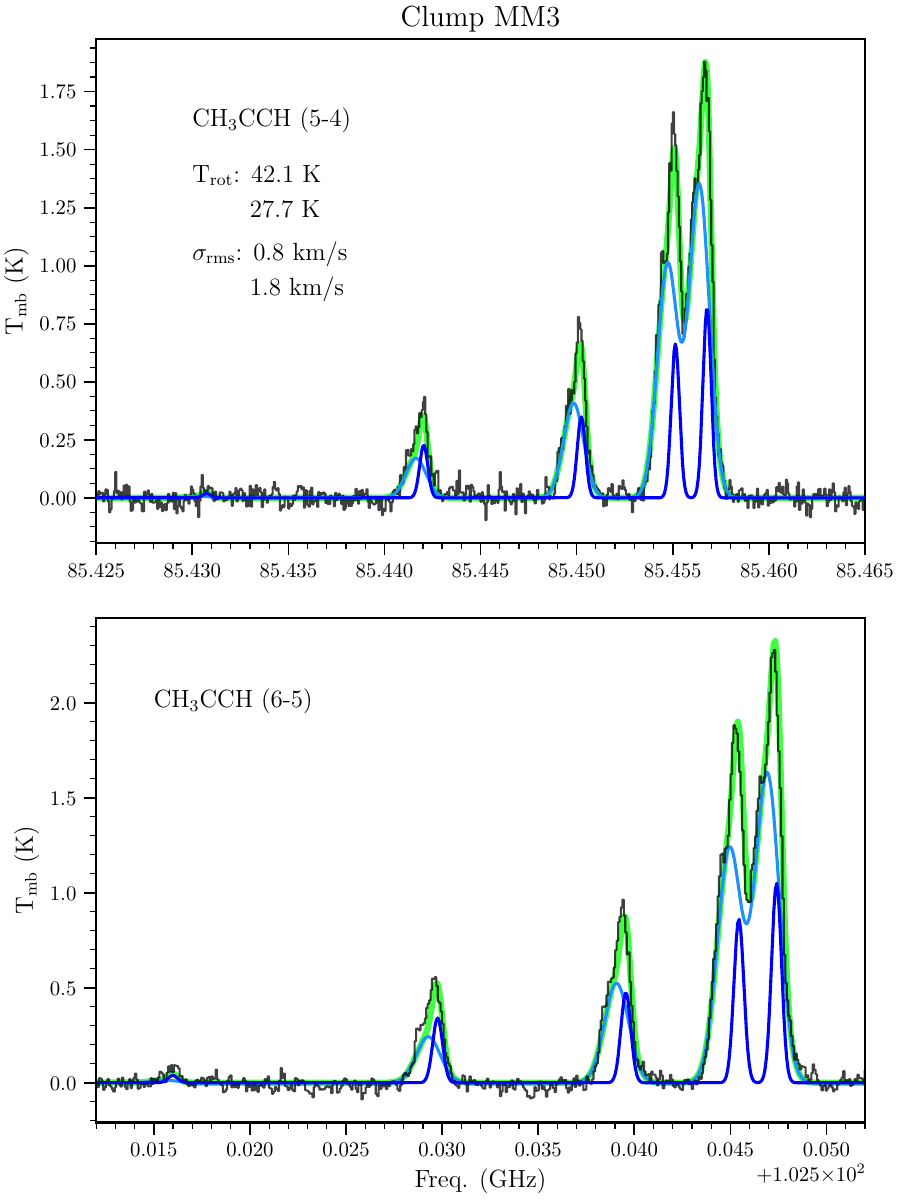}\\
\end{tabular}
\caption{CH$_{3}$CCH spectra at the peak emission of clumps C, MM2, and MM3. For clump MM2, the CH$_{2}$CHCN emission is detected, with red line showing the fitting result. In all plots, whenever there are more than one velocity component, both the composite line profile of the fits (green line) and the individual fit for each component (blue lines) are shown. The parameters of the peak intensity ($T_{\mathrm{mb}}$) and linewidth ($\Delta V$) are listed in Table \ref{tab:line_peak}.}
\label{fig:ch3cch_cen_sps}
\end{figure*}

\begin{figure*}
\begin{tabular}{p{0.15\linewidth}p{0.3\linewidth}p{0.3\linewidth}p{0.15\linewidth}}
&\includegraphics[scale=0.35]{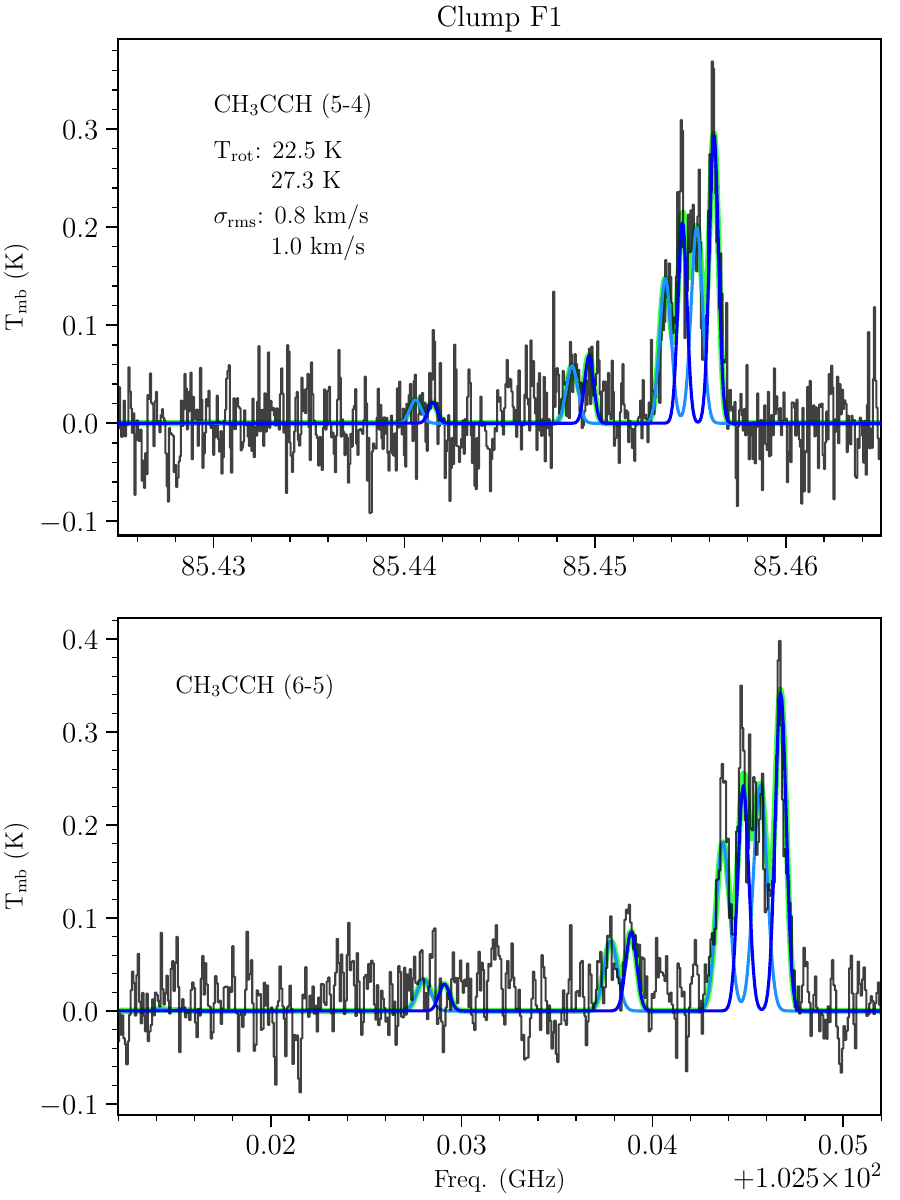}&\includegraphics[scale=0.35]{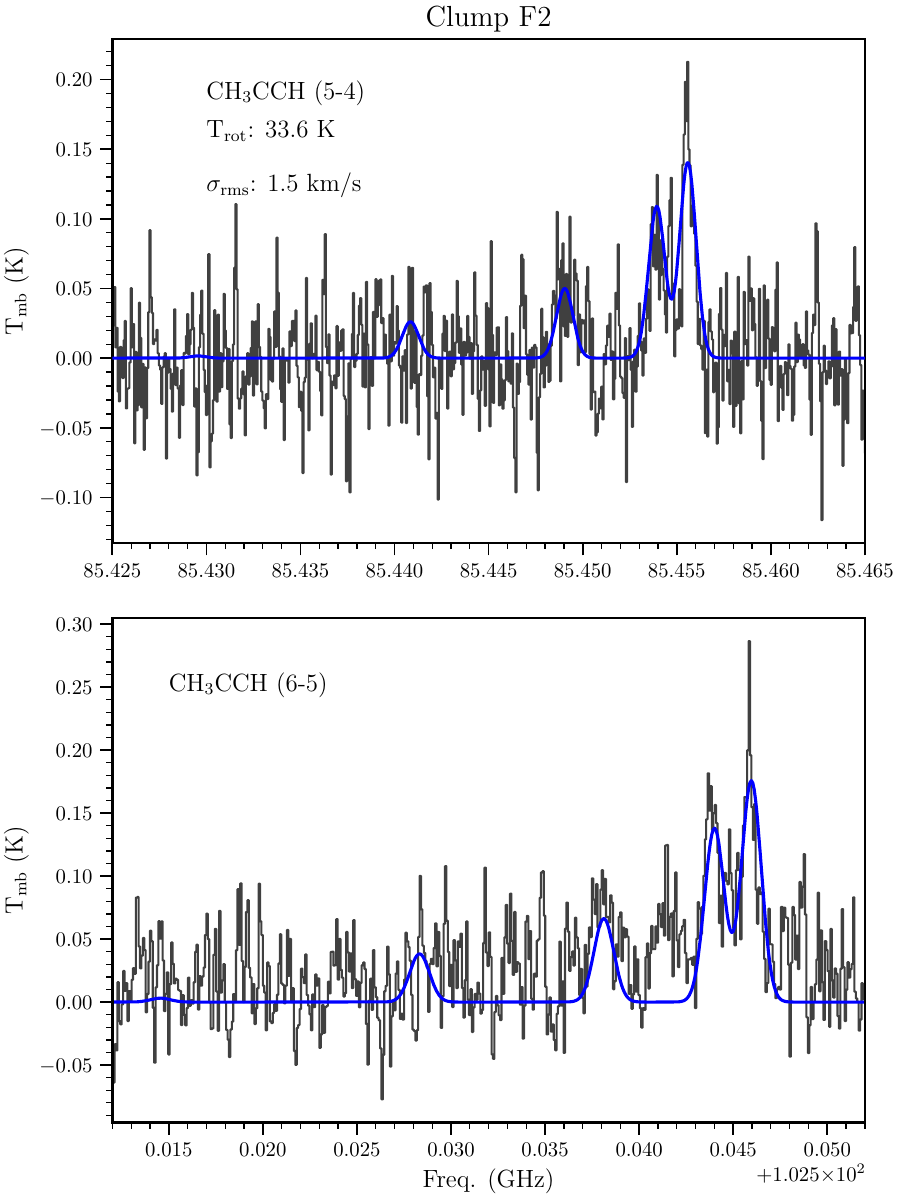}&\\
\end{tabular}
\caption{Same as Figure \ref{fig:ch3cch_cen_sps}, but for clumps F1 and F2.}
\label{fig:ch3cch_cen_sps1}
\end{figure*}

\begin{figure*}
\includegraphics[scale=0.45]{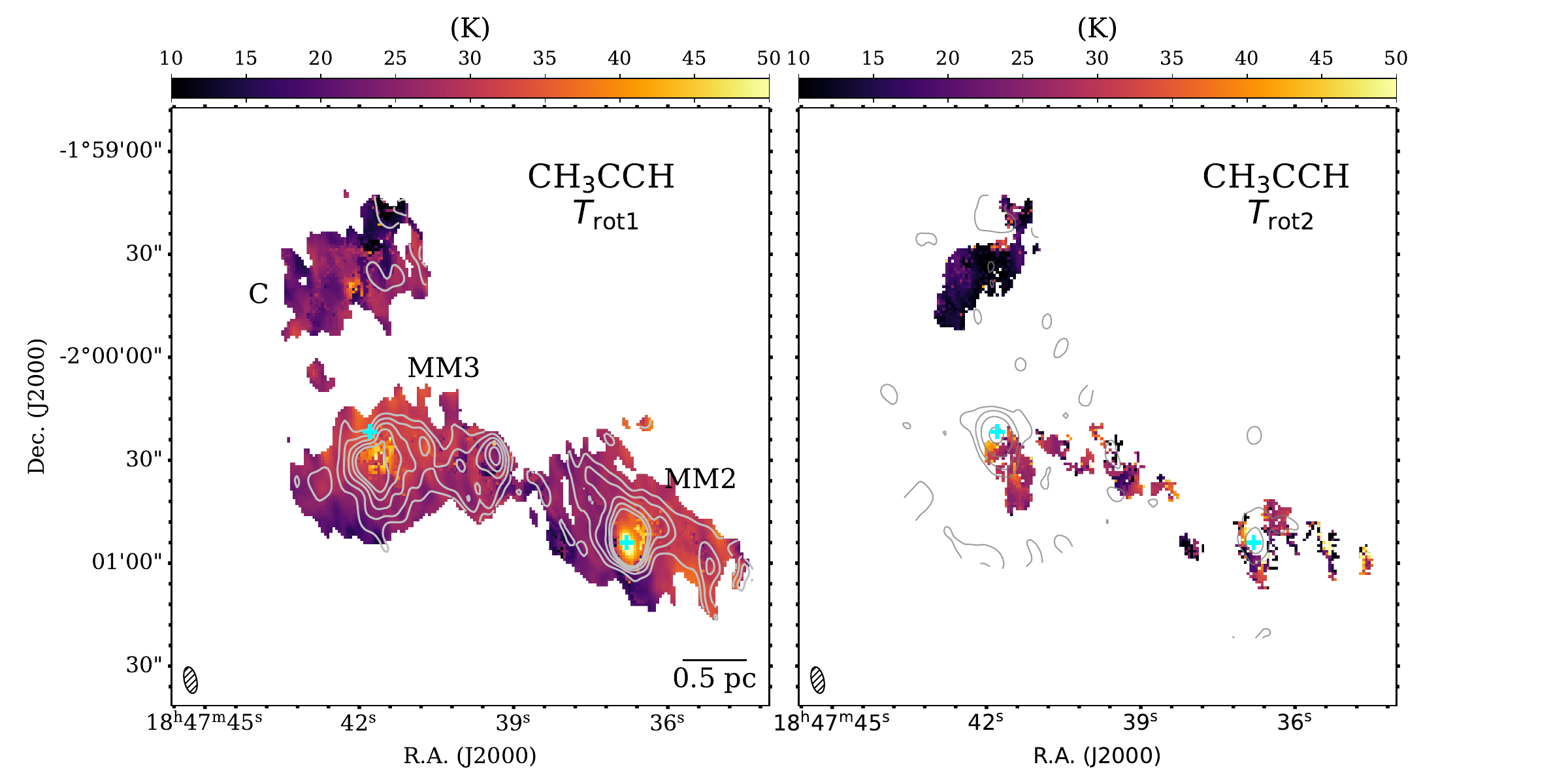}
\caption{Rotational temperature maps derived by CH$_{3}$CCH (6-5) and (5-4) lines. The left and right panel show the temperatures from the two-component LTE models separately. In the left panel the contours represent integrated intensity map of CH$_{3}$CCH 5$_{2}$-4$_{2}$ line. The velocity range for integration is 80-100 km s$^{-1}$, and the contour levels are from 1.0 K km s$^{-1}$ (5 $\sigma$) to 2 K km s$^{-1}$ (0.3 times peak emission value of MM2) with 5 uniform intervals. In the right panel, the contours represent the 3 mm continuum emission, and are logrithmic-spaced from 2$\sigma_{\mathrm{v}}$ ($\sigma_{\mathrm{v}}$ = 1.6 mJy~beam$^{-1}$) to 225.0 mJy~beam$^{-1}$ with 5 intervals (same as Figure \ref{fig:rgb_cont}). In both plots, the cyan crosses indicate the position of the peak intensity of 3 mm emission.}
\label{fig:ch3cch_2comp_trot}
\end{figure*}

To estimate $T_{\mathrm{rot}}$ of the aforementioned thermometer lines, we utilise the XCLASS package (\citealt{Moeller17}) to establish single-component LTE models in a pixel by pixel manner. We keep the parameters of molecular column density, $N_{\mathrm{mol}}$, rotational temperature $T_{\mathrm{rot}}$ and line-width $\Delta V$, and centroid velocity $V_{\mathrm{LSR}}$ as free parameters in the fitting (Figs. \ref{fig:ch3cch_cen_sps}-\ref{fig:ch3cch_cen_sps1}). 
We visually inspect the datacube first and find that there are subregions where more than one velocity component are present, we therefore use LTE model composed of two independent velocity components to describe the line emission. These subregions are identified by comparing the Akaike Information Criterion (AIC) of the one-component and two-component best-fit models. 
Such regions include the central regions of clumps MM3, C, and F1. After performing the fits, the set of two-component parameter maps are generated as follows: 
We examine the $V_{\mathrm{c}}$ map composed of only pixels where one-component model applies, with the pixels of two-component fits initially masked, and then calculate a weighted (by distance) centroid velocity from adjacent unmasked pixels to interpolate their masked neighboring pixels. The velocity values of these neighboring pixels are filled with the one of centroid velocities from the two-component model that is closer to the weighted velocity. The procedure is done progressively until all masked pixels with valid two-component fits are filled and produces two velocity component maps. The other parameter maps are generated correspondingly. The maps of the fitted $T_{\mathrm{rot1}}$ and $T_{\mathrm{rot2}}$, are shown in Figure \ref{fig:ch3cch_2comp_trot}.

For clump MM3 and C we resolve a narrow velocity component of hotter gas and a broad velocity component of relatively cold gas (Figure \ref{fig:ch3cch_cen_sps}).  Clump MM2 also exhibits two-component velocity features around the region of the most prominent emission. For clump F1, it seems there are more than two velocity components (Figure \ref{fig:ch3cch_cen_sps1}), however, limited by the achieved sensitivity, especially for the higher $K$ ladder of these lines, a model of more than two velocity components cannot be established robustly.

\begin{figure*}
\includegraphics[scale=0.45]{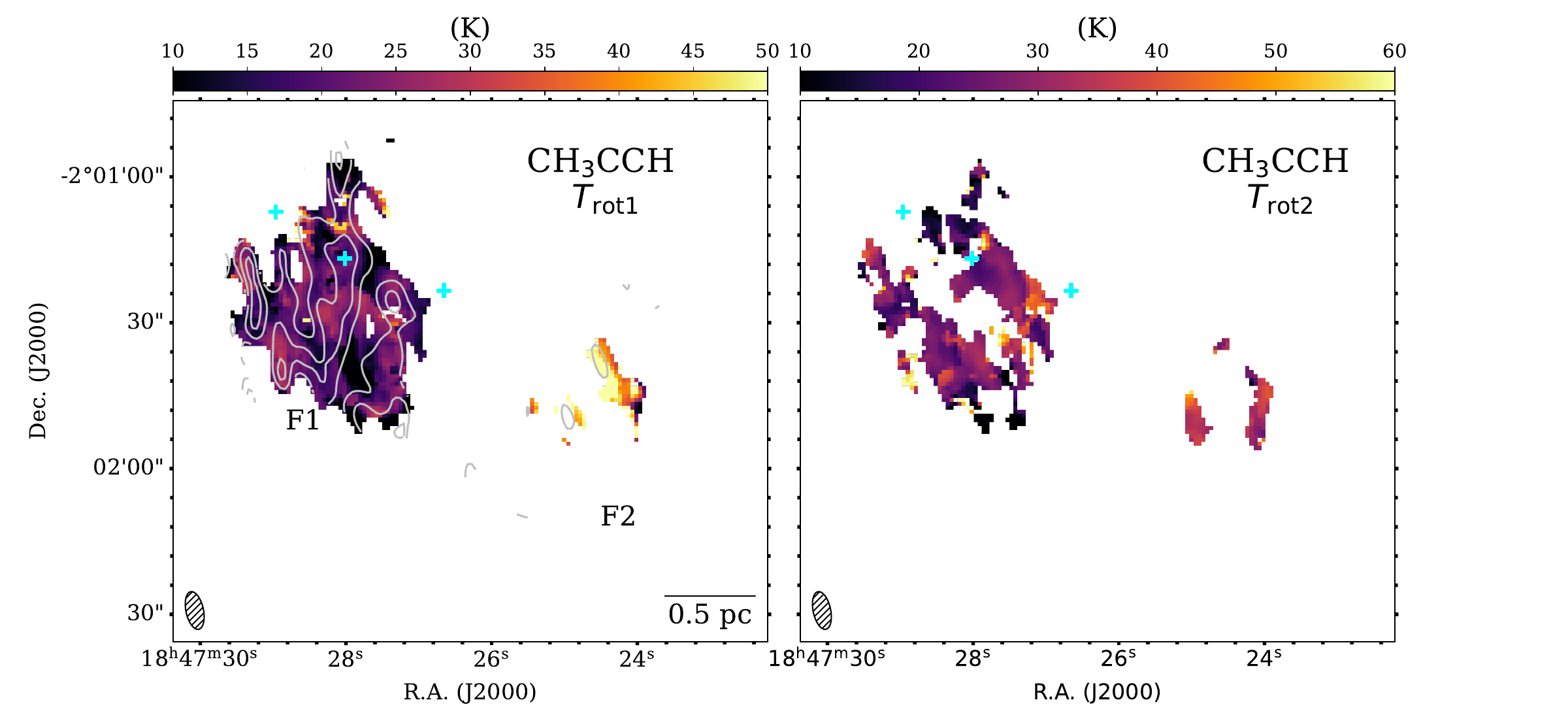}
\caption{Same as Figure \ref{fig:ch3cch_2comp_trot}, but for the clumps F1 and F2. In the left panel the contours represent integrated intensity map of CH$_{3}$CCH 5$_{2}$-4$_{2}$ line. The velocity range for integration is 75-105 km s$^{-1}$, and the contour levels are from 0.45 K km s$^{-1}$ (3 $\sigma$) to 0.8 K km s$^{-1}$ (0.9 times peak emission value of F1) with 3 uniform intervals. Cyan crosses mark the 3 cores from 3 mm continuum of F1 (as in Figure \ref{fig:rgb_cont}, right panel).}
\label{fig:ch3cch_2comp_trots}
\end{figure*}

The derived $T_{\mathrm{rot}}$ maps of CH$_{3}$CCH and H$_{2}$CS lines are shown in Figures \ref{fig:ch3cch_2comp_trot}-\ref{fig:ch3cch_2comp_trots} and \ref{fig:h2cs_2comp_trot}. 
In general, the two $T_{\mathrm{rot}}$ distributions show similar features. Clump C has relatively low temperature mostly $<$25 K, with one gas component having temperature $\sim$ 10 K (right panel of Figure \ref{fig:ch3cch_2comp_trot}). Clump MM3 has an overall warmer condition of 30-45 K although it does not exhibit a strong peaking of gas temperature, and the highest temperature seen by CH$_{3}$CCH is offset from the continuum peak. 
Clump MM2 has a significantly higher temperature in the central continuum peak, where the $T_{\mathrm{rot}}$ of CH$_{3}$CCH reaches $\gtrsim$50 K and that of CH$_{2}$CHCN is up to $\sim$130 K. 
The difference between the two $T_{\mathrm{rot}}$ is expected as CH$_{3}$CCH emission is mostly coming from the extended gas of the colder envelope of the clumps (\citealt{Molinari16a}, \citealt{Giannetti17}, \citealt{Lin22a}). 
Clumps F1 and F2 show a rather uniform $T_{\mathrm{rot}}$ distribution, mostly below 25 K, comparable to that of clump C.
\subsubsection{Deriving the hydrogen gas density distribution from CH$_{3}$OH $J$ = 2-1 $A/E$ lines}\label{sec:dens}
To derive the hydrogen gas density, we first conducted one-component Gaussian fits of CH$_{3}$OH (2-1) $A/E$ lines (for each $K$ component) in a pixel-by-pixel manner, and then used one-component non-LTE RADEX models (\citealt{vdt07}) to predict the hydrogen volume density ($n(\mathrm{H_{2}})$) together with methanol column density $N_{\mathrm{CH_{3}OH}}$, assuming an $A$-to-$E$ ratio of 1, with $T_{\mathrm{kin}}$ assumed to (one-component) follow a normal distribution with mean value of $T_{\mathrm{rot}}$ derived from CH$_{3}$CCH lines (for available pixels) and a standard deviation of 10\,K. In general, the intensity ratios of CH$_{3}$OH J-J-1 line series have a weak dependence on $T_{\mathrm{kin}}$, so this assumption does not impact the derived parameters significantly while aids in a better convergence.  

CH$_{3}$OH (2-1) lines may also consist of more than one velocity component (Appendix \ref{sec:m0s}), but the confusion between different $K$ ladders in this line series inhibits a robust multi-component Gaussian decomposition. Such uncertain decomposition of line intensities will propagate to the RADEX modeling results and introduce a large bias in the estimating n(H$_{2}$) and $N_{\mathrm{CH_{3}OH}}$. We therefore retain a one-component Gaussian fit to describe the line profile of CH$_{3}$OH (2-1) lines. 

\begin{figure*}
\begin{tabular}{p{0.95\linewidth}}
\includegraphics[scale=0.45]{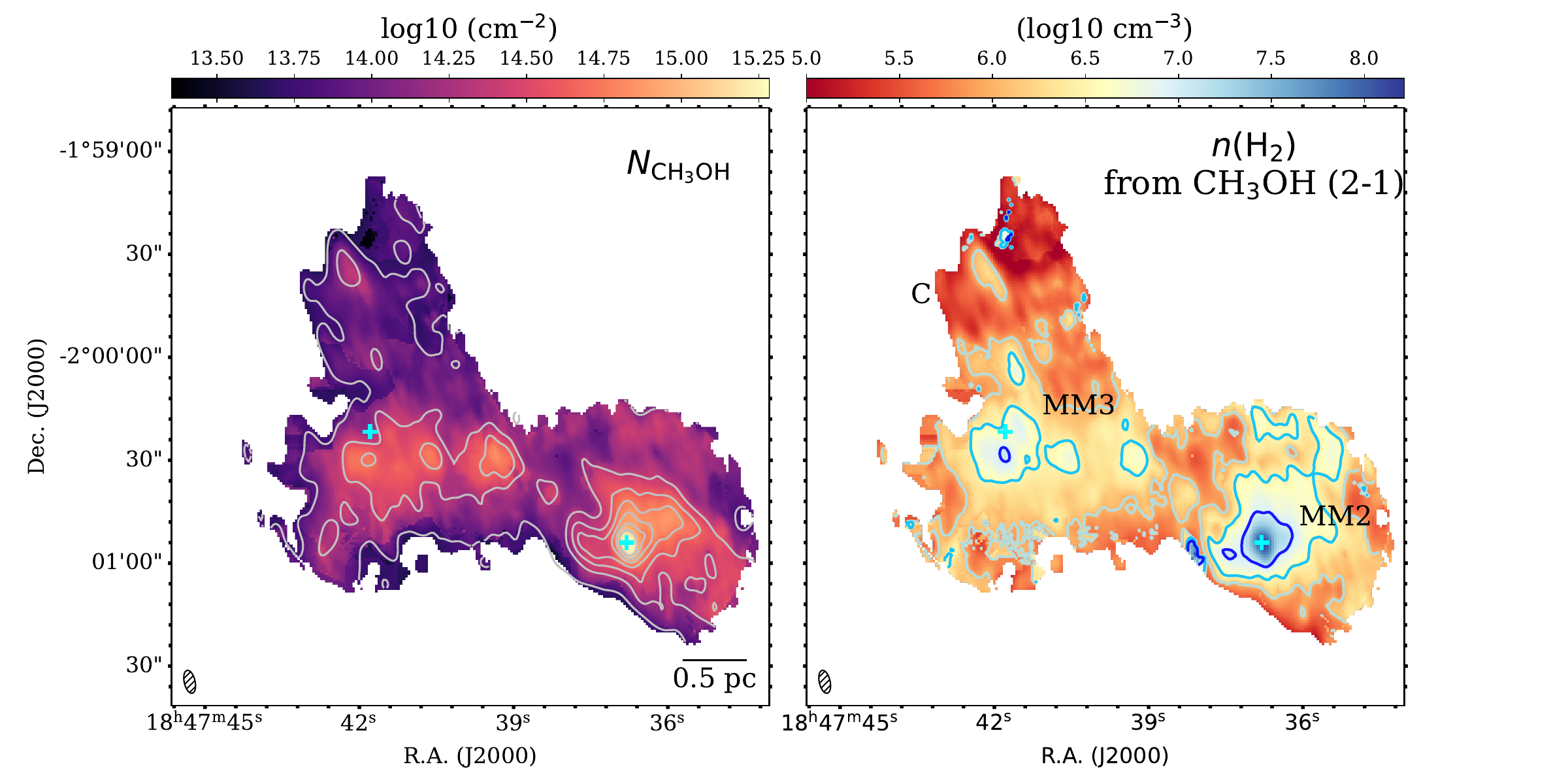}\\
\includegraphics[scale=0.45]{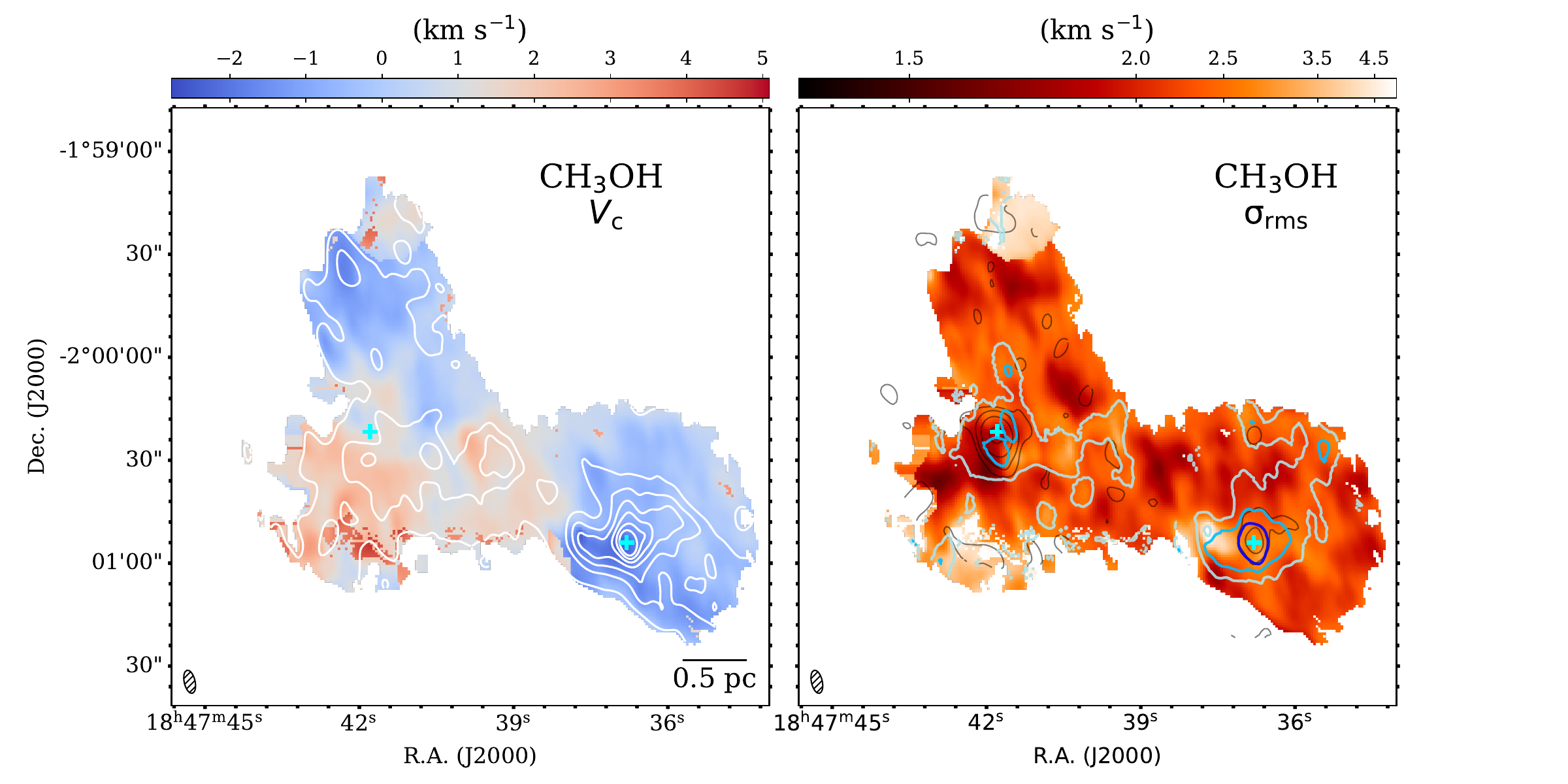}\\
\end{tabular}
\caption{{\emph{Upper panel:}} Derived CH$_{3}$OH column density ({\emph{E}}-type) and hydrogen volume density distribution from RADEX modeling for clumps MM2, MM3 and C. {\emph{Lower panel:}} One-component Gaussian fit result of CH$_{3}$OH (2-1) line series. The centroid velocity (with respect to $V_{\mathrm{LSR}}$ = 91.7 km s$^{-1}$) and velocity dispersion ($\sigma_{\mathrm{v}}$ = $\Delta V/2.355$) distribution. In left panels, the contours represent integrated intensity levels of CH$_{3}$OH $E$-type 2$(0,1)$-1$(0,1)$ line.  The velocity range for integration is 85-100 km s$^{-1}$, and the contour levels are from 1.3 K km s$^{-1}$ (7 $\sigma$) to 37 K km s$^{-1}$ (0.75 times peak emission value of MM2) with 7 uniform intervals. In right panels, the gray contours represent the 3 mm continuum emission, and are logrithmic-spaced from 2$\sigma_{\mathrm{v}}$ ($\sigma_{\mathrm{v}}$ = 1.6 mJy~beam$^{-1}$) to 225.0 mJy~beam$^{-1}$ with 5 intervals (same as Figure \ref{fig:rgb_cont}). Cyan crosses mark the 3 mm continuum peak of MM2 and MM3. The green contours indicate n(H$_{2}$) levels of 10$^{6.1}$ cm$^{-3}$ and 10$^{7}$ cm$^{-3}$.}
\label{fig:ch3oh_1comp}
\end{figure*}

\begin{figure*}
\begin{tabular}{p{0.95\linewidth}}
\includegraphics[scale=0.45]{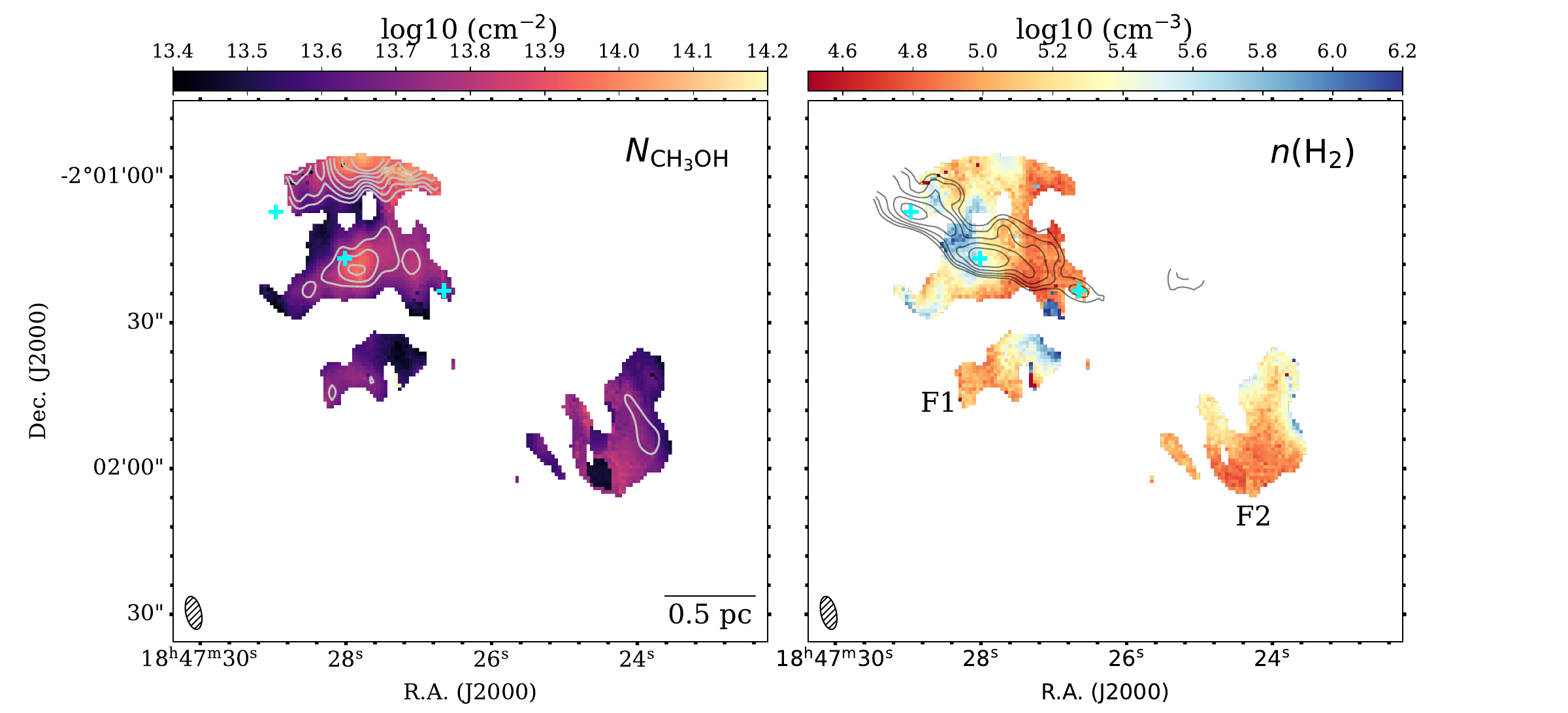}\\
\includegraphics[scale=0.45]{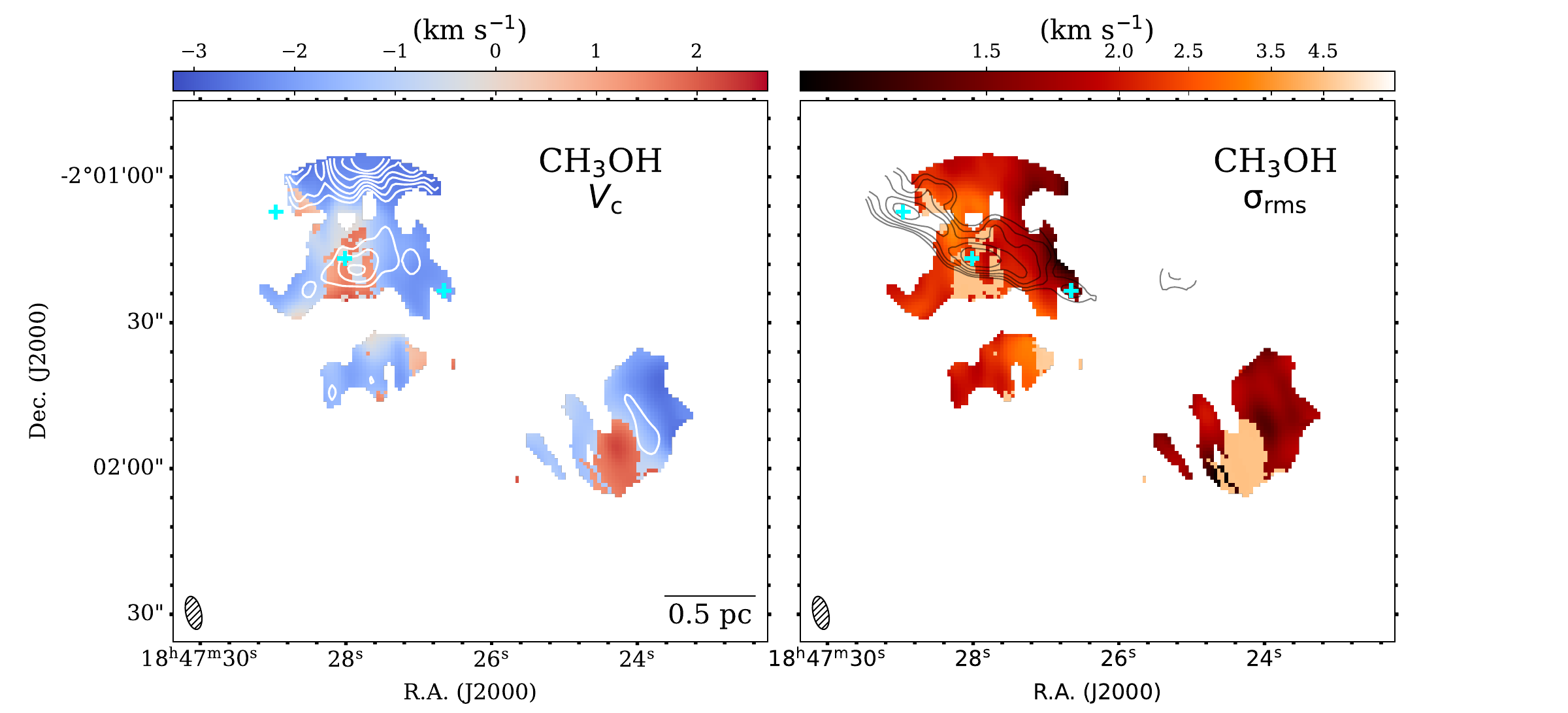}\\
\end{tabular}
\caption{Same as Figure \ref{fig:ch3oh_1comp}, but for the clumps F1 and F2. The centroid velocity is with respect to $V_{\mathrm{LSR}}$ = 85.5 km s$^{-1}$ of clump F2. In left panels, the contours represent integrated intensity levels of CH$_{3}$OH $E$-type 2$(0,1)$-1$(0,1)$ line.  The velocity range for integration is 80-95 km s$^{-1}$, and the contour levels are from 1.2 K km s$^{-1}$ (7 $\sigma$) to 2.6 K km s$^{-1}$ (0.75 times peak emission value of F1) with 7 uniform intervals.  Cyan crosses mark the 3 cores in 3 mm continuum of F1 (as in Figure \ref{fig:rgb_cont}, right panel).}
\label{fig:ch3oh_1comps}
\end{figure*}

We adopt a Markov-Chain Monte-Carlo procedure with RADEX modelling, following the method described in \citet{Lin22a}.
The obtained maps of $n(\mathrm{H_{2}})$ and $N_{\mathrm{CH_{3}OH}}$ ($A/E$) from RADEX modeling, and of the centroid velocity $V_{c}$ and linewidth $\sigma_{\mathrm{v}}$ from the one-component Gaussian fit, are shown in Figures \ref{fig:ch3oh_1comp} and \ref{fig:ch3oh_1comps} (right panel). 
The most prominent dense gas enhancement reside in clumps MM2 and MM3, with maximum gas densities reaching $\gtrsim$10$^{7}$ cm$^{-3}$.
In clump C, the gas densities also display increment around the peak emission but the density level (10$^{5.5}$-10$^{6}$ cm$^{-3}$) is close to the bulk gas density of MM2 and MM3.  
The central density enhancement of MM2 ($\sim$10$^{7.5}$ cm$^{-3}$) coincides with its continuum peak and extends towards the west-east direction by an elongated structure of lesser density enhancement ($\sim$10$^{6.5}$ cm$^{-3}$) showing a curved morphology. 
\citet{Cortes19} resolved several cores in the northwest of the central cores in MM2 (more in Section \ref{sec:fmtt_prop}), which lie close to the attaching region of the elongated structure and the central core. 
In clump MM3, the central density enhancement ($\sim$10$^{6.5}$-10$^{7}$ cm$^{-3}$) is less concentrated. It peaks immediately to the south of the continuum peak. There are several smaller pockets of dense gas enhancement distributed in MM3 (smaller regions enclosed by green contours in Fig. \ref{fig:ch3oh_1comp}), which are $\sim$0.1-0.2 pc in size and remain mostly unresolved. In Section \ref{sec:vo_vw}, we compare the distribution of the dense gas enhancement of clumps MM2 and MM3 with the velocity distribution (linewidths and velocity field) of the CH$_{3}$OH 2$_{-1, 1}$-1$_{-1, 0}$, 5$_{1, 5}$-4$_{0, 4}$ and SiO (2-1) lines.

The $n(\mathrm{H_{\mathrm 2}})$ distribution of clump F1 shows an enhancement at the CH$_{3}$OH emission peak near core F1-c1 and to the north, which is not fully covered by our observations. It also shows distributed density peaks at the edge of the extended CH$_{3}$OH emission. The maximum gas densities, however, reach only $\lesssim$10$^{6}$ cm$^{-3}$. 
The F1-c1 core has an averaged gas density of 10$^{5.5}$ cm$^{-3}$. 
Clump F2 shows a rather uniform distribution of gas densities, having values mostly below 10$^{5}$ cm$^{-3}$.

\subsection{Characteristics of SiO emission and gas density enhancement}\label{sec:vo_vw}
SiO is commonly used as a tracer of shock activities since relatively high-velocity shocks (shock velocity $\gtrsim$ 20 km\,s$^{-1}$) are needed to sputter SiO from dust grains in the gas phase (\citealt{Schilke97a}, \citealt{Caselli97}, \citealt{JS08}) unless some SiO is trapped in the ice mantles. In the latter case, lower velocities can also allow some SiO to go back to the gas phase by erosion (e.g., \citealt{JS08}), although the abundance of SiO is much lower than in the case of faster shocks. 
To understand the shock activities and its induced gas dynamics at 0.1 pc scale, we analyse the velocity features of the SiO (2-1) line, and examine the spatial correlations between the excessive line wings and dense gas enhancement based on the $n(\mathrm{H_{2}})$ map (Figure \ref{fig:ch3oh_1comp}).

\begin{figure*}[htb]
\begin{tabular}{p{0.95\linewidth}}
\vspace{-.25cm}
\hspace{1.cm}\includegraphics[scale=0.43]{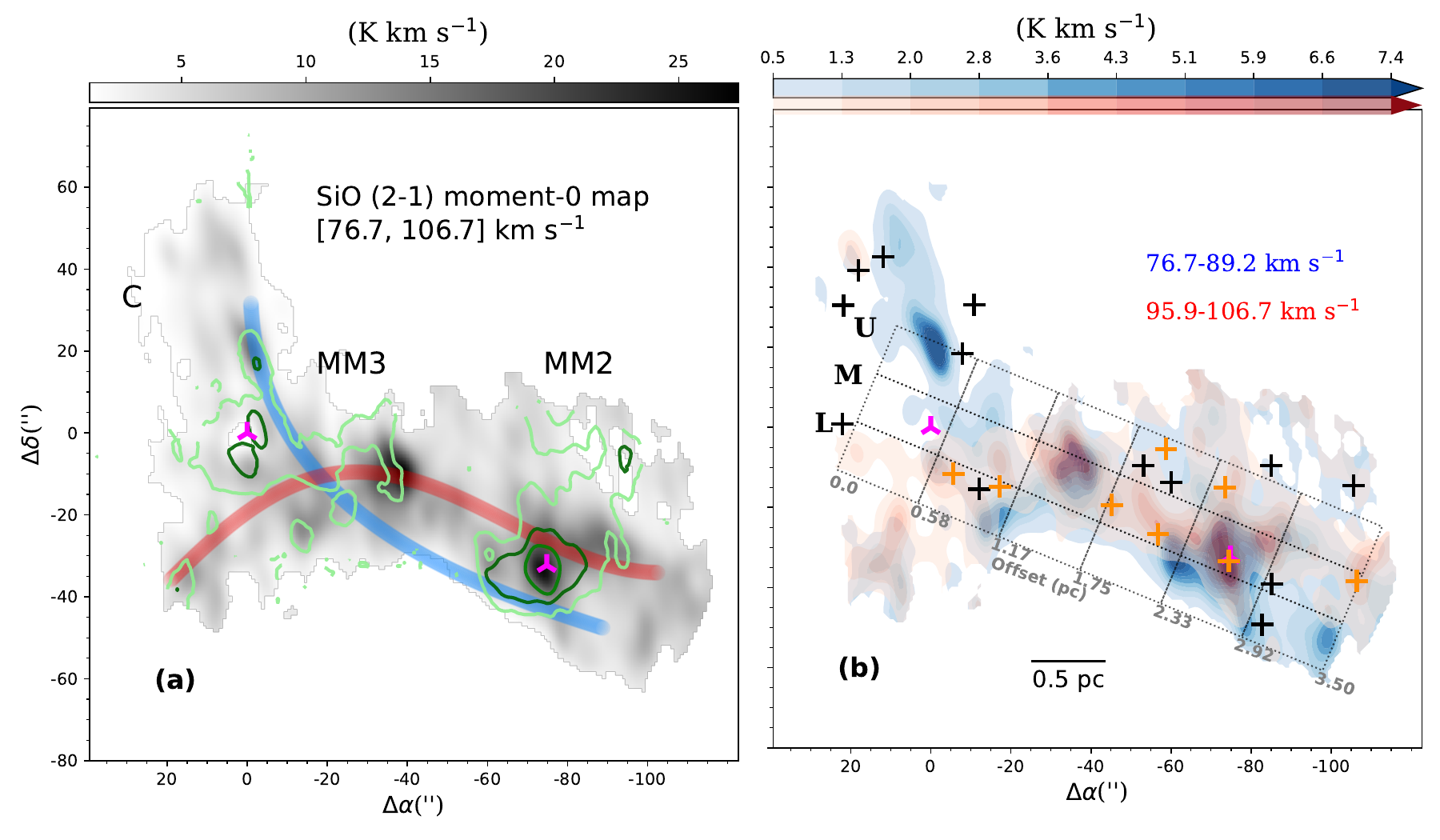}\\
\hspace{1.cm}\includegraphics[scale=0.32]{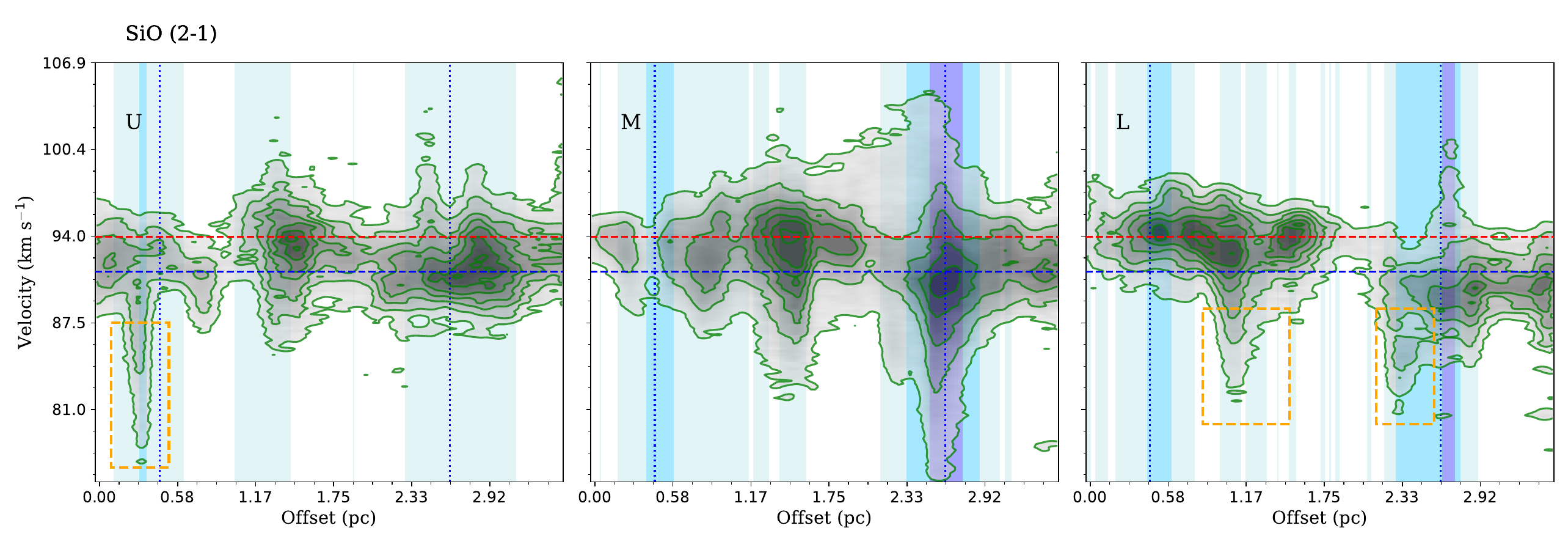}\\
\hspace{1.cm}\includegraphics[scale=0.32]{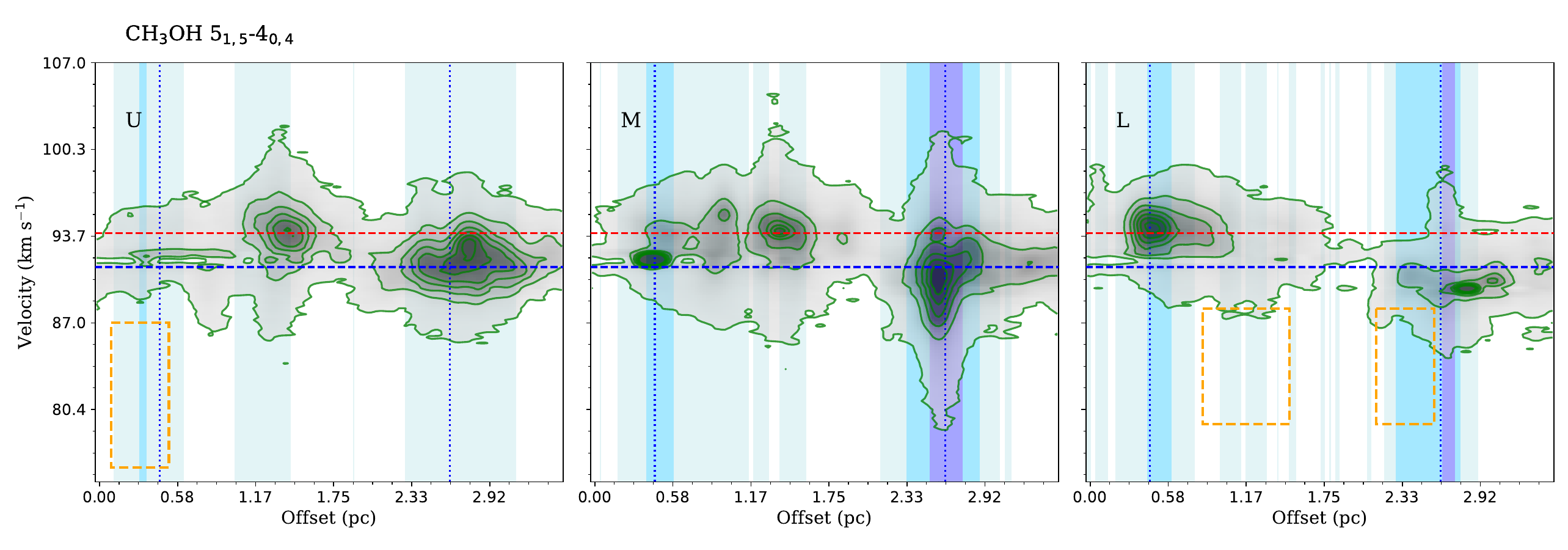}\\
\end{tabular}
\caption{\footnotesize{{\emph{Upper panel:}}(a). Integrated intensity map of SiO (2-1) (76.7-106.7 km s$^{-1}$, gray scale) overlaid with the n(H$_{2}$) levels of 10$^{6}$, 10$^{6.5}$, 10$^{7}$ cm$^{-3}$ (in contours of various shades of green, respectively). The 3 mm continuum peaks of MM2 and MM3 are indicated as magenta crosses. The red and blue thick lines indicate the ridge of the emission of the blue-shifted and red-shifted SiO in panel (b). (b). The integrated intensity maps (velocity ranges indicated in the plot) of the SiO (2-1) blue-shifted and red-shifted emission. The crosses in orange (with at least one broader velocity component of $\sigma_{\mathrm{v}}$$\,>\,$0.85 km s$^{-1}$, see Sect.~\ref{sec:nh2d}) and black indicate the NH$_{2}$D cores. For plot (a) and (b), R.A. and Decl. offsets correspond to the relative position with respect to 3 mm continuum peak of MM3. {\emph{Lower panel:}} Position-velocity diagrams of SiO (2-1) (contour levels of 0.2-1.4\,K~km~s$^{-1}$ with 6 intervals) and CH$_{3}$OH 5$_{1,5}$-4$_{0,4}$ (contour levels of 0.1-10.3\,K~km~s$^{-1}$ with 6 intervals) line along the three cuts U, M and L, as in the upper panel. The color-shaded regions indicate the n(H$_{2}$) levels of 10$^{6}$, 10$^{6.5}$, 10$^{7}$ cm$^{-3}$ (in light blue, blue and purple). The $V_{\mathrm{LSR}}$ of MM2 and MM3 are indicated as red and blue horizontal lines (93.9 km $^{-1}$ and 91.3 km $^{-1}$). Blue vertical dotted lines indicate the positions of the 3 mm continuum peak of MM2 and MM3.}}
\label{fig:sio_maps1}
\end{figure*}

The distribution of the SiO emission (integrated intensity map, of velocity range of 76.7-106.7 km s$^{-1}$) is shown in the left panel of Figure \ref{fig:sio_maps1}. The two sub-regions showing the most intense SiO emission are located in the central region of clump MM2, and at the position of (35$''$, -10$''$) ($\Delta\alpha$, $\Delta\delta$) offset from MM2, respectively. The latter sub-region is located in the connection region of MM2 and MM3. 
The gas velocity field traced by SiO shows a pattern of complementary distribution: moving from clump C to MM3 and MM2 in the northeast to southwest direction, the velocity changes from being blue-shifted to red-shifted, and again to blue-shifted (compared to the $V_{\mathrm{LSR}}$ of MM3). A similar picture is also revealed in the integrated intensity maps of the same velocity ranges of other lines shown in Figure \ref{fig:channs_g307}. To further illustrate this pattern, we separated the blue-shifted and red-shifted velocity ranges and integrated the two velocity ranges individually, which are shown together in Figure \ref{fig:sio_maps1} (panel (b)). The two distributions of blue-shifted and red-shifted emission compose a ``Y''-shaped morphology. Particularly, the density enhancement of clump MM3 (shown as colored contours in panel (a)) seems to follow the vertex region of the ``Y'' tightly, while the central region of clump MM2 lies in-between the two distributions.

In the lower panel of Figure \ref{fig:sio_maps1} the sub-regions showing density enhancement of $>$10$^{6}$ cm$^{-3}$ (indicated by contours in panel (a) of Figure \ref{fig:sio_maps1} based on the n(H$_{2}$) map), are shown together as shaded regions with the position-velocity (PV) diagrams following the three cuts (U, M and L). 
The three position-velocity cuts are chosen to encompass the ``Y"-shape structure of the SiO emission, to illustrate the emission associated along the ridge of most prominent density enhancements (linking central regions of MM3 and MM2) and the two offsets along the same direction. 
Each of the three cuts is averaged over a width of 0.35 pc as indicated in panel (b) of Figure \ref{fig:sio_maps1}. 

The PV diagrams are then extracted from the SiO (2-1) and CH$_{3}$OH 5$_{1,5}$-4$_{0,4}$ spectral cubes. The line wings of both emission mostly appear within 8 km/s around the $V_{\mathrm{LSR}}$ (low-velocity regime). Along the three cuts, SiO emission shows more extended line wings in both spatial and velocity regimes than that of CH$_{3}$OH 5$_{1,5}$-4$_{0,4}$. There is a broad red-shifted SiO velocity wing at offset 1.0-2.4 pc in the PV cut of M, which also extends to higher terminal velocities. 
The terminal velocities of SiO reach up to $V_{\mathrm{LSR}}$$\pm$15 km s$^{-1}$, as seen from the U and M cuts. 
There are also several prominent wing components that appear asymmetric in the PV maps of SiO (indicated by orange rectangles in middle and lower panel of Fig. \ref{fig:sio_maps1}), with no counterparts in opposite velocity regime. 
These single, excessive line wings appear mostly blue-shifted and do not have counterparts in the CH$_{3}$OH emission; they are likely outcome of high-velocity shocks from outflow activities sputtering the cores of dust grains to release SiO (\citealt{Snow96}, \citealt{JS08}). 
It seems the density enhancement at the central region of clump MM2, at offset 1.8-3.2 pc (illustrated as blue and purple shaded regions around the blue dotted line) coincides with double-peaked velocity wings, while the density enhancement associated with the immediate vicinity of the continuum of MM3 is not associated  with significant velocity wing features, as seen from the M cut of SiO PV map. 
The sub-regions of secondary density enhancement (indicated by light-blue shaded regions) are mostly correlated with prominent line wings. 

\begin{figure*}
\begin{tabular}{p{0.975\linewidth}}
\hspace{1.cm}\includegraphics[scale=0.35]{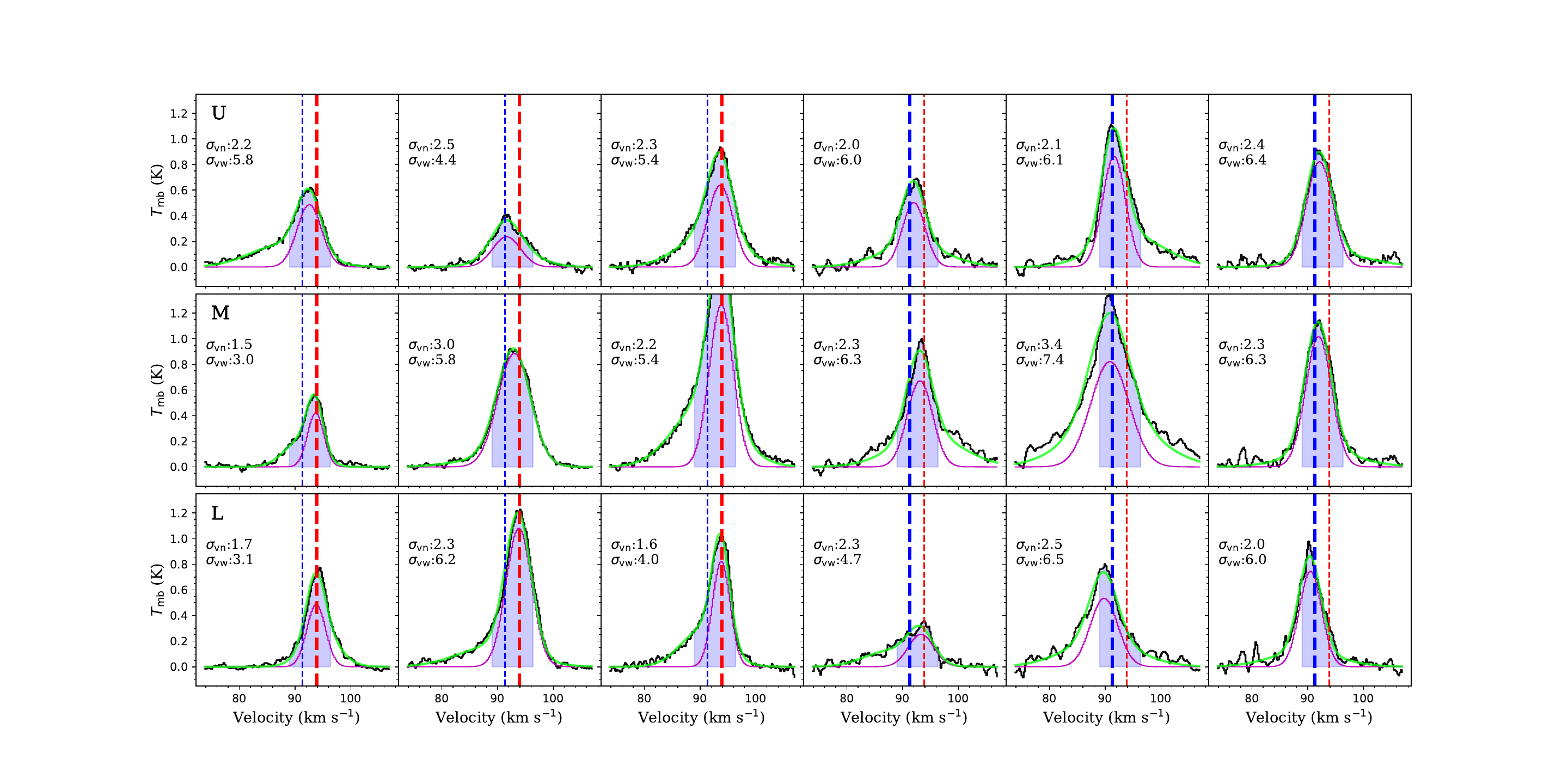}\\
\hspace{1.cm}\includegraphics[scale=0.35]{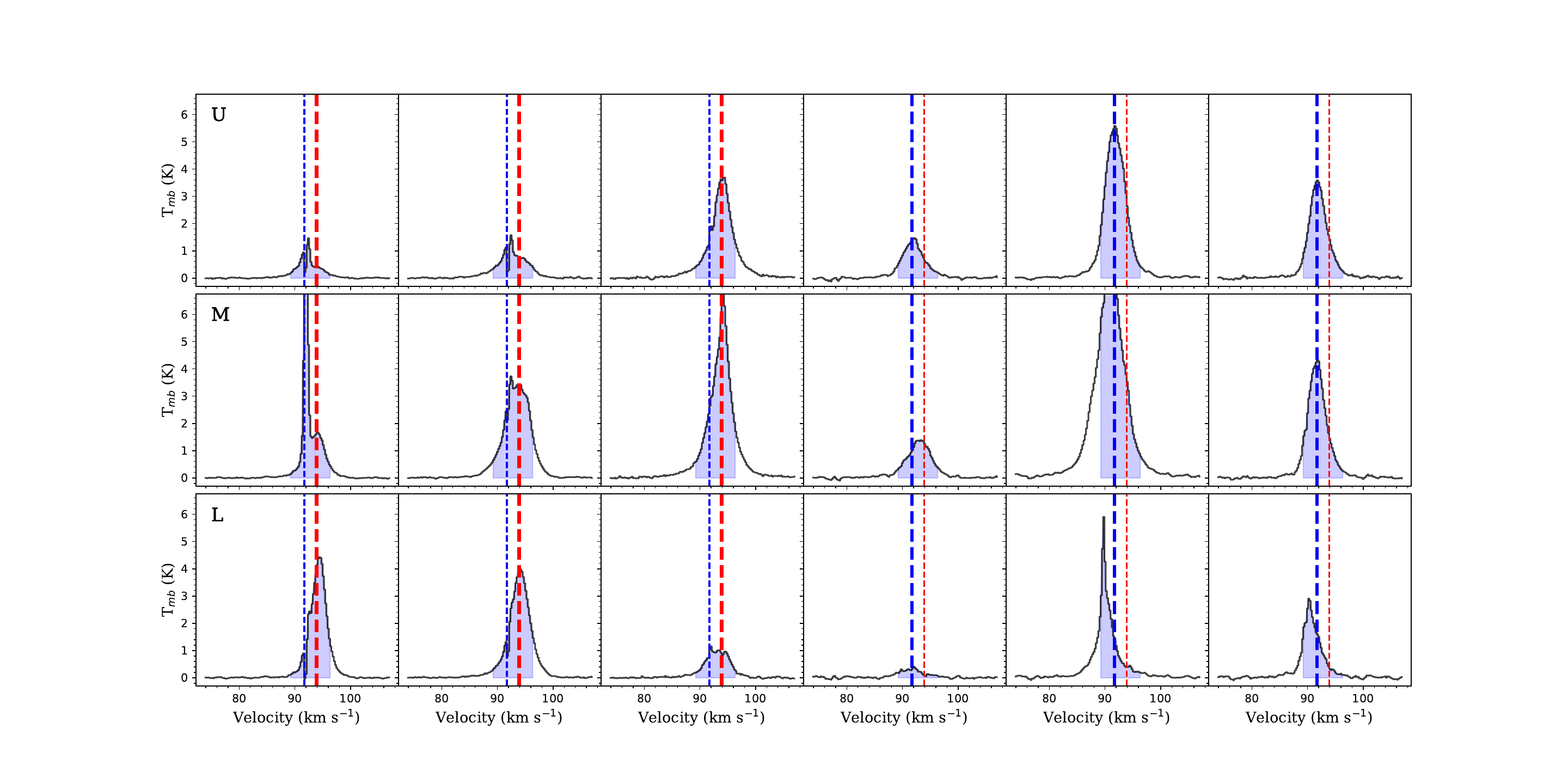}\\
\end{tabular}
\caption{\footnotesize{Average spectra of SiO (2-1) (upper panel) and CH$_{3}$OH 5$_{1,5}$-4$_{0,4}$ (lower panel) 
along the three PV cuts U, M, L as in panel (b) of Figure \ref{fig:sio_maps1}. The spectra are arranged from northeast to southwest along the cuts. The blue and red dashed lines indicate the $V_{\mathrm{LSR}}$ of clump MM3 and MM2 (93.9 km s$^{-1}$ and 91.3 km s$^{-1}$), with the thicker line corresponding to the $V_{\mathrm{LSR}}$ of either clump MM2 or MM3 at this position. The blue shaded regions indicate the line central velocity range from (91.3-2.5) km s$^{-1}$ to (93.9+2.5) km s$^{-1}$, which is excluded from the integration of the intensity maps of panel (b) in Figure \ref{fig:sio_maps1}. In the upper plot, the two-component Gaussian fit to the SiO line is shown additionally, as green line. The broad-component subtracted spectrum (narrow-component line profile) is shown as magenta line. The velocity dispersions for the two components are indicated in the figure, in unit of km s$^{-1}$.}}
\label{fig:sio_ch3oh_pv_sps}
\end{figure*}

For each PV cut, we separate the sampled area into 6 parts (panel (b) of Figure \ref{fig:sio_maps1}) and plot the average spectra of SiO (2-1) and CH$_{3}$OH 5$_{1,5}$-4$_{0,4}$ lines in Figure \ref{fig:sio_ch3oh_pv_sps}. 
We use a two-component Gaussian model to characterize the SiO (2-1) line profile. We obtained velocity dispersion of the narrow component ($\sigma_{\mathrm n}$) ranging between 1.5-3.4 km s$^{-1}$ with a mean value of $\sim$1.7$\pm$1.5 km s$^{-1}$, and the $\sigma_{\mathrm{v}}$ of the broad component ($\sigma_{\mathrm w}$) between 3.0-7.4 km s$^{-1}$ with a mean value of $\sim$4.6$\pm$1.5 km s$^{-1}$. 
We compare the broad-component subtracted narrow line profiles (in red) with 
the $V_{\mathrm{LSR}}$ of clumps MM2 and MM3 in Figure \ref{fig:sio_ch3oh_pv_sps}. 
It can be seen that the majority of the peak velocities of the narrow component are consistent with the peak velocity of the whole line profile, and within the range of the $V_{\mathrm{LSR}}$ of MM2 and MM3.

\subsection{Velocity distribution of different molecules: systematic velocity shift}\label{sec:veloshift}

\begin{figure*}
    \begin{tabular}{p{0.95\linewidth}}

    \hspace{1.cm} \includegraphics[scale=0.4]{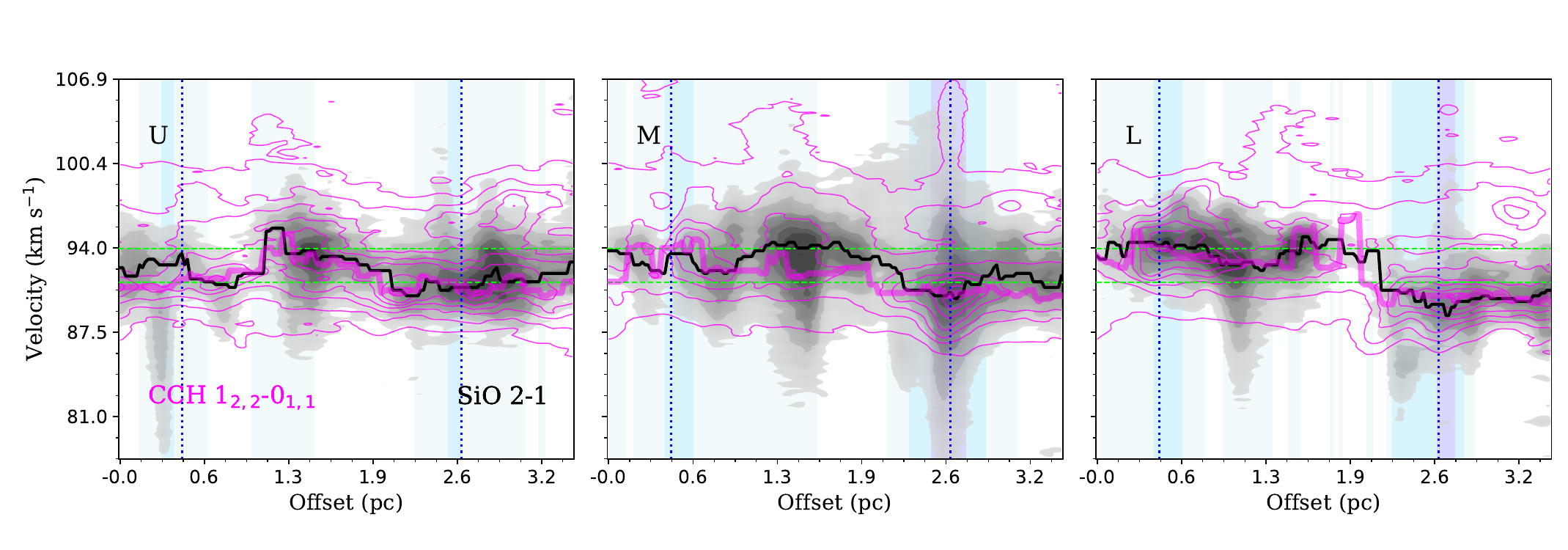}\\
  \hspace{1.cm} \includegraphics[scale=0.4]{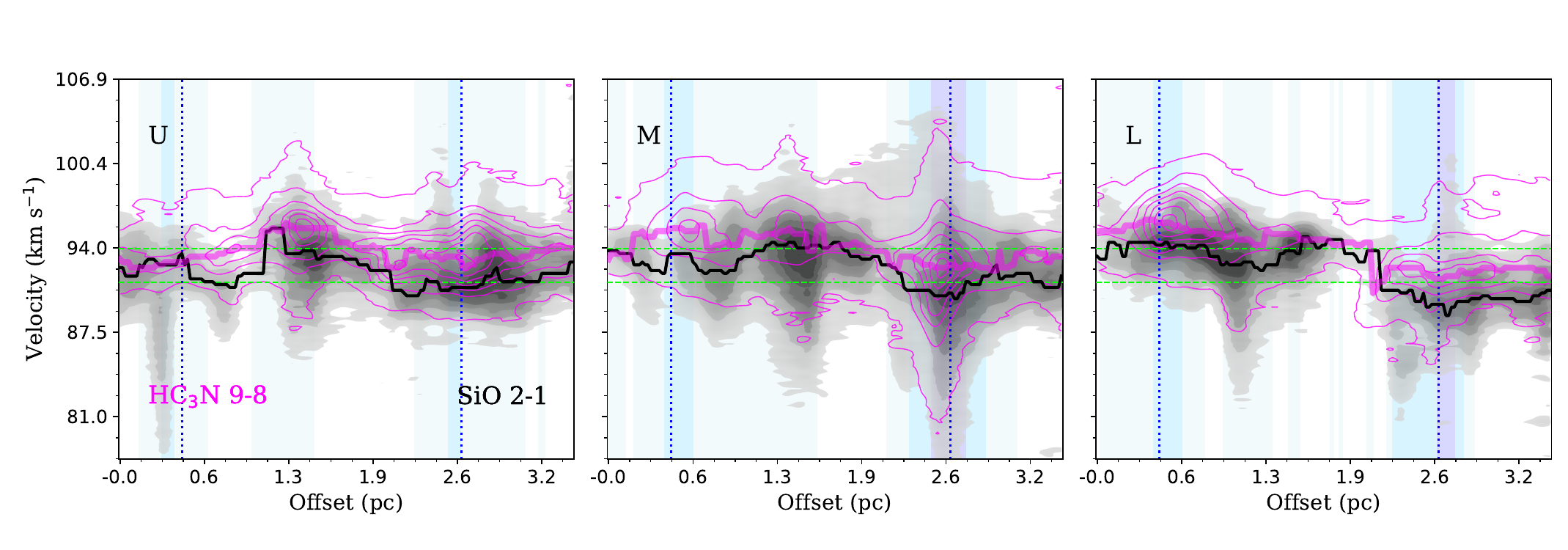}\\
\end{tabular}
    \caption{Position-velocity diagram for CCH 1$_{2,2}$-0$_{1,1}$, HC$_{3}$N (9-8) 
    (in magenta contours) in comparison to that of SiO (2-1) (gray filled contours), along the U, M, L cuts as illustrated in panel (b) of Figure \ref{fig:sio_maps1}. The $V_{\mathrm{LSR}}$ of MM2 and MM3 are indicated as green horizontal lines (93.9 km $^{-1}$ and 91.3 km $^{-1}$). The vertical dotted line and shaded regions follow those as in Figure \ref{fig:sio_maps1}. The peak velocity along the spatial offset for SiO (2-1) and the respective molecular line (that have extended emission) in each plot are indicated as black and magenta lines.}
    \label{fig:pv_sio_cch_ch3cch_hcn15_hc3n}
\end{figure*}

\begin{figure*}
    \begin{tabular}{p{0.95\linewidth}}
 \hspace{1.cm} \includegraphics[scale=0.4]{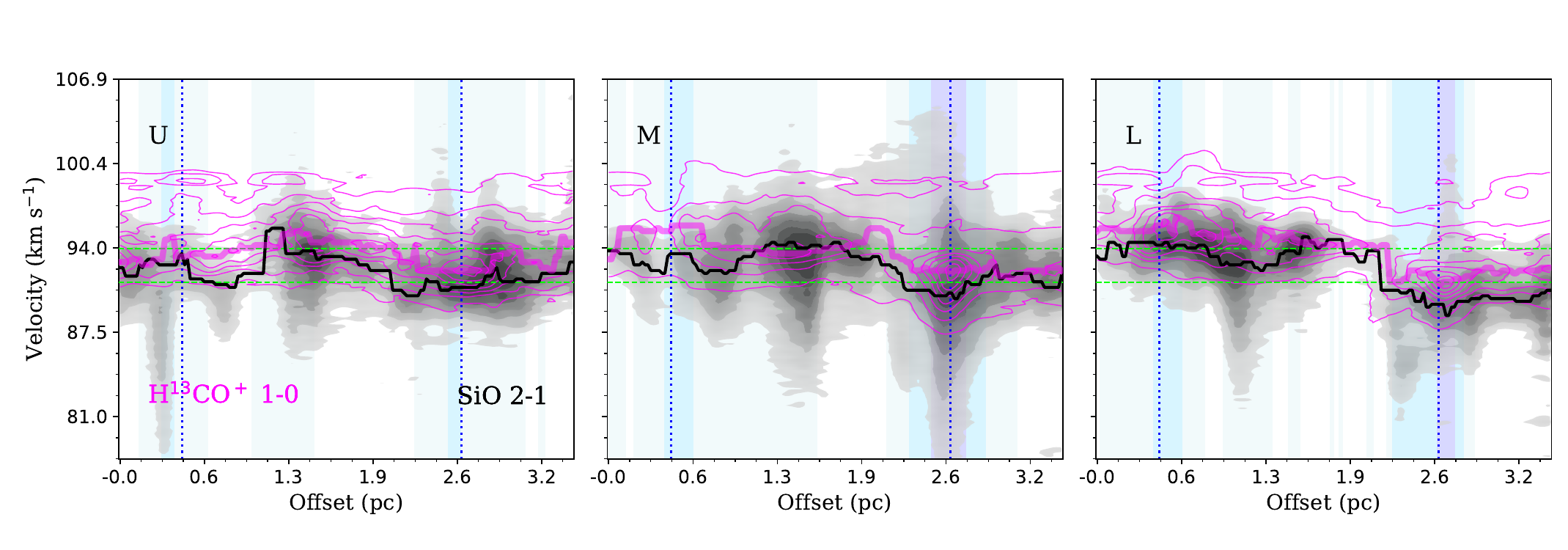}\\
    \hspace{1.cm} \includegraphics[scale=0.4]{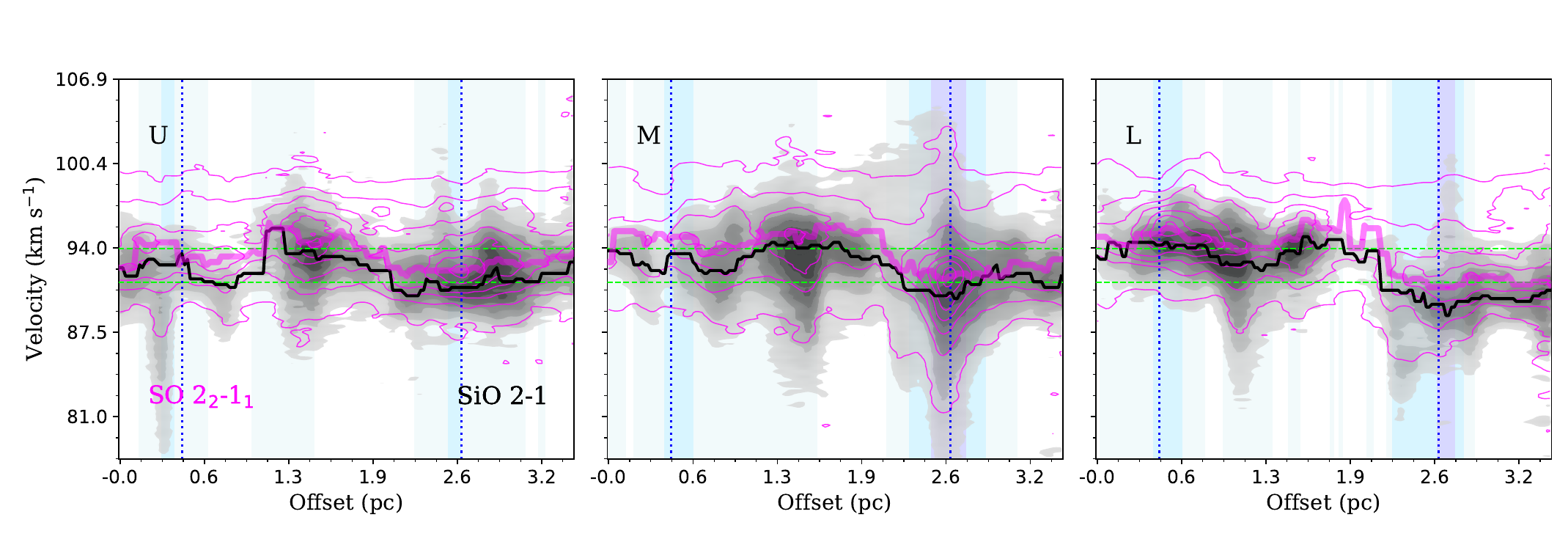}\\
    \end{tabular}
   \caption{Same as Figure \ref{fig:pv_sio_cch_ch3cch_hcn15_hc3n}, but for H$^{13}$CO$^{+}$ (1-0) and SO 2$_{2}$-1$_{1}$ lines.}
    \label{fig:pv_sio_so}
\end{figure*}

In Figures \ref{fig:pv_sio_cch_ch3cch_hcn15_hc3n} and \ref{fig:pv_sio_so} we plot the PV diagrams along the three cuts U, M, L in the northern region (Figure \ref{fig:sio_maps1}) for molecular lines CCH 1$_{1,0}$-0$_{1,1}$, 
HC$_{3}$N (9-8), 
H$^{13}$CO$^+$ (1-0), 
SO 2$_{2}$-1$_{1}$, 
together with SiO (2-1) line. The emission of some of these transitions is similar to or even more extended than SiO.

In general the peak velocities of the CCH line coincide well with that of SiO along all three PV cuts, except the offset region at (0.8,2.4) pc in M cut, where peak velocities of CCH appear more blue-shifted. 
The peak velocities of the SO 2$_{2}$-1$_{1}$ line, on the other hand, show a slightly red-shifted tendency along the cuts. The velocity shift between both HC$_{3}$N and H$^{13}$CO$^+$ w.r.t. SiO is most prominent and exist everywhere except the offset region of (1.3, 2.0) pc in the M and L cut. The peak velocities of HC$_{3}$N and H$^{13}$CO$^+$ appear in the more red-shifted regime, with the velocity shifts reaching up to $\sim$2 km s$^{-1}$. 

From the PV maps of the CCH line, it is also clearly seen that at spatial offsets around clump MM2 in the L cut, there is a prominent red-shifted component which is not present in the SiO PV map. 
This component also exists in the HC$_{3}$N and H$^{13}$CO$^+$ PV map of cut L, but of smaller spatial extension. 

\begin{figure*}
    \begin{tabular}{p{0.975\linewidth}}
    \hspace{0.25cm}\includegraphics[scale=0.35]{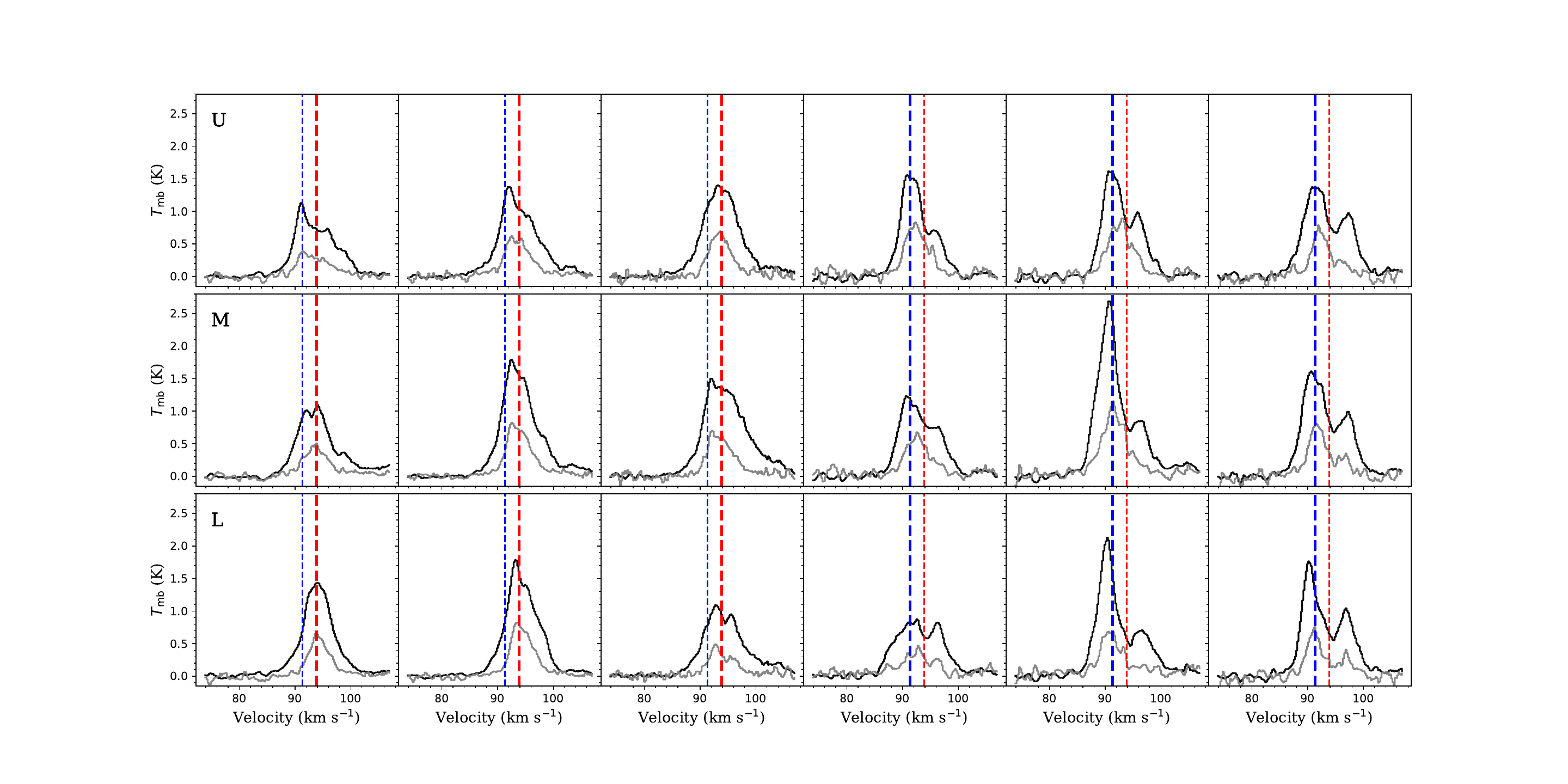}
    \end{tabular}
    \caption{Average spectrum of the main line of CCH $\emph{hfs}$ component 1$_{2,2}$-0$_{1,1}$ (black line) and the satellite line 1$_{1,1}$-0$_{1,0}$ (gray line, the intensity is artificially enlarged by 3 times for comparison) along the three PV cuts U, M, L. From left to right, the spectra are arranged from northeast to southwest along the cuts. The blue and red dashed lines follow that in Figure \ref{fig:sio_ch3oh_pv_sps}.}
    \label{fig:cch_pv_sps}
\end{figure*}

To further demonstrate this, we present the spectra of the CCH line along the three PV cuts in Figure \ref{fig:cch_pv_sps}. 
The two velocity components are seen both in the main line of CCH 1$_{1,0}$-0$_{1,1}$, which is slightly optically thick ($\tau_{m}$$\sim$0.5-2.5 from hyperfine fitting, see Sect. \ref{sec:cch}) in this region, and the least intense satellite line CCH 1$_{1,1}$-0$_{1,1}$ which is optically thin ($\tau$$<$0.5, more in Sect.~\ref{sec:nh2d}). 
This suggests that the two-component line profile is not an optical depth effect but produced by (at least) two velocity gas components.  Also from the CCH PV maps, there are several prominent red-shifted wings at offset $\sim$1.3 pc along U, M and L cut. These structures have terminal velocities reaching beyond $V_{\mathrm{LSR}}$+10 km s$^{-1}$, and do not have clear counterparts in the PV maps of the other lines. We compared the velocity of the second component with the large-scale $^{13}$CO\,(2-1) data obtained from the IRAM 30m legacy program HERO (\citealt{Carlhoff}) and found that it has a counterpart of similar velocity, confirming that this additional velocity component represents an outer, less dense gas layer in the cloud merging/collision process.

\subsection{Hyperfine fitting of CCH lines}\label{sec:cch}
CCH (1-0) lines are split into hyperfine structures ({\emph{hfs}}) that allow to measure the excitation temperatures and optical depths. We conduct one-component and two-component fits and chose in-between the best-fit model. 

In Figure \ref{fig:cch_2comp_nmol_tex} and Figure \ref{fig:cch_2comp_vwvo} the fitted excitation temperature $T_{\mathrm{ex}}$, total CCH column density $N_{\mathrm{CCH}}$, and the centroid velocity and linewidth of CCH (1-0) lines towards clumps MM2, MM3 and C are shown. Figure \ref{fig:cch_2comp_nmol_texs} and Figure \ref{fig:cch_2comp_vwvos} show the same maps towards clump F1 and F2. The emission of CCH is very extended especially in clumps MM2 and MM3. Clump C has CCH emission mostly in the north-west region. For clump MM3, particularly, the CCH emission around the UCH\textsc{ii} regions has a sharp decrease of emission in the north-east direction, exhibiting an arc-like morphology. This is consistent with previous observations and predictions that CCH in active star-forming regions is efficiently converted to other molecules (e.g. \citealt{Beuther08}, \citealt{Jiang15}). The derived $N_{\mathrm{CCH}}$ ranges between 10$^{15.5}$-10$^{16.5}$ cm$^{-2}$.
The $T_{\mathrm{ex}}$ of CCH is mostly $\sim$5-10 K, with local maxima reaching 25 K at the periphery of the central region of clump MM2 and at intersection region of MM2 and MM3. 
Comparing the velocity dispersion ($\sigma_{\mathrm{v_{1}}}$ and $\sigma_{\mathrm{v_{2}}}$) to $T_{\mathrm{ex}}$, we again resolve, similarly as CH$_{3}$CCH lines, that the regions of high  $T_{\mathrm{ex}}$ in clumps MM2 and MM3 are displaying larger line-widths, while in clump C, $T_{\mathrm{ex}}$ and line-widths seem anti-correlated. 

The velocity field of CCH in the regions where CH$_{3}$CCH is detected has a generally consistent distribution. There are prominent red-shifted components of CCH reaching $+$8 km s$^{-1}$ in the intersection region of clumps MM2 and MM3, and in the southern part of MM2 (as illustrated also in Figure \ref{fig:pv_sio_cch_ch3cch_hcn15_hc3n}). As discussed in Sect.~\ref{sec:veloshift}, these red-shifted components peaking at 96-98 km~s$^{-1}$ may represent an outer layer of the clouds in collision. 

\section{NH$_{2}$D emission and NH$_{2}$D-traced cores}\label{sec:nh2d}

NH$_{2}$D 1(1,1)-1(0,1) line also exhibits hyperfine structures. Compared to CCH, emission of NH$_{2}$D 1(1,1)-1(0,1) is more localised and compact. 
The integrated intensity maps of NH$_{2}$D 1(1,1)-1(0,1) are shown in Figures \ref{fig:nh2d_core_map} (left panel), for the northern and southern region, respectively. 
In comparison to the 3 mm dust continuum, we see that only clump MM2 and core F1-c1 has strong emission of NH$_{2}$D. 


To better characterise the localised emission of the NH$_{2}$D line, 
we run the {\tt{dendrogram}} (\citealt{Rosolowsky08}) algorithm on NH$_{2}$D spectral cube to extract physical parameters that define these NH$_{2}$D-traced subregions. For the {\tt{dendrogram}} input parameters, we set both {\tt{min$\_$value}} and {\tt{min$\_$delta}} to be 3$\sigma$ of the noise level, and {\tt{min$\_$npix}} to be the number of pixels inside one beam, which means we consider structures of peak emission larger than 3$\sigma$ and size larger than one beam and are at least 3$\sigma$ more significant than the emission of their lower level parental structures. Since {\tt{dendrogram}} algorithm works optimally for 2D hierarchical structures, after this initial step, we then generate clusters of the ``leaves'' by merging adjacent leaves that have a position distance less than the square root of the {\tt{min$\_$npix}}, specifically by using the linkage matrix and {\tt{fcluster}} method in {\tt{scipy.cluster}} module with a distance threshold. These clusters are then regarded as representing independent, compact structures from the 3D PPV cube. This method essentially means that we do not distinguish cores by the emission difference in the velocity axis as long as they are located in similar projected positions.
The parameters of the extracted compact structures are listed in Table \ref{tab:dendro_nh2d}-\ref{tab:dendro_nh2ds}. The locations of the extracted structures are marked in Figure \ref{fig:nh2d_core_map}. Hereafter we refer to these compact structures as NH$_{2}$D cores. 

Average spectra towards all the identified cores are shown in Figure \ref{fig:nh2d_sps}-\ref{fig:nh2d_spss}.  We use one and two-component {\emph{hfs}} fits to describe the observed line profiles, and the parameter uncertainties are estimated by MCMC method (Appendix \ref{app:nh2d}). Considering a gas temperature of 50\,K (upper limit seen by extended CH$_{3}$CCH emission), the corresponding sound speed is $\sim$0.42\,km\,s$^{-1}$. Based on the linewidths, these NH$_{2}$D cores can be classified into two categories, ones having borad, supersonic linewidth ($\sigma_{\mathrm{v}}$$\gtrsim$2$c_{\mathrm{s}}$\,=\,0.85\,km s$^{-1}$), and the others having narrow, only subsonic to transonic line-width component(s) ($\sigma_{\mathrm{v}}$$\lesssim$2$c_{\mathrm{s}}$\,=\,0.85\,km s$^{-1}$). 

Assuming a uniform density profile, the virial mass of NH$_{2}$D cores can be estimated by $M_{\mathrm{vir}}$ = $\frac{5\sigma_{\mathrm{v}}^{2}R}{G}$,
with $\sigma_{\mathrm{v}}$ obtained from the {\emph{hfs}} fits and $R$ the effective radius defined by core area from {\tt{dendrogram}} source extraction. For NH$_{2}$D cores that have two velocity components, we used $\sigma_{\mathrm{narrow}}$ as $\sigma_{\mathrm{v}}$, assuming these narrow linewidth structures are stabilised. The derived excitation temperature $T_{\mathrm{ex}}$ and the optical depth $\tau_{\mathrm m}$ of the main line, and the $M_{\mathrm{vir}}$ are also listed in Table \ref{tab:dendro_nh2d}-\ref{tab:dendro_nh2ds}. 
The virial mass of NH$_{2}$D cores range from 11\,$M_{\odot}$ to 300\,$M_{\odot}$. The column densities of NH$_{2}$D are derived to be 0.5-5$\times$10$^{13}$ cm$^{-2}$ with a mean value of 2.4$\times$10$^{13}$ cm$^{-2}$, which are consistent with the column density values of NH$_{2}$D observed in massive star-forming regions (e.g., \citealt{Pillai11}, \citealt{Fontani15}).

\section{Discussion}\label{sec:diss}
\subsection{Gas dynamics dominated by shock activities in massive clumps in W43-main}\label{sec:cc_scenario}
Previous single-dish observations revealed wide-spread emission of SiO (2-1) throughout the ridges of the ``Z''-shape structures of W43-main, which is interpreted as arising from low-velocity shocks created by cloud collisions (\citealt{NL13}, see also \citealt{Louvet16} for interferometric observations towards clump MM1). 
However there is ambiguity as to whether the detected low-velocity SiO emission originates from outflows of embedded YSOs (or outflowing gas mixing back to the cloud) or cloud collisions. \citet{JS10} found that SiO emission towards an infrared dark cloud composes of an extended low-velocity (line-width$\sim$0.8 km s$^{-1}$) component, which is interpreted as a result of the following scenarios: large-scale shock induced at the formation of the cloud; outflows of a population of low-mass protostars; recently processed gas
by the magnetic precursor of young magneto-hydrodynamic shocks (\citealt{JS04}). In our target clumps, \citet{Nony23} resolved numerous outflow features from $\sim$0.015 pc resolution observations originating from embedding cores in MM2 and MM3, which may as well be accompanied by SiO emission. 
Despite the possible confusion, in Sect. \ref{sec:vo_vw} we show that the $\sim$0.1 pc scale SiO emission is distributed nearly continuously across the region encompassing the three clumps, hinting at the existence of a global origin from large-scale shocks, together with contribution from multiple outflows of both high-mass and low-mass protostars.

The spatio-kinematic offsets between different tracers have been suggested as indicator of cloud-cloud collision or merging for early-stage clouds (\citealt{JS10}, \citealt{Henshaw13}, \citealt{Bisbas18}, \citealt{Priestley21}). On the PV diagrams, broad bridging features between two velocity components (e.g. \citealt{Haworth15}, \citealt{Priestley21}), as well as peculiar ``V''-shaped structures (e.g. \citealt{Takahira14}, \citealt{Fukui18}) also suggest cloud-cloud collisions. 
In the case of the studied region, the velocity difference between the two clouds in collision is small ($V_{\mathrm{LSR}}$ of 91.3 and 93.9 km s$^{-1}$). The bridging features are therefore not spatially significant in the velocity domain, but they are present in all tracers that have prominent extended emission, e.g., H$^{13}$CO$^+$, CCH, HC$_{3}$N, and SO, as shown in the PV diagrams of Figs. \ref{fig:pv_sio_cch_ch3cch_hcn15_hc3n}-\ref{fig:pv_sio_so}. In particular, the broader CCH lines and the additional more red-shifted velocity component around clump MM2 may reflect the less dense outer layer in the cloud merging process.


\citet{Pouteau22} reveal that the core mass function inside MM2 and MM3 appear top-heavy, and there is an evolutionary trend of pre-stellar vs. protostellar core mass function possibly resulting from accretion (\citealt{Nony23}).
The very dense and turbulent gas is common feature of massive clouds ($\sim$100\,$M_{\odot}$) in collision (\citealt{IF13}, \citealt{Fukui16}). Indeed, besides the central regions of MM2 and MM3 where the most massive cores reside in, the secondary dense gas enhancements probed by CH$_{3}$OH (2-1) lines are also correlated with line wings of SiO emission, as shown in Fig. \ref{fig:sio_maps1}. Although the terminal velocities of the SiO emission are relatively moderate, these excessive velocity wings likely result from the outflow activities from the embedding protostellar cores, especially in clump MM2 and MM3 (\citealt{Nony23}). There is also possibility that the much higher velocity SiO wings, which fit better in typical high-velocity jets, still remain undetected with our achieved sensitivity. In any case, this association provides the direct link of supersonic shocks and enhancements of gas density.

The gas kinematics seen from these 0.1-1 pc scale observations is consistent with the larger scale $^{13}$CO observations (see also left panel of Fig. 1) which reveal that a 82 km s$^{-1}$ and a 94 km s$^{-1}$ cloud are intersecting at the southern ridge of W43-main (\citealt{Kohno21}). Our observations likely represent dense gas tips of the shock-compressed layer close to the red-shifted 94 km s$^{-1}$ cloud.  

\subsection{Column density of NH$_{2}$D and comparison with chemical models}\label{sec:nh2d_chem}

\begin{figure*}
\begin{tabular}{p{0.5\linewidth}p{0.5\linewidth}}
   \hspace{-0.3cm}\includegraphics[scale=0.55]{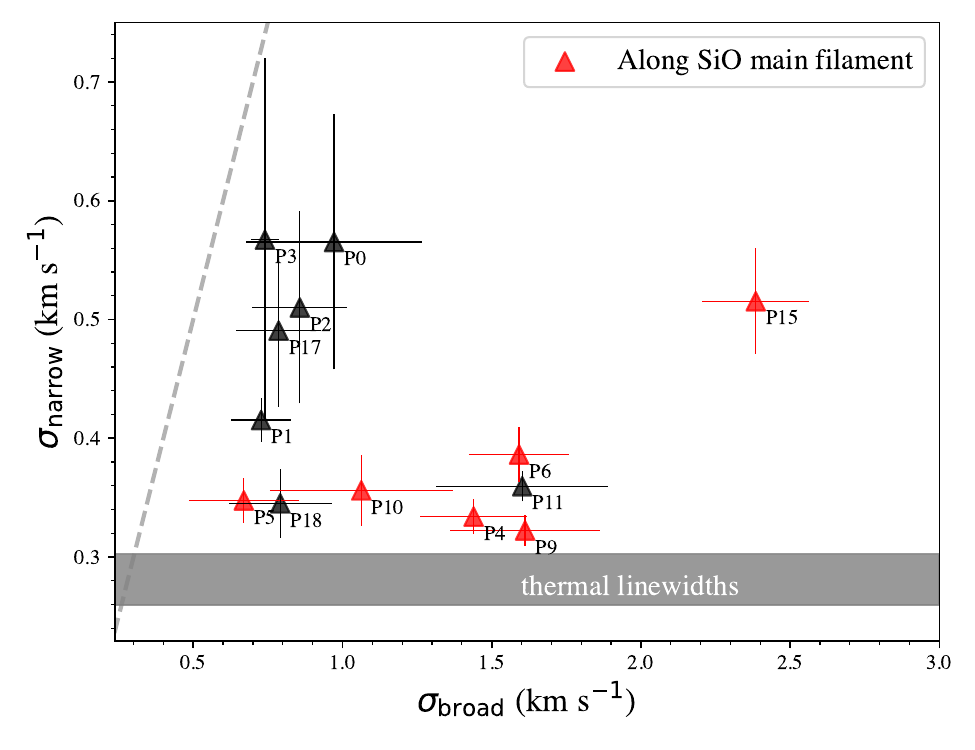}&\hspace{-0.3cm}\includegraphics[scale=0.55]{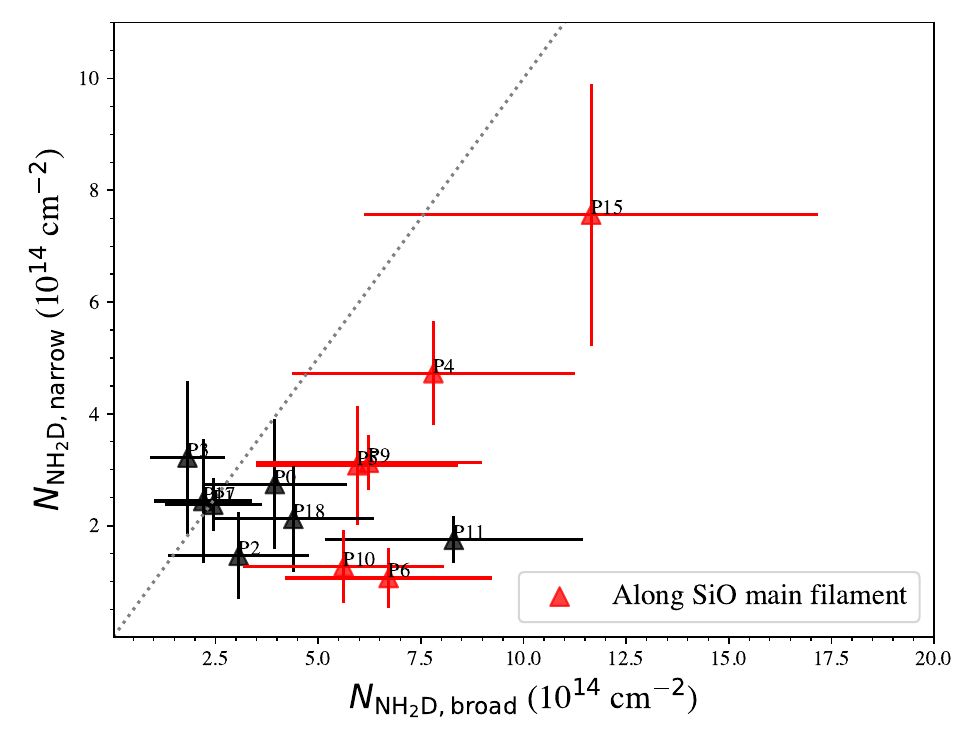}\\
    \end{tabular}
    \caption{{\it{Left: }}Relation between the narrow linewidths and broad linewidths of spectra of NH$_{2}$D cores residing in clumps MM2, MM3 and C. The gray dashed line indicates equal linewidths. NH$_{2}$D cores that reside along the main filament of SiO emission are marked in red, and others in black. 
    The shaded region indicates the thermal linewidths based on $T_{\mathrm{rot}}$ map derived by CH$_{3}$CCH lines (Section \ref{sec:temp_dens}). 
    {\it{Right:}} Comparison between column density of NH$_{2}$D of the narrow linewidth and broad linewidth components. The gray dashed line indicates equal column density.}
    \label{fig:nh2d_core_velo_statistics}
    \end{figure*}

The formation of deuterated ammonia favors cold ($<$\,20\,K) and dense environments, where the reaction to form H$_{2}$D$^+$ which then transfers to NH$_{3}$ to form NH$_{2}$D can proceed without the H$_{2}$D$^+$ being destroyed by gas-phase CO (\citealt{Caselli08}, \citealt{Sipila15}), as CO depletes by freeze-out onto dust grains (\citealt{Tafalla02}, \citealt{Crapsi05}). Thus NH$_{2}$D emission is expected to preferentially trace the cold and dense gas, e.g., prior to protostellar activity (\citealt{Pillai11}) or other heating mechanisms. 
As can be seen in Figure \ref{fig:nh2d_core_map} (right panel), the NH$_{2}$D emission is mostly uncorrelated with CH$_{3}$OH emission peak and the localised density enhancements on the $n(\mathrm{H_{\mathrm 2}})$ map, which are likely induced by intense shocks and subject to heating. Meanwhile, there are widely distributed gas pockets of compact NH$_{2}$D emission in MM2, which compose a more extended emission area than that in clump MM3 and C.

We calculated the NH$_{2}$D column densities based on the MCMC {\it{hfs}} modeling of the two velocity components (Sect. \ref{sec:nh2d}), taking into consideration the uncertainties of all the fitted parameters and error propagation. 
The uncertainty of the column density associated with the broader linewidth component is larger, which is mostly due to the large uncertainty of $\tau_{m\star}$ (as shown in Fig. D. 3, which then propagates into uncertainties of $T_{\mathrm{ex}}$) as a result of the blending of the satellite components.  
However, within the uncertainties, we find that the column densities of NH$_{2}$D of the broad linewidth component are comparable or even larger than the narrow linewidth counterparts (Fig. \ref{fig:nh2d_core_velo_statistics}). 

 We perform chemical simulations to quantitatively understand the variations of NH$_{2}$D abundance in the shock environments typical to W43-main. For this, we use the gas-grain chemical code pyRate which includes extensive deuterium chemical networks and has been previously used to explain ammonia observations (\citealt{Sipila13, Sipila15, Sipila19a, Sipila19b}). Here we run simple zero-dimensional models and adopt three sets of physical conditions characterising pre-shock, shocked, and post-shock gas. Specifically, the gas density, temperature and extinction level of the gas are listed in Table \ref{tab:shock_cond}. We assume the outer envelopes of the three clumps MM2, MM3 and C represent pre-shock gas, having a gas temperature of $\sim$20 K and gas density 10$^{4.5}\,$cm$^{-3}$. After the shock processing, the gas density is enhanced, approaching typical densities of 10$^{5.5}\,$cm$^{-3}$ as probed by CH$_{3}$OH lines (Sect. \ref{sec:dens}), together with an elevated while moderate temperature of $\sim$40 K, as seen by H$_{2}$CS (Sect. \ref{sec:xclass}). We also assume the post-shock gas efficiently cools down to a temperature similar to pre-shock gas while the gas density maintains the high level. The extinction levels are estimated from the $N(\mathrm{H_{2}})$ map shown in Fig. \ref{fig:w43_npts} (middle panel), namely the environmental extinction of the clumps, shielding the external UV radiation.

\begin{figure}
    \centering
    \includegraphics[width=0.95\linewidth]{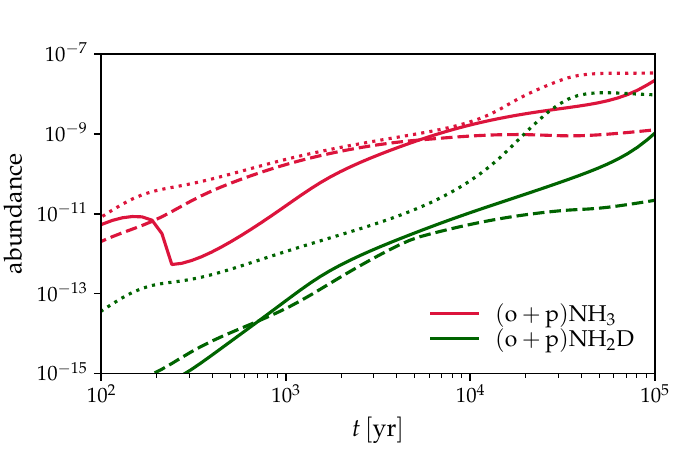}
    \caption{Abundance of (ortho $+$ para) NH$_{3}$ and NH$_{2}$D in shock environment, predicted by chemical models. Static physical conditions of pre-shock (solid line), shocked (dashed line) and post-shock gas (dotted line) are characterised by gas density and temperature, and extinction levels listed in Table \ref{tab:shock_cond}. }
    \label{fig:chem_nh3_noevo}
\end{figure}

\begin{figure*}
\hspace{.5cm}\includegraphics[width=0.95\linewidth]{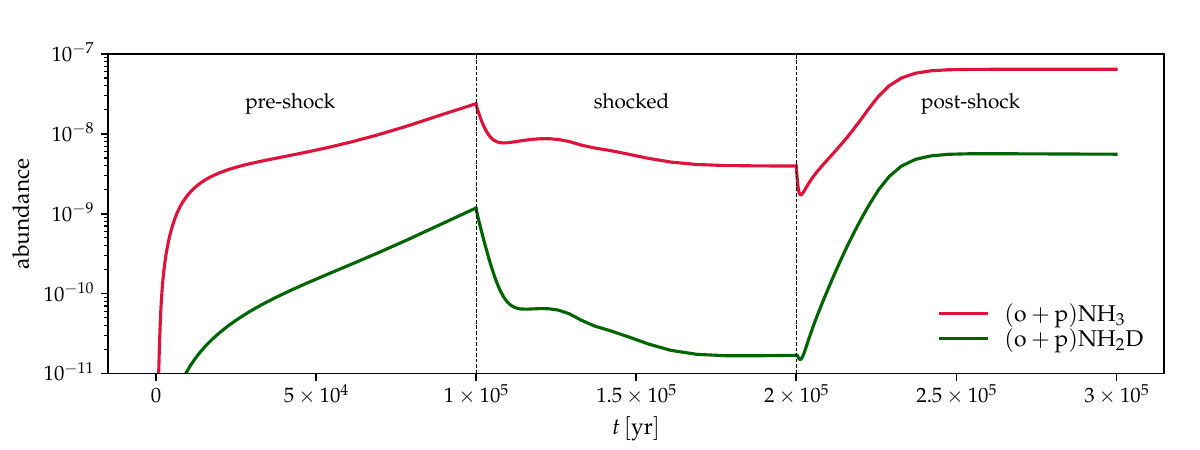}
\caption{Evolution of the abundance of (ortho $+$ para) NH$_{3}$ and NH$_{2}$D in shock environment, predicted by chemical models. Similar to Fig. \ref{fig:chem_nh3_noevo}, three conditions of pre-shock, shocked and post-shock gas are characterised by gas density and temperature, and extinction levels listed in Table \ref{tab:shock_cond}; Contrary to models shown in Fig. \ref{fig:chem_nh3_noevo}, each physical state is evolved for 0.1 Myr, with the latter two states having abundance levels initialised by the end values of the prior state. }
\label{fig:chem_nh3_evo}
\end{figure*}

We run two different sets of chemical models. In the first set, the three physical conditions are allowed to develop independently for 0.1 Myr, from the same set of
(mostly atomic) initial abundances (\citealt{Sipila19a}). We chose 0.1 Myr for the chemicals evolution epoch as it is a typical timescale of the lifetime of the massive protostellar phase for an object of luminosity 10$^{4}$-10$^{5}$ $L_{\odot}$ (e.g., \citealt{Mottram11}), which is also consistent with the longest formation timescale we estimate assuming shock-enhanced accretion to form the stablised NH$_{2}$D cores (Sect. \ref{sec:fmtt_prop}).
This set of static chemical models allows us to spot the variation of NH$_{2}$D abundance which is solely determined by the three distinct physical conditions. The resultant evolution of the NH$_{2}$D abundance is shown in Fig. \ref{fig:chem_nh3_noevo}, together with that of the main isotopologue NH$_{3}$. The lowest NH$_{2}$D
abundances are seen in the shocked gas, and the highest abundances in
the post-shock gas: these results are sensible given the favored conditions for deuteration --high density and low temperature environment. The abundance in the shocked gas is smaller but still comparable to that of the pre-shock gas in the first 0.01 Myr of evolution in this set of chemical models.  

In the second set of chemical model, we consider the evolution of the three conditions consecutively: the initial stage of pre-shock gas
is let evolve for 0.1 Myr, followed by evolution in
the shock stage for another 0.1 Myr which adopts the end result of the first
stage as initial abundances for all the species, and finally another 0.1 Myr with post-shock conditions
with initial abundances collected again from the shock stage. Fig. \ref{fig:chem_nh3_evo} shows the results of this
simulation; there is drastic decrease of the abundances when the shock conditions kick in and also a steady increment in the beginning of the post-shock stage. In fact, after the first $\sim$0.03 Myr in the post-shock stage, the
NH$_{2}$D abundance achieves and then maintains the highest level, among all epochs. 
The observed comparable or even higher NH$_{2}$D column densities associated with the broader linewidth component is compatible with the chemical model predictions: we are likely probing the 
post-shock enhancement of NH$_{2}$D with the broad linewidth velocity component. We acknowledge that our current investigation solely allows us to examine NH$_{2}$D column densities. However, it is important to highlight that the chemical models provide predictions for NH$_{2}$D abundance, introducing a non-direct one-to-one comparison between the observed data and model predictions. On the other hand, the narrow linewidth velocity components observed may predominantly signify pre-shock gas or potentially post-shock gas from the preceding shock event (formed in the previous, lower density-enhanced environment), which have undergone decay through turbulence dissipation. In the context of core formation within shock conditions, uncertainty arises regarding whether these cores serve as remnants from the shock waves preceding the current one or represent the concluding outcomes of the second-to-last shock wave. The centroid velocities of the narrow linewidth components are more widely distributed and appear oscillatory (see more in Sect. \ref{sec:nh2d_core_dyn}) compared to that of the SiO, indicating that their location in spatial and velocity domain is further away from the current shock front. 

Similar results of the enhancement of N$_{2}$H$^+$ deuterium fractionation in post-shock region is suggested in \citet{Cosentino23} towards an infrared-dark cloud adjacent to a supernovae remnant, as an outcome of the fast chemistry in density-enhanced gas. In addition, thermal heating and/or shock sputtering that cause desorption of the grain-surface NH$_{2}$D may also contribute to the observed enhancement. We also note that in our present simulations, which do not take into account the actual shock events, the temperature is never high enough for efficient thermal NH$_{3}$ or NH$_{2}$D desorption.

\subsection{Kinematics of NH$_{2}$D cores}\label{sec:nh2d_core_dyn}

To understand the origin of the NH$_{2}$D cores from a kinematic perspective, we compare the position-velocity distribution of the NH$_{2}$D cores with respect to emission of SiO (2-1), H$^{13}$CO$^+$ (1-0), and CCH 1$_{\mathrm{2,2}}$-0$_{\mathrm{1,1}}$ lines. 
Following three positional cuts that are illustrated in Figure \ref{fig:nh2d_core_cch_pv}, we show the PV diagrams of CCH and SiO lines in the right panel (in colorscale), in which the peak velocities of H$^{13}$CO$^+$, SiO and CCH are shown as reference lines. 
The centroid velocities marking the position of NH$_{2}$D cores on the velocity axis are derived from the one-component or two-component {\emph{hfs}} fits, shown as red and green markers, respectively. 
In the PV diagrams, the zero offsets correspond to the northernmost position of the three cuts. The centroid velocities of the broader velocity component (darker green markers) of NH$_{2}$D cores mostly coincide with the peak velocities of SiO, whereas centroid velocities of the narrow velocity components or of the one-component fit can either be blue-/red-shifted with respect to that of SiO. This indicates that the broader line-widths of NH$_{2}$D are more closely connected to the shock activities. Although associated with large uncertainties, the column density of the broader linewidth components also appear to be larger than their narrow linewidth counterparts, as shown in Fig. \ref{fig:nh2d_core_velo_statistics} right panel. 
These gas components are dominated by cloud collisions, whose turbulent energy has not been dissipated, despite their spatial closeness (along line-of-sight) with the dense gas ridge (main filament of SiO) linking MM2 and MM3. 

To further illustrate the relation of the NH$_{2}$D cores and the bulk gas structures in these massive clumps, we show in Figure \ref{fig:nh2d_core_vp} the distribution of velocity offsets between the centroid velocities of NH$_{2}$D lines and the peak velocity of other molecular lines (SiO, CCH, CH$_{3}$OH, HC$^{15}$N, HC$_{3}$N and H$^{13}$CO$^+$). All lines are extracted and averaged from the position of the NH$_{2}$D cores. 
It can be seen that the centroid velocities of the broad-linewidth velocity components of NH$_{2}$D cores are more consistent with the peak velocities of SiO, CCH and CH$_{3}$OH, with velocity offsets centralised around zero. Interestingly, most of the centroid velocities of the narrow-line-width components are red-shifted or blue-shifted compared to peak velocities of H$^{13}$CO$+$, HC$_{3}$N or HC$^{15}$N. From the PV diagrams in Figure \ref{fig:nh2d_core_cch_pv}, this behaviour can also be seen clearly, and the distribution of the centroid velocity of the narrow-line-width components along the cuts appear oscillatory with respect to the peak velocities of H$^{13}$CO$+$: they alternate from being blue-shifted to to red-shifted and blue-shifted again, from north to south.  
The oscillatory distribution in intensity space is resolved in simulations of cloud-cloud collisions in \citet{Takahira14, Takahira18} which appear orthogonal to the collision axis, and may represent outer gas layers perturbed by the shock activities.

We make a comparison of the line-widths of the two velocity components of the NH$_{2}$D cores identified in clumps MM2, MM3 and C, as shown in the left panel of Figure \ref{fig:nh2d_core_velo_statistics}. We classify the NH$_{2}$D cores into two categories: those that are located close ($\lesssim$0.2 pc) to the main filament of SiO (in-between the thick red and blue lines in the upper panel of Figure \ref{fig:sio_maps1}, mainly the dense gas ridge of MM2 and MM3) and those that are further away. Apparently the difference between the broad and narrow line-width in the second class is smaller, while the cores along the SiO main filament have narrow line-width comparable to the sonic value. There is a large difference between the broad and narrow linewidth of the two velocity components. 

The broad linewidths NH$_{2}$D emission in high mass star-forming regions have been observed with single-dish observations (e.g., \citealt{Fontani15}, \citealt{Li23}). However, compared to other NH$_{2}$D cores identified by interferometric observations in high-mass star forming regions which mostly show narrow linewidths ($\sigma_{\mathrm{v}}$$\lesssim$0.5 km\,s$^{-1}$, \citealt{Pillai11}, \citealt{Zhang20}, \citealt{Li21}) and also H$_{2}$D$^+$ cores (\citealt{Redaelli21}), our discovery of supersonic linewidth NH$_{2}$D gas components appear peculiar and may reflect a recent shock processing of their embedded massive clumps.   


\begin{table}
\begin{threeparttable}
\caption{Physical parameters used for three set of chemical models}
    \label{tab:shock_cond}
    \begin{tabular}{ccccc}
    \toprule

        &$n(\mathrm{H_{2}})$ & $T_{\mathrm{dust}}$$^{a}$ &
$A_{\mathrm{v}}$ \\
             &cm$^{-3}$& K &   (1.1$\times$$10^{-21}$$N_{\mathrm{H_{2}, env}}$)\\
                \midrule
     pre-shock gas& $10^{4.5}$     & 20&        30      \\
     shocked gas&      $10^{5.5}$    & 40&       160\\
     post-shock gas&      $10^{5.5}$    &20 &     160     \\
          \bottomrule
     \end{tabular}
     \begin{tablenotes}
      \item[$^a$]: We assume gas temperature is equal to dust temperature.
    \end{tablenotes}
  \end{threeparttable}
  
\end{table}

\subsection{Fragmentation of the NH$_{2}$D cores}\label{sec:fmtt_prop}

As an assessment of the possible fragmentation process leading to the formation of the NH$_{2}$D cores, we made comparisons between the observed core separation i.e., distance to each core's nearest neighbour, core virial mass and the critical length and mass scales based on Jeans fragmentation predictions.  Towards the region of clumps MM2, MM3 and C, we adopt the clump envelope gas density and the turbulent line-width (the broader velocity component) as one set of parameters, and core gas density and thermal line-widths as another, for Jeans fragmentation assessment. We regard these two parameter sets as representing roughly the properties of large-scale clump bulk gas, and localised dense gas. The core densities are yielded from the CH$_{3}$OH derived n(H$_{2}$) maps, and the envelope density is adopted as a characteristic bulk gas density of 10$^{5}$ cm$^{-3}$ for these clumps (\citealt{Lin16}). \citet{Takahira18} find that the core formation from cloud-cloud collisions is dominated by accretion of shocked gas. We regard the core virial masses based on estimate from the narrow linewidth components represent typical core mass formed in this shock environments (although we mention in Sect. \ref{sec:nh2d_chem} that these narrow linewidth objects may represent cores formed in last shock activities), and following formulations in \citet{Takahira18}, we can then calculate core formation timescale by assuming a characteristic accretion rate and isotropic infall. The accretion rate is estimated by $\dot{M}$ = $\pi$$r_{\mathrm{acc}}^{2}$$\sigma_{\mathrm{eff}}$$\rho_{\mathrm{acc}}$, where $r_{\mathrm{acc}}$ = $\frac{2GM_{\mathrm{core}}}{\sigma_{\mathrm{eff}}^{2}}$ + $r_{\mathrm{core}}$, and $\rho_{\mathrm{acc}}$ is the average ambient gas density (within $r_{\mathrm{acc}}$. excluding the core). Taking $\rho_{\mathrm{acc}}$ from the $n(\mathrm{H_{2})}$ map, and $M_{\mathrm{core}}$ as $M_{\mathrm{vir}}$, $\sigma_{\mathrm{eff}}$ as the $\sigma_{\mathrm{v}}$ as that of the broad velocity component, we calculate a core formation timescale as $t_{\mathrm{form}}$ = $\frac{M_{\mathrm{vir}}}{\dot{M}}$.
In the left panel of Figure \ref{fig:nh2d_fmtt} we show the relation between the core virial mass and separation, and the ratio of the separations to the estimated critical Jeans lengths. In the right panel of Figure \ref{fig:nh2d_fmtt}, the comparison between core virial mass and core formation timescale, and the ratio of virial masses and critical Jeans masses are shown.

The observed separation between cores and the core virial mass does not particularly favour predictions of either turbulent fragmentation of the envelope, or thermal fragmentation of the in-situ dense gas. In general, the separations seem to be more close to predictions of turbulent fragmentation of the envelope, but the mass scales lie in-between the two scenarios. We note again that the virial mass estimated from the narrow linewidth NH$_{2}$D line is used as a proxy for core mass, which may be several times underestimated (\citealt{Pillai11}, \citealt{Zhang20}), compared to the dust mass. This will make the difference between the core mass and that of the turbulent fragmentation smaller, but still cannot single out this scenario. The formation timescale $t_{\mathrm{form}}$ of these NH$_{2}$D cores is weakly correlated with core mass, with a Spearman correlation coefficient of 0.37 (p-value $\sim$ 0.1), ranging between 1$\times$10$^{3}$-10$^{5}$ yr.

\begin{figure*}
\centering
\begin{tabular}{p{0.95\linewidth}}
\includegraphics[scale=0.4]{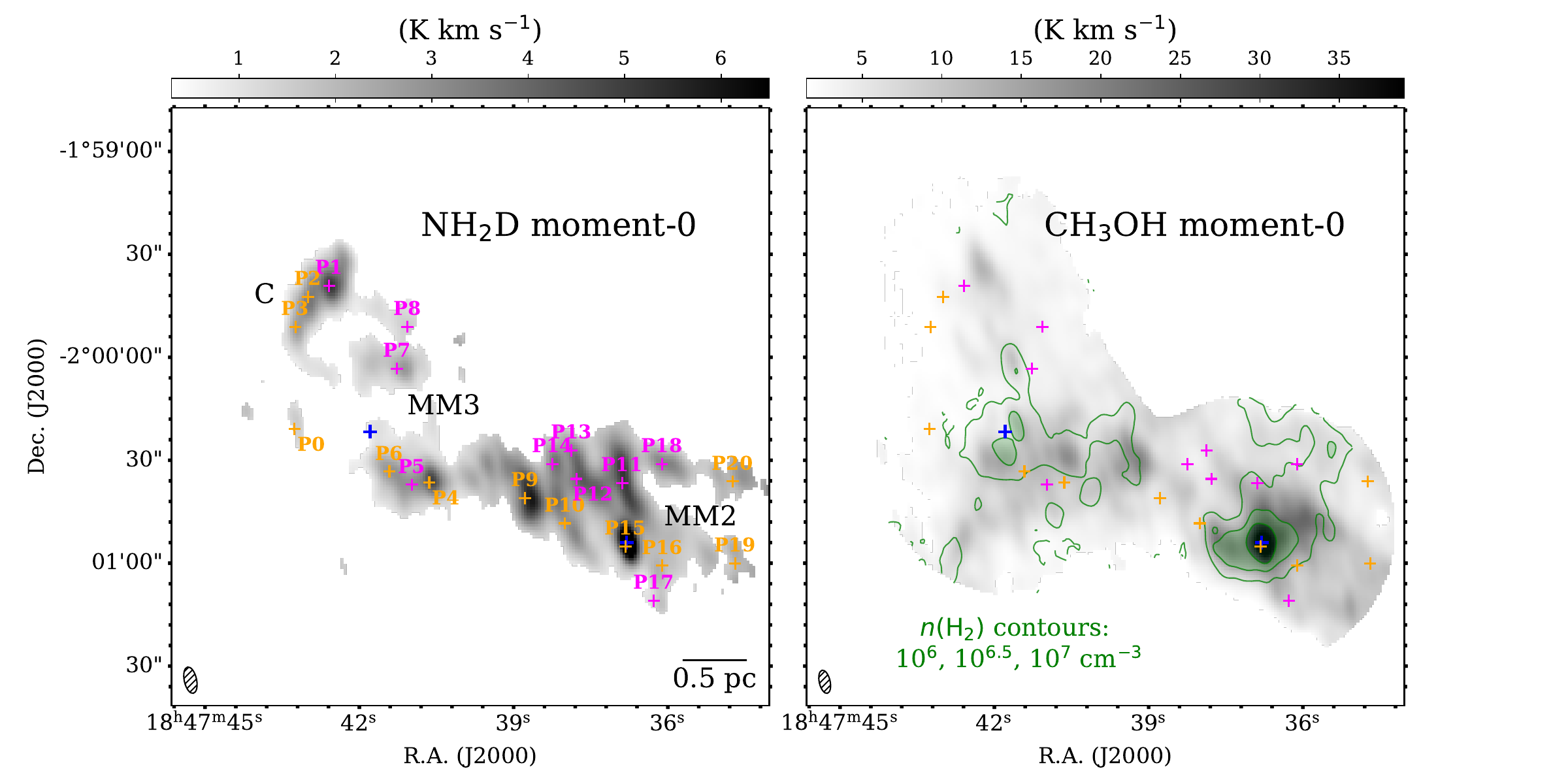}\\
\includegraphics[scale=0.4]{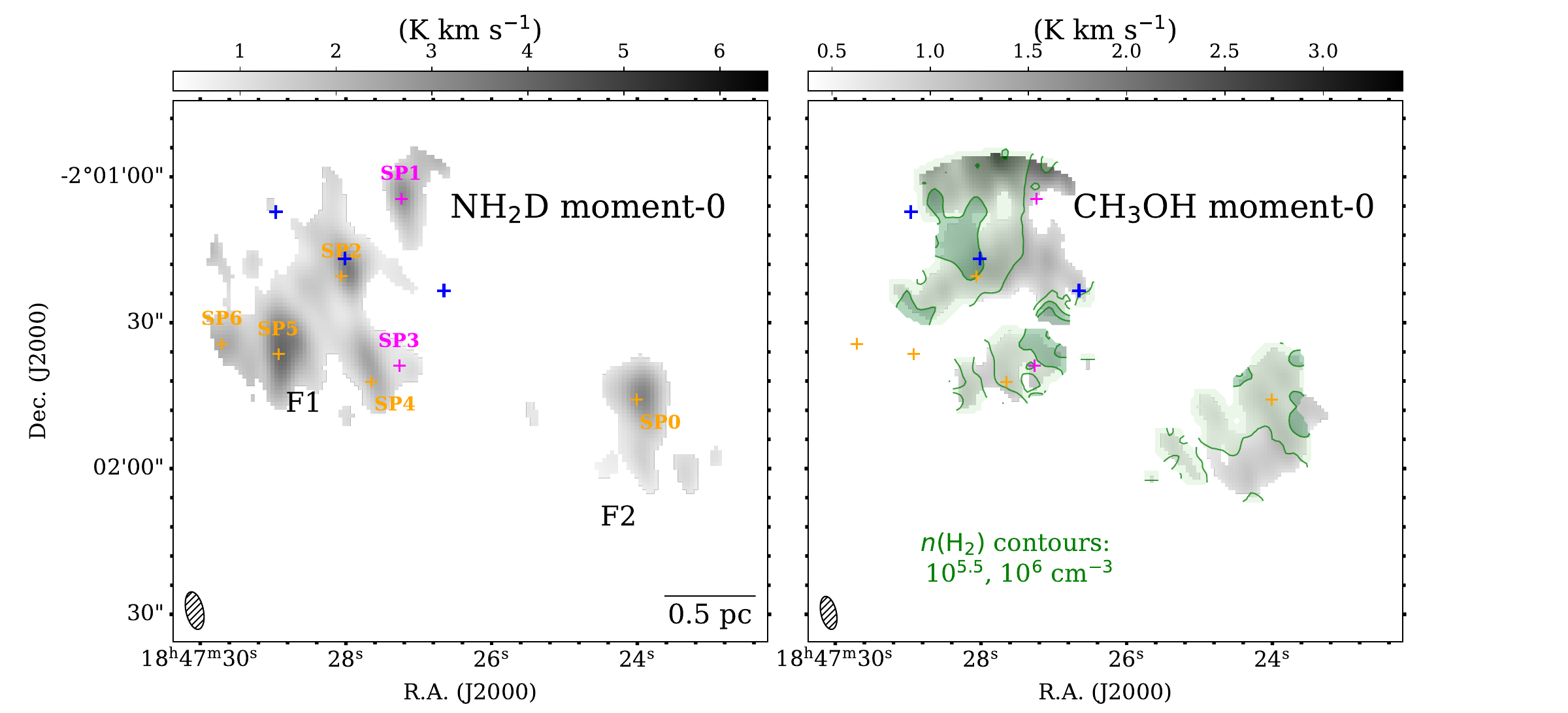}\\
\end{tabular}
\caption{{\emph{Upper left:}} Integrated intensity map of NH$_{2}$D line emission of clumps MM2, MM3 and C (colorscale, of velocity range 80-115 km s$^{-1}$) marked with locations of NH$_{2}$D cores by crosses of different colors: the orange crosses indicate the cores which have a $\sigma_{\mathrm{v}}$$>$0.85 km s$^{-1}$ velocity components and magenta crosses represent cores that only have narrow velocity components, $\sigma_{\mathrm{v}}$$\lesssim$0.5 km s$^{-1}$. {\emph{Upper right}}: The integrated intensity map of CH$_{3}$OH 2(0,2)-1(0,1) line (colorscale, of velocity range 85-100 km s$^{-1}$) overlaid by contours of n(H$_{2}$) map (green). In both plots, the blue crosses mark the position of the peak intensity of the 3 mm continuum of clumps MM2 and MM3. {\emph{Lower panels:}} same as upper panels, but for the clumps F1 and F2. In both plots, the blue crosses mark the position of the 3 cores in 3 mm continuum of clump F1 (as in Figure \ref{fig:rgb_cont}, right panel).}
\label{fig:nh2d_core_map}
\end{figure*}


\begin{figure*}
\begin{tabular}{p{0.32\linewidth}p{0.68\linewidth}}
\vspace{-6.5cm}\hspace{-.5cm}\includegraphics[width=\linewidth]{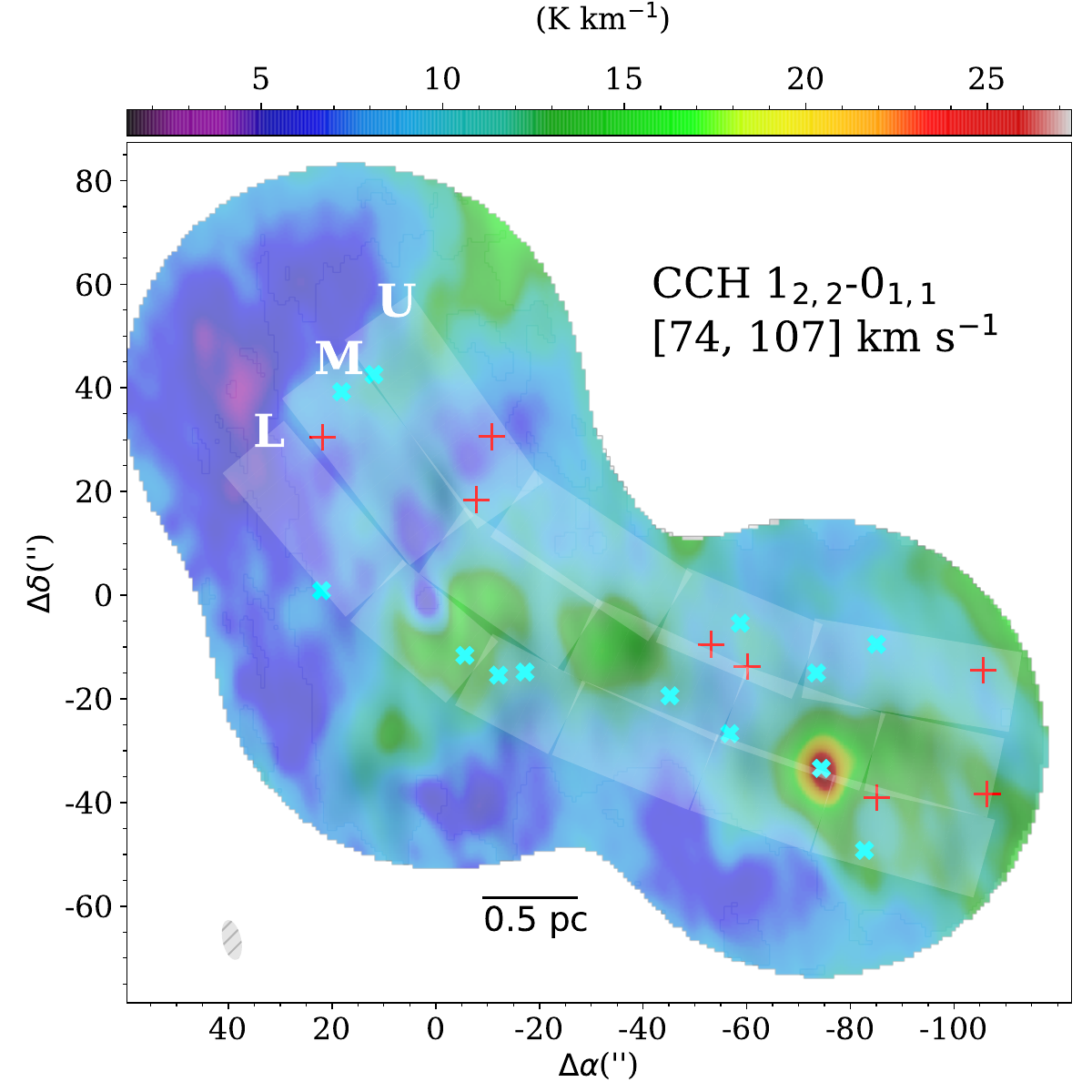}&\hspace{-1cm}\includegraphics[width=1.05\linewidth]{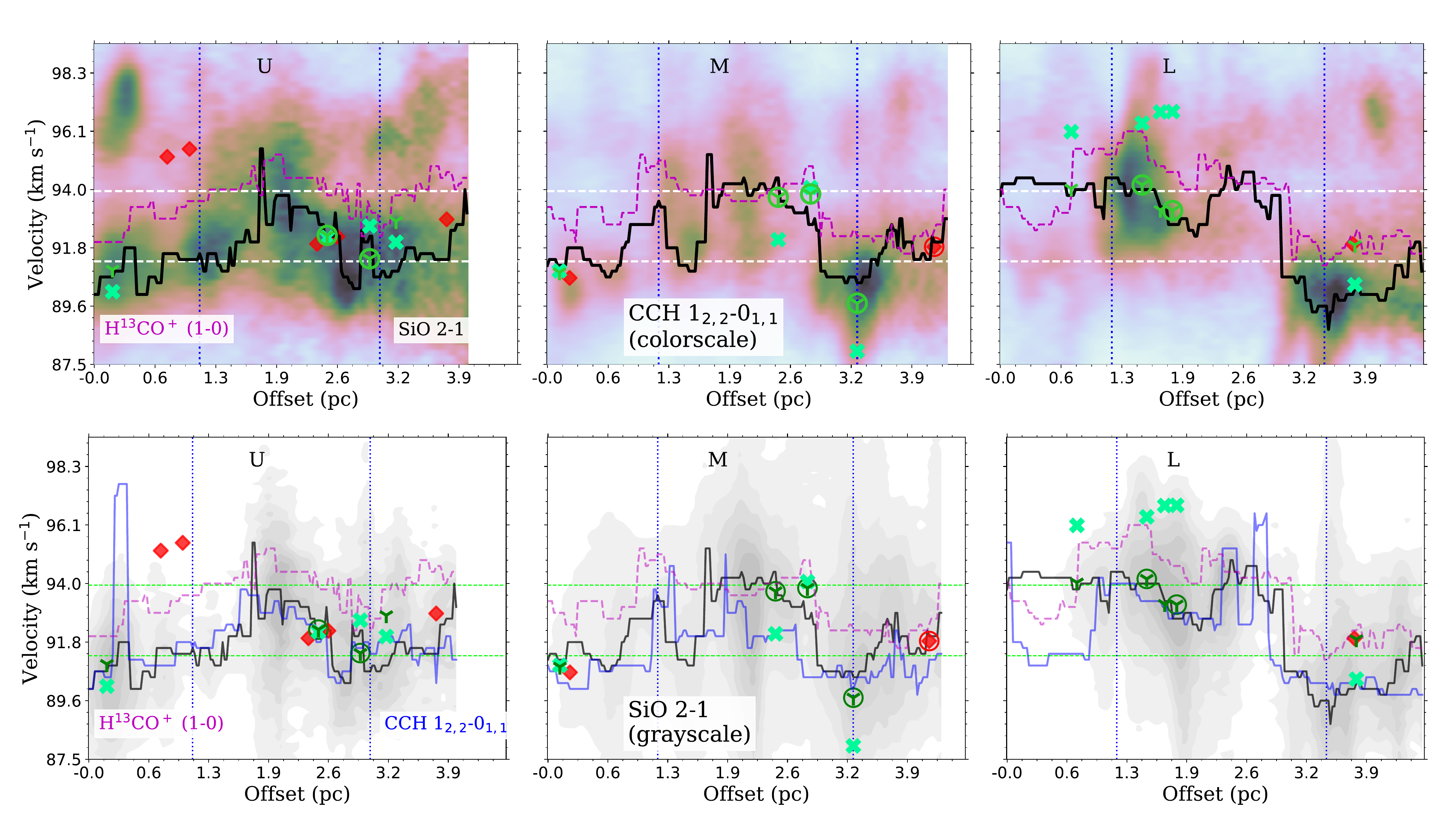}\\

\end{tabular}
\caption{{\it{Left:}} Position of NH$_{2}$D cores on CCH integrated intensity map. Red crosses mark the NH$_{2}$D cores that have two velocity components, and cyan crosses mark those of one velocity component.{\it{Right:}} Relation of NH$_{2}$D cores with CCH and H$^{13}$CO$^+$ emission, shown as position-velocity diagram. The three panels are position-velocity cuts along the shaded bands U, M and L shown in {\it{left}} figure. Peak velocities along the spatial offset for different lines are shown as dashed or solid lines with respective colors (H$^{13}$CO$^+$: magenta, SiO: black, CCH: blue). Background color scale show the PV diagram of CCH (in upper panel) and SiO (in lower panel). NH$_{2}$D cores with two velocity components are marked as green X and three-branched triangles, and those with only one velocity component are shown as red diamonds. NH$_{2}$D cores that have $\sigma_{\mathrm{v}}$>0.85 km s$^{-1}$ are marked with an additional hollow circle around the aforementioned markers.}
\label{fig:nh2d_core_cch_pv}
\end{figure*}

\begin{figure*}
\begin{tabular}{p{0.975\linewidth}}
\hspace{0.25cm}\includegraphics[scale=0.475]{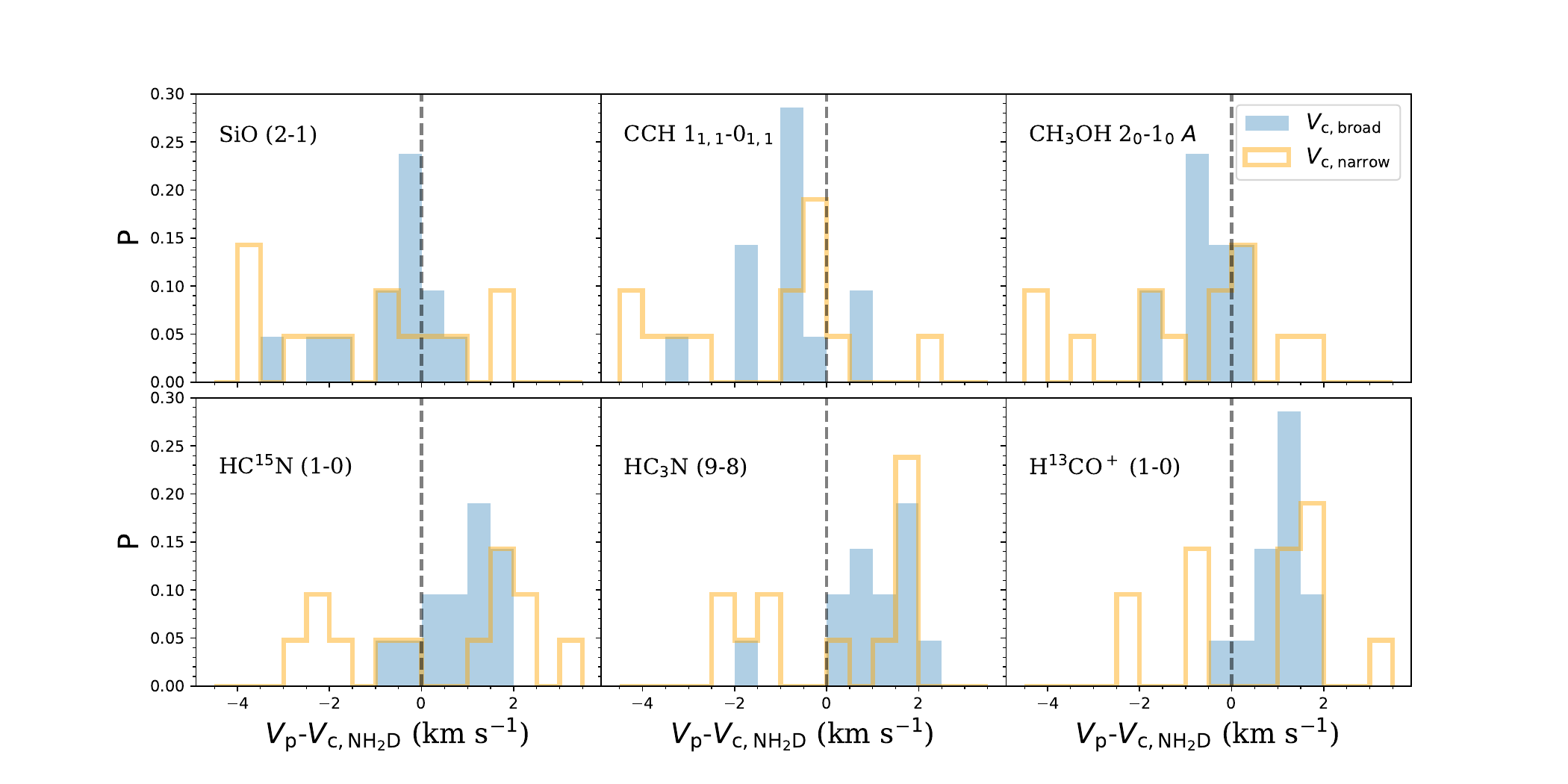}\\
\end{tabular}
\caption{Probability distribution of velocity offset between peak velocities of multiple lines with respect to the centroid velocity of NH$_{2}$D lines of NH$_{2}$D cores. The $V_{\mathrm{p}}$ stands for peak velocity of the corresponding molecular line noted in upper left corner of each subplot; $V_{\mathrm{c, NH_{2}D}}$ indicate the centroid velocity of the broad-linewidth component or the narrow-linewidth component of the two-component fits to NH$_{2}$D lines (Figure \ref{fig:nh2d_sps}).}
\label{fig:nh2d_core_vp}
\end{figure*}

\begin{figure*}
\begin{tabular}{p{0.5\linewidth}p{0.5\linewidth}}
   \hspace{-0.3cm}\includegraphics[scale=0.48]{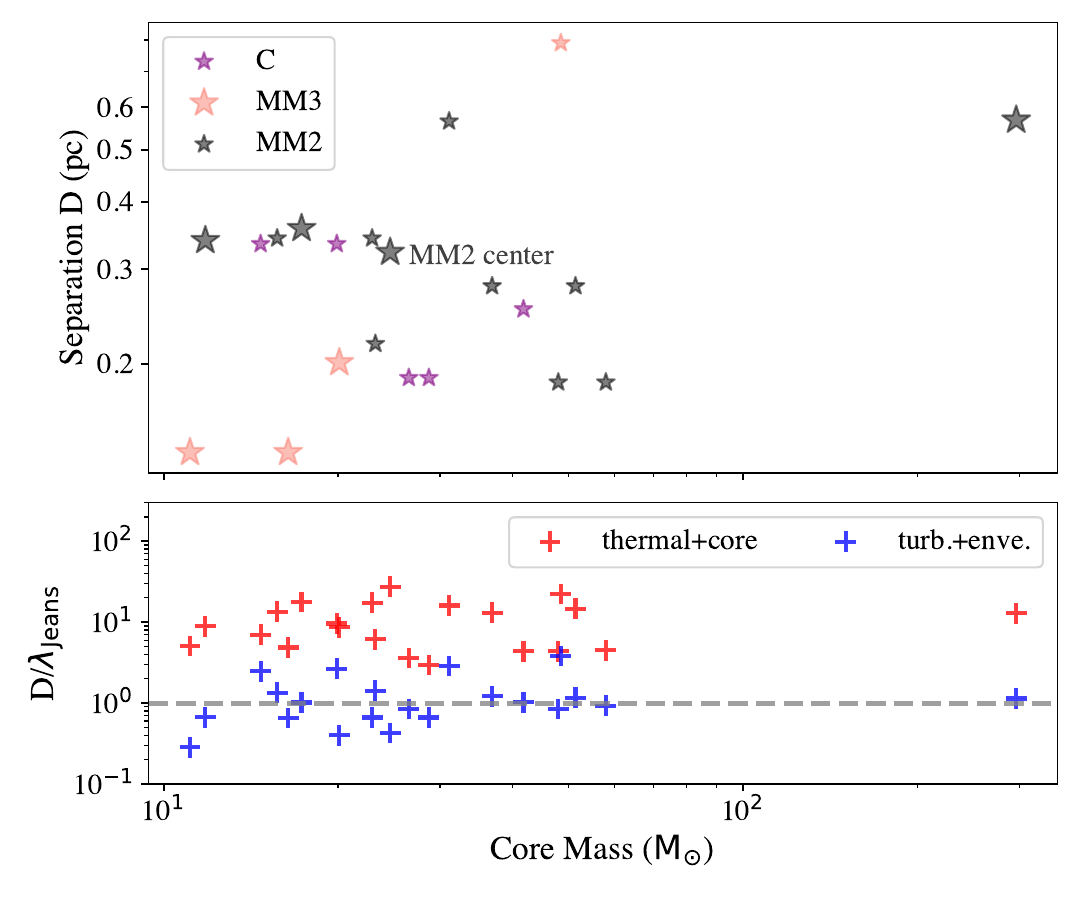}&
   \hspace{-.35cm}\includegraphics[scale=0.48]{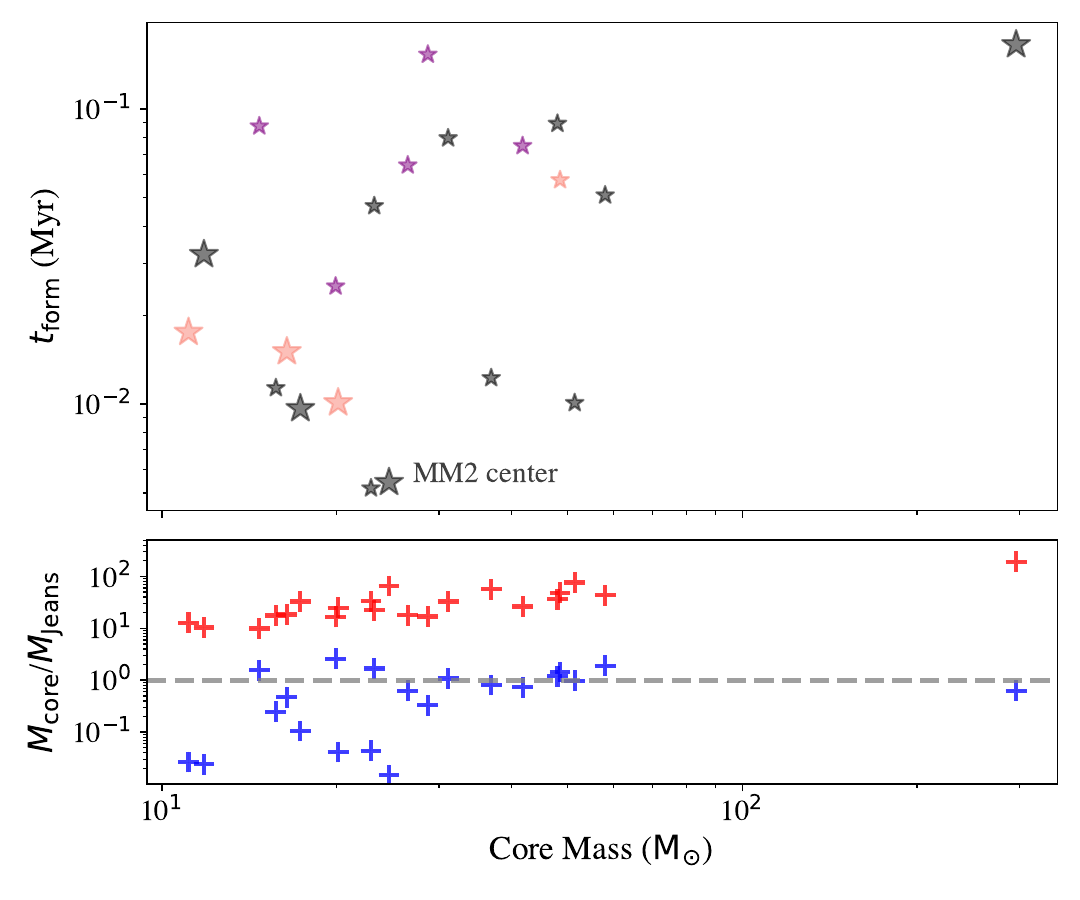}\\
    \end{tabular}
   \caption{{\emph Left panel:} The relation of mass of NH$_{2}$D cores (virial mass) in G30.7 and their separation (to the nearest neighbor) (upper panel); the ratio between the separation and the critical length scale estimated from Jeans fragmentation (lower panel). {\emph Right panel}: The relation of mass of NH$_{2}$D core (virial mass) and formation timescale (upper panel). The ratio between the separation and the critical mass scale estimated from Jeans fragmentation (lower panel). The plots in the right panel share the same labels as in the legend of the left panel. In both upper plots, the cores that have $\sigma_{\mathrm{v}}$ > 0.85 km s$^{-1}$ in at least one velocity component is shown as enlarged markers.}
    \label{fig:nh2d_fmtt}
    \end{figure*}


\begin{table*}
\begin{threeparttable}
\scriptsize
\caption{Properties of {\tt{dendrogram}} extracted NH$_{2}$D cores that reside in clumps MM2, MM3 and C.}
    \label{tab:dendro_nh2d}
    \begin{tabular}{cccccccccc}
    \toprule
        Core & R.A. & Decl. & $R_{\mbox{\scriptsize{eff}}}$$^{a}$&$\Delta V$$^b$ & $V_{\mbox{\scriptsize{c}}}$$^b$&$\tau_{\mbox{\scriptsize{m}}}$&$T_{\mbox{\scriptsize{ex}}}$&$N_{\mathrm{NH_{2}D}}$& $M_{\mbox{\scriptsize{vir}}}$$^{c}$\\
                &(J2000)&(J2000)& (pc) & (km s$^{\mbox{\scriptsize{-1}}}$)&(km s$^{\mbox{\scriptsize{-1}}}$)&&(K)&$\times 10^{\mbox{\scriptsize{13}}}$ (cm${\mbox\scriptsize{^{-2}}}$)&($M_{\odot}$)\\
                \midrule
 P0 & 18\rah47\ram43\ras.20 & -02\decd00\decm20\decs0 &   0.10 &          1.5(0.1) &           96.0(0.1) &          0.9(0.0) &          3.0(0.0) &           11.2(2.3) &   48.5 \\
         P1 & 18\rah47\ram42\ras.52 & -01\decd59\decm38\decs3 &   0.13 & 1.6(0.5),1.0(0.1) & 91.1(0.4),90.2(0.0) & 0.4(0.2),0.6(0.1) & 3.5(0.5),4.7(0.4) &   2.4(2.3),2.2(0.8) &   26.5 \\
         P2 & 18\rah47\ram42\ras.94 & -01\decd59\decm41\decs6 &   0.10 & 2.1(0.5),1.2(0.3) & 90.9(0.2),90.9(0.1) & 0.3(0.2),0.4(0.2) & 3.8(0.9),4.5(1.0) &   2.8(3.2),1.3(1.4) &   28.7 \\
         P3 & 18\rah47\ram43\ras.18 & -01\decd59\decm50\decs3 &   0.16 & 1.8(0.2),1.1(0.5) & 90.7(0.1),95.7(2.5) & 0.4(0.2),0.4(0.3) & 4.2(0.9),3.1(0.4) &   2.4(2.8),2.2(1.8) &   41.8 \\
         P4 & 18\rah47\ram40\ras.58 & -02\decd00\decm35\decs7 &   0.08 & 3.5(0.9),0.8(0.1) & 93.2(0.3),96.8(0.0) & 0.5(0.2),0.8(0.1) & 3.4(0.4),3.7(0.1) &   7.6(6.7),4.7(1.7) &   11.1 \\
         P5 & 18\rah47\ram40\ras.92 & -02\decd00\decm36\decs3 &   0.12 & 1.5(0.8),0.8(0.1) & 93.2(0.2),96.8(0.0) & 0.7(0.2),0.7(0.2) & 3.4(0.2),3.7(0.3) &   5.8(4.7),3.2(2.2) &   16.4 \\
         P6 & 18\rah47\ram41\ras.35 & -02\decd00\decm32\decs4 &   0.12 & 3.7(0.8),0.9(0.1) & 94.2(0.4),96.4(0.0) & 0.4(0.2),0.4(0.2) & 3.2(0.3),4.3(0.8) &   6.6(5.0),1.1(1.2) &   20.1 \\
         P7 & 18\rah47\ram41\ras.21 & -02\decd00\decm02\decs4 &   0.11 &          0.9(0.0) &           95.5(0.0) &          0.8(0.0) &          3.5(0.0) &            6.1(0.8) &   19.9 \\
         P8 & 18\rah47\ram41\ras.01 & -01\decd59\decm50\decs2 &   0.07 &          1.0(0.0) &           95.2(0.0) &          0.5(0.1) &          3.8(0.3) &            2.1(1.1) &   14.7 \\
         P9 & 18\rah47\ram38\ras.72 & -02\decd00\decm40\decs2 &   0.10 & 3.7(1.0),0.8(0.1) & 93.6(0.6),92.1(0.0) & 0.4(0.2),0.8(0.1) & 3.3(0.4),4.4(0.2) &   7.0(6.0),3.3(1.1) &   11.8 \\
        P10 & 18\rah47\ram37\ras.95 & -02\decd00\decm47\decs5 &   0.11 & 2.6(1.3),0.8(0.1) & 94.0(1.0),94.0(0.0) & 0.5(0.3),0.5(0.2) & 3.1(0.2),4.7(0.9) &   5.4(4.6),1.2(1.2) &   17.3 \\
        P11 & 18\rah47\ram36\ras.83 & -02\decd00\decm35\decs8 &   0.15 & 3.8(1.2),0.8(0.1) & 91.3(0.7),92.6(0.0) & 0.5(0.3),0.6(0.1) & 3.1(0.2),5.0(0.5) &   8.2(6.1),1.7(0.9) &   22.9 \\
        P12 & 18\rah47\ram37\ras.72 & -02\decd00\decm34\decs6 &   0.09 &          1.1(0.0) &           92.2(0.0) &          0.7(0.0) &          4.5(0.1) &            3.5(0.5) &   23.2 \\
        P13 & 18\rah47\ram37\ras.81 & -02\decd00\decm26\decs2 &   0.13 &          1.5(0.1) &           92.2(0.0) &          0.6(0.1) &          3.9(0.2) &            4.1(1.3) &   58.0 \\
        P14 & 18\rah47\ram38\ras.19 & -02\decd00\decm30\decs3 &   0.09 &          1.6(0.1) &           91.9(0.0) &          0.7(0.1) &          3.8(0.1) &            6.2(1.4) &   48.0 \\
        P15 & 18\rah47\ram36\ras.77 & -02\decd00\decm54\decs2 &   0.08 & 5.6(0.9),1.2(0.2) & 89.8(0.2),88.0(0.1) & 0.5(0.2),0.8(0.2) & 4.0(0.5),3.2(0.1) & 12.5(11.3),7.7(4.5) &   24.6 \\
        P16 & 18\rah47\ram36\ras.05 & -02\decd00\decm59\decs9 &   0.12 & 1.8(0.9),1.4(0.4) & 91.0(0.5),92.3(0.2) & 0.5(0.3),0.3(0.2) & 3.2(0.3),4.0(0.8) &   3.7(3.5),1.4(1.6) &   51.4 \\
        P17 & 18\rah47\ram36\ras.21 & -02\decd01\decm10\decs0 &   0.12 & 1.7(0.7),1.2(0.3) & 91.9(0.4),90.5(0.1) & 0.3(0.2),0.4(0.2) & 3.5(0.5),3.7(0.5) &   2.5(2.6),2.2(2.2) &   36.9 \\
        P18 & 18\rah47\ram36\ras.06 & -02\decd00\decm30\decs3 &   0.11 & 1.9(0.8),0.8(0.1) & 92.8(0.5),92.0(0.1) & 0.5(0.3),0.6(0.2) & 3.2(0.3),3.9(0.4) &   4.7(4.0),2.4(2.0) &   15.7 \\
        P19 & 18\rah47\ram34\ras.63 & -02\decd00\decm59\decs2 &   0.10 &          3.7(0.7) &           91.8(0.2) &          0.1(0.6) &         5.2(13.2) &            0.8(8.1) &  296.4 \\
        P20 & 18\rah47\ram34\ras.68 & -02\decd00\decm35\decs3 &   0.07 &          1.4(0.1) &           92.9(0.1) &          0.8(0.1) &          3.3(0.1) &            8.5(3.4) &   31.1 \\
          \bottomrule
     \end{tabular}
      \begin{tablenotes}
      \item[$^a$]: $R_{\mathrm{eff}}$ is the effective radius of the core area, i.e. $\pi$$R_{\mathrm{eff}}$$^{2}$ = Area;
      \item[$^b$]: $\Delta V$ and $V_{\mathrm{c}}$ are the FWHM linewidth and centroid velocity from Gaussian line profile, listed for two-component fits and one-component fit accordingly; \item[$^c$]: For the two-component fits, $M_{\mathrm{vir}}$ is calculated using the narrower linewidth.
    \end{tablenotes}
  \end{threeparttable}
\end{table*}

\begin{table*}[]
\begin{threeparttable}
\scriptsize
\caption{Properties of {\tt{dendrogram}} extracted NH$_{2}$D cores that reside in clump F1 and F2.}
    \label{tab:dendro_nh2ds}
    \begin{tabular}{cccccccccc}
    \toprule

        Core & R.A. & Decl. & $R_{\mbox{\scriptsize{eff}}}$$^{a}$&$\Delta V$$^b$ & $V_{\mbox{\scriptsize{c}}}$$^b$&$\tau_{\mbox{\scriptsize{m}}}$&$T_{\mbox{\scriptsize{ex}}}$&$N_{\mathrm{NH_{2}D}}$& $M_{\mbox{\scriptsize{vir}}}$$^{c}$\\
                &(J2000)&(J2000)& (pc) & (km s$^{\mbox{\scriptsize{-1}}}$)&(km s$^{\mbox{\scriptsize{-1}}}$)&&(K)&$\times 10^{\mbox{\scriptsize{13}}}$ (cm${\mbox\scriptsize{^{-2}}}$)&($M_{\odot}$)\\
                \midrule
                          SP0 & 18\rah47\ram23\ras.95 & -02\decd01\decm45\decs0 &   0.21 &          2.4(0.2) &            189.3(0.1) &          0.5(0.1) &          3.3(0.1) &          5.7(2.2) &  244.8 \\
        SP1 & 18\rah47\ram27\ras.18 & -02\decd01\decm03\decs6 &   0.12 & 2.1(1.5),1.4(0.2) & 192.3(1.2),195.7(0.1) & 0.5(0.3),0.5(0.2) & 3.0(0.2),3.8(0.6) & 4.8(4.5),2.8(2.8) &   48.3 \\
        SP2 & 18\rah47\ram28\ras.00 & -02\decd01\decm19\decs6 &   0.10 & 3.0(0.8),1.1(0.1) & 197.1(0.6),198.1(0.0) & 0.4(0.2),0.5(0.2) & 3.2(0.3),4.1(0.6) & 5.2(3.7),1.8(1.5) &   24.0 \\
        SP3 & 18\rah47\ram27\ras.21 & -02\decd01\decm38\decs0 &   0.12 &          1.1(0.1) &            197.7(0.0) &          0.6(0.1) &          3.3(0.1) &          3.7(1.1) &   29.0 \\
        SP4 & 18\rah47\ram27\ras.59 & -02\decd01\decm41\decs3 &   0.18 &          5.5(1.7) &            194.9(0.1) &          0.1(0.7) &         4.2(10.2) &         1.6(20.2) & 1156.2 \\
        SP5 & 18\rah47\ram28\ras.86 & -02\decd01\decm35\decs6 &   0.14 & 1.1(2.6),0.8(0.1) & 194.8(1.0),198.2(0.0) & 0.6(0.3),0.7(0.1) & 3.2(0.3),4.0(0.2) & 3.0(7.6),2.7(1.1) &   19.8 \\
        SP6 & 18\rah47\ram29\ras.64 & -02\decd01\decm33\decs5 &   0.10 & 2.8(1.6),1.0(0.2) & 195.9(1.0),199.2(0.1) & 0.5(0.3),0.5(0.3) & 3.0(0.2),3.4(0.4) & 6.1(4.8),2.1(2.0) &   21.2 \\
          \bottomrule
     \end{tabular}
     \begin{tablenotes}
      \item[$^a$$^b$$^c$]: Same as Table \ref{tab:dendro_nh2d}.
    \end{tablenotes}
  \end{threeparttable}
  
\end{table*}

\section{Conclusion}
We conduct 3 mm NOEMA and IRAM 30m observations towards massive clumps in W43-main molecular complex, with the primary goal of constraining the physical structures. The clumps are residing in a highly turbulent environment and may be formed coevally as indicated by their similar luminosity to mass ratios. Our results are summarized as follows: 
\begin{itemize}
    \item We derive the rotational temperature maps using H$_{2}$CS (3-2), CH$_{3}$CCH (6-5), (5-4), and C$_{2}$H$_{3}$CN (9-8) lines as thermometers, and the hydrogen volume density maps from the CH$_{3}$OH (2-1) lines. The gas temperatures of clump MM3 show the most centrally peaked profile, reaching 130 K in the center and levelling to 30 K at a radius of 0.3 pc. Clump MM3 has temperatures of $\sim$40-50 K at the peak of the line emission, which is offset from the peak of the 3\,mm continuum emission. There are two distinct velocity components towards clump C, one with bulk temperature 25-35 K and one with $\lesssim$20 K, as traced by CH$_{3}$CCH. The two rotational temperatures derived by H$_{2}$CS show less contrast, which may indicate that H$_{2}$CS is mostly tracing the lower density, shocked envelope gas surrounding the CH$_{3}$CCH emission.
    \item The position-velocity maps of SiO (2-1) and CH$_{3}$OH 5$_{1,5}$-4$_{0,4}$ along the main filament linking MM2 and MM3 show multiple double-wing velocity profiles, with linewidths $\lesssim$8 km s$^{-1}$. The line wings in CH$_{3}$OH 5$_{1,5}$-4$_{0,4}$ all have counterparts in SiO (2-1) profiles, while SiO (2-1) exhibits additionally single blue-shifted line wings, reaching terminal velocities of $\gtrsim$10 km s$^{-1}$ around the $V_{\mathrm{LSR}}$. With possible confusion from outflows driven by embedded clusters in the formation, the relatively narrow-wing, spatially extended and highly asymmetric SiO line emission likely arises from cloud collisions. Double-peaked CCH (1-0) emission likely traces the larger scale merging clouds of less dense gas.
    \item The gas density enhancements ($\gtrsim$10$^{6}$\,cm$^{-3}$) seen by CH$_{3}$OH across clumps MM2, MM3 and C are strongly correlated with SiO line wings. The most significant density enhancements ($\sim$10$^{7}$\,cm$^{-3}$) likely represent strongly self-gravitating regions that are actively forming stars. The secondary density enhancements are more extended and widely distributed, showing the direct link between dense gas concentration and the impact of supersonic turbulence.
    \item Systematic velocity shifts are observed among HC$_{3}$N, H$^{13}$CO$^+$, SO, H$_{2}$CS, and SiO, CH$_{3}$OH, CCH, CH$_{3}$CCH along the main filament. SiO peak velocities align well with CH$_{3}$OH, CCH, and CH$_{3}$CCH, while SO and H$_{2}$CS consistently exhibit a red-shift of 1 km s$^{-1}$ compared to SiO. HC$_{3}$N, H$^{13}$CO$^+$, and HCN show a further red-shift with a $\sim$2 km s$^{-1}$ difference. The distinct and systematic layering observed among these tracers provides an important evidence of cloud collisions and merging.
    \item We discover a population of NH$_{2}$D cores which can be categorised into two classes: those that lie close to the main SiO filament, having broad, supersonic velocity components of $\sigma_{\mathrm{v}}$ > 2$c_{\mathrm{s}}$\,=\,0.85 km s$^{-1}$, and the others that are located further away from the main filament, only having narrow velocity component of $\sigma_{\mathrm{v}}$$\sim$0.5 km s$^{-1}$, in the range of transonic ($\sim$2$c_{\mathrm s}$) to subsonic ($\lesssim$$c_{\mathrm s}$) regimes. The enhancement of NH$_{2}$D abundance in a shock environment, which peaks in the post-shock gas, is predicted by our chemical models. The NH$_{2}$D cores located close to the main filament may represent evolutionary stages still prior or in the stage of turbulence dissipation. 
    \item The virial masses of these NH$_{2}$D cores range from 3 to 100 M$_{\odot}$, estimated by the narrow linewidth component. The existence of these cores may hint at a new generation of stars being formed in the post-shock gas compressed by cloud collisions. The formation timescales of these cores roughly scale with their masses, and the fragmentation properties are more close to turbulent fragmentation of envelope gas, than to the in-situ fragmentation of dense gas.  
\end{itemize}
Our results reveal the direct link between shock activities and enhanced gas densities inside these massive star-forming clumps. The broad linewidth ``cold'' and high-density gas tracers, such as NH$_{2}$D, may possess important information to pin down the relation of new-generation core formation, and timescales of shock activities. Core formation under the shock environment can be quick and chaotic, facilitated by enhanced accretion of gas channelling, and/or local density-driven infall. The prevalence of such systems in favor of massive cluster formation may be revealed by future observations, towards a statistically significant sample of massive star-forming clouds.

\begin{acknowledgements} %
This work was partly funded by the Deutsche Forschungsgemeinschaft (DFG, German Research Foundation) under SFB 956. H.B.L. is supported by the National Science and Technology Council (NSTC) of Taiwan (Grant Nos. 111-2112-M-110-022-MY3).

\end{acknowledgements}

\bibliography{zref}

\begin{thebibliography}{97}
\expandafter\ifx\csname natexlab\endcsname\relax\def\natexlab#1{#1}\fi

\bibitem[{{Anathpindika}(2010)}]{Ana10}
{Anathpindika}, S.~V. 2010, \mnras, 405, 1431

\bibitem[{{Balan{\c{c}}a} {et~al.}(2018){Balan{\c{c}}a}, {Dayou}, {Faure},
  {Wiesenfeld}, \& {Feautrier}}]{Balanca18}
{Balan{\c{c}}a}, C., {Dayou}, F., {Faure}, A., {Wiesenfeld}, L., \&
  {Feautrier}, N. 2018, \mnras, 479, 2692

\bibitem[{{Balfour} {et~al.}(2017){Balfour}, {Whitworth}, \&
  {Hubber}}]{Balfour17}
{Balfour}, S.~K., {Whitworth}, A.~P., \& {Hubber}, D.~A. 2017, \mnras, 465,
  3483

\bibitem[{{Balfour} {et~al.}(2015){Balfour}, {Whitworth}, {Hubber}, \&
  {Jaffa}}]{Balfour15}
{Balfour}, S.~K., {Whitworth}, A.~P., {Hubber}, D.~A., \& {Jaffa}, S.~E. 2015,
  \mnras, 453, 2471

\bibitem[{{Bally} {et~al.}(2010){Bally}, {Anderson}, {Battersby}, {Calzoletti},
  {Digiorgio}, {Faustini}, {Ginsburg}, {Li}, {Nguyen Luong}, {Molinari},
  {Motte}, {Pestalozzi}, {Plume}, {Rodon}, {Schilke}, {Schlingman},
  {Schneider-Bontemps}, {Shirley}, {Stringfellow}, {Testi}, {Traficante},
  {Veneziani}, \& {Zavagno}}]{Bally10}
{Bally}, J., {Anderson}, L.~D., {Battersby}, C., {et~al.} 2010, \aap, 518, L90

\bibitem[{{Bertoldi}(1989)}]{Bertoldi89}
{Bertoldi}, F. 1989, \apj, 346, 735

\bibitem[{{Beuther} {et~al.}(2008){Beuther}, {Semenov}, {Henning}, \&
  {Linz}}]{Beuther08}
{Beuther}, H., {Semenov}, D., {Henning}, T., \& {Linz}, H. 2008, \apjl, 675,
  L33

\bibitem[{{Bisbas} {et~al.}(2018){Bisbas}, {Tan}, {Csengeri}, {Wu}, {Lim},
  {Caselli}, {G{\"u}sten}, {Ricken}, \& {Riquelme}}]{Bisbas18}
{Bisbas}, T.~G., {Tan}, J.~C., {Csengeri}, T., {et~al.} 2018, \mnras, 478, L54

\bibitem[{{Bisbas} {et~al.}(2009){Bisbas}, {W{\"u}nsch}, {Whitworth}, \&
  {Hubber}}]{Bisbas09}
{Bisbas}, T.~G., {W{\"u}nsch}, R., {Whitworth}, A.~P., \& {Hubber}, D.~A. 2009,
  \aap, 497, 649

\bibitem[{{Bisbas} {et~al.}(2011){Bisbas}, {W{\"u}nsch}, {Whitworth}, {Hubber},
  \& {Walch}}]{Bisbas11}
{Bisbas}, T.~G., {W{\"u}nsch}, R., {Whitworth}, A.~P., {Hubber}, D.~A., \&
  {Walch}, S. 2011, \apj, 736, 142

\bibitem[{{Blum} {et~al.}(1999){Blum}, {Damineli}, \& {Conti}}]{Blum99}
{Blum}, R.~D., {Damineli}, A., \& {Conti}, P.~S. 1999, \aj, 117, 1392

\bibitem[{{Brunthaler} {et~al.}(2021){Brunthaler}, {Menten}, {Dzib}, {Cotton},
  {Wyrowski}, {Dokara}, {Gong}, {Medina}, {M{\"u}ller}, {Nguyen},
  {Ortiz-Le{\'o}n}, {Reich}, {Rugel}, {Urquhart}, {Winkel}, {Yang}, {Beuther},
  {Billington}, {Carrasco-Gonzalez}, {Csengeri}, {Murugeshan}, {Pandian}, \&
  {Roy}}]{AB21}
{Brunthaler}, A., {Menten}, K.~M., {Dzib}, S.~A., {et~al.} 2021, \aap, 651, A85

\bibitem[{{Carey} {et~al.}(2009){Carey}, {Noriega-Crespo}, {Mizuno}, {Shenoy},
  {Paladini}, {Kraemer}, {Price}, {Flagey}, {Ryan}, {Ingalls}, {Kuchar},
  {Pinheiro Gon{\c{c}}alves}, {Indebetouw}, {Billot}, {Marleau}, {Padgett},
  {Rebull}, {Bressert}, {Ali}, {Molinari}, {Martin}, {Berriman}, {Boulanger},
  {Latter}, {Miville-Deschenes}, {Shipman}, \& {Testi}}]{Carey09}
{Carey}, S.~J., {Noriega-Crespo}, A., {Mizuno}, D.~R., {et~al.} 2009, PASP,
  121, 76

\bibitem[{{Carlhoff} {et~al.}(2013){Carlhoff}, {Nguyen Luong}, {Schilke},
  {Motte}, {Schneider}, {Beuther}, {Bontemps}, {Heitsch}, {Hill}, {Kramer},
  {Ossenkopf}, {Schuller}, {Simon}, \& {Wyrowski}}]{Carlhoff}
{Carlhoff}, P., {Nguyen Luong}, Q., {Schilke}, P., {et~al.} 2013, \aap, 560,
  A24

\bibitem[{{Caselli} {et~al.}(1997){Caselli}, {Hartquist}, \&
  {Havnes}}]{Caselli97}
{Caselli}, P., {Hartquist}, T.~W., \& {Havnes}, O. 1997, \aap, 322, 296

\bibitem[{{Caselli} {et~al.}(2008){Caselli}, {Vastel}, {Ceccarelli}, {van der
  Tak}, {Crapsi}, \& {Bacmann}}]{Caselli08}
{Caselli}, P., {Vastel}, C., {Ceccarelli}, C., {et~al.} 2008, \aap, 492, 703

\bibitem[{{Churchwell}(2002)}]{Churchwell02}
{Churchwell}, E. 2002, ARA\&A, 40, 27

\bibitem[{{Cortes} {et~al.}(2019){Cortes}, {Hull}, {Girart}, {Orquera-Rojas},
  {Sridharan}, {Li}, {Louvet}, {Cortes}, {Le Gouellec}, {Crutcher}, \&
  {Lai}}]{Cortes19}
{Cortes}, P.~C., {Hull}, C. L.~H., {Girart}, J.~M., {et~al.} 2019, \apj, 884,
  48

\bibitem[{{Cosentino} {et~al.}(2022){Cosentino}, {Jim{\'e}nez-Serra}, {Tan},
  {Henshaw}, {Barnes}, {Law}, {Zeng}, {Fontani}, {Caselli}, {Viti}, {Zahorecz},
  {Rico-Villas}, {Meg{\'\i}as}, {Miceli}, {Orlando}, {Ustamujic}, {Greco},
  {Peres}, {Bocchino}, {Fedriani}, {Gorai}, {Testi}, \&
  {Mart{\'\i}n-Pintado}}]{Cosentino22}
{Cosentino}, G., {Jim{\'e}nez-Serra}, I., {Tan}, J.~C., {et~al.} 2022, \mnras,
  511, 953

\bibitem[{{Cosentino} {et~al.}(2023){Cosentino}, {Tan}, {Jim{\'e}nez-Serra},
  {Fontani}, {Caselli}, {Henshaw}, {Barnes}, {Law}, {Viti}, {Fedriani}, {Hsu},
  {Gorai}, \& {Zeng}}]{Cosentino23}
{Cosentino}, G., {Tan}, J.~C., {Jim{\'e}nez-Serra}, I., {et~al.} 2023, \aap,
  675, A190

\bibitem[{{Crapsi} {et~al.}(2005){Crapsi}, {Caselli}, {Walmsley}, {Myers},
  {Tafalla}, {Lee}, \& {Bourke}}]{Crapsi05}
{Crapsi}, A., {Caselli}, P., {Walmsley}, C.~M., {et~al.} 2005, \apj, 619, 379

\bibitem[{{Daniel} {et~al.}(2014){Daniel}, {Faure}, {Wiesenfeld}, {Roueff},
  {Lis}, \& {Hily-Blant}}]{Daniel14}
{Daniel}, F., {Faure}, A., {Wiesenfeld}, L., {et~al.} 2014, \mnras, 444, 2544

\bibitem[{{Dyson} \& {Williams}(1997)}]{Dyson97}
{Dyson}, J.~E. \& {Williams}, D.~A. 1997, {The physics of the interstellar
  medium}

\bibitem[{Endres {et~al.}(2016)Endres, Schlemmer, Schilke, Stutzki, \&
  Müller}]{Endres16}
Endres, C.~P., Schlemmer, S., Schilke, P., Stutzki, J., \& Müller, H.~S. 2016,
  Journal of Molecular Spectroscopy, 327, 95, new Visions of Spectroscopic
  Databases, Volume II

\bibitem[{{Faure} {et~al.}(2016){Faure}, {Lique}, \& {Wiesenfeld}}]{Faure16}
{Faure}, A., {Lique}, F., \& {Wiesenfeld}, L. 2016, \mnras, 460, 2103

\bibitem[{{Fontani} {et~al.}(2015){Fontani}, {Busquet}, {Palau}, {Caselli},
  {S{\'a}nchez-Monge}, {Tan}, \& {Audard}}]{Fontani15}
{Fontani}, F., {Busquet}, G., {Palau}, A., {et~al.} 2015, \aap, 575, A87

\bibitem[{{Fukui} {et~al.}(2021){Fukui}, {Habe}, {Inoue}, {Enokiya}, \&
  {Tachihara}}]{Fukui21}
{Fukui}, Y., {Habe}, A., {Inoue}, T., {Enokiya}, R., \& {Tachihara}, K. 2021,
  \pasj, 73, S1

\bibitem[{{Fukui} {et~al.}(2018){Fukui}, {Kohno}, {Yokoyama}, {Nishimura},
  {Torii}, {Hattori}, {Sano}, {Ohama}, {Yamamoto}, \& {Tachihara}}]{Fukui18}
{Fukui}, Y., {Kohno}, M., {Yokoyama}, K., {et~al.} 2018, PASJ, 70, S44

\bibitem[{{Fukui} {et~al.}(2016){Fukui}, {Torii}, {Ohama}, {Hasegawa},
  {Hattori}, {Sano}, {Ohashi}, {Fujii}, {Kuwahara}, {Mizuno}, {Dawson},
  {Yamamoto}, {Tachihara}, {Okuda}, {Onishi}, \& {Mizuno}}]{Fukui16}
{Fukui}, Y., {Torii}, K., {Ohama}, A., {et~al.} 2016, \apj, 820, 26

\bibitem[{{Giannetti} {et~al.}(2017){Giannetti}, {Leurini}, {Wyrowski},
  {Urquhart}, {Csengeri}, {Menten}, {K{\"o}nig}, \& {G{\"u}sten}}]{Giannetti17}
{Giannetti}, A., {Leurini}, S., {Wyrowski}, F., {et~al.} 2017, \aap, 603, A33

\bibitem[{{Habe} \& {Ohta}(1992)}]{HO92}
{Habe}, A. \& {Ohta}, K. 1992, \pasj, 44, 203

\bibitem[{{Haworth} {et~al.}(2015){Haworth}, {Shima}, {Tasker}, {Fukui},
  {Torii}, {Dale}, {Takahira}, \& {Habe}}]{Haworth15}
{Haworth}, T.~J., {Shima}, K., {Tasker}, E.~J., {et~al.} 2015, \mnras, 454,
  1634

\bibitem[{{Hennebelle} \& {Andr{\'e}}(2013)}]{Hennebelle13}
{Hennebelle}, P. \& {Andr{\'e}}, P. 2013, \aap, 560, A68

\bibitem[{{Henshaw} {et~al.}(2013){Henshaw}, {Caselli}, {Fontani},
  {Jim{\'e}nez-Serra}, {Tan}, \& {Hernandez}}]{Henshaw13}
{Henshaw}, J.~D., {Caselli}, P., {Fontani}, F., {et~al.} 2013, \mnras, 428,
  3425

\bibitem[{{Hill} \& {Hollenbach}(1978)}]{Hill78}
{Hill}, J.~K. \& {Hollenbach}, D.~J. 1978, \apj, 225, 390

\bibitem[{{H{\"o}gbom}(1974)}]{hogbom}
{H{\"o}gbom}, J.~A. 1974, \aaps, 15, 417

\bibitem[{{Inoue} \& {Fukui}(2013)}]{IF13}
{Inoue}, T. \& {Fukui}, Y. 2013, \apjl, 774, L31

\bibitem[{{Inoue} {et~al.}(2018){Inoue}, {Hennebelle}, {Fukui}, {Matsumoto},
  {Iwasaki}, \& {Inutsuka}}]{Inoue18}
{Inoue}, T., {Hennebelle}, P., {Fukui}, Y., {et~al.} 2018, \pasj, 70, S53

\bibitem[{{Jiang} {et~al.}(2015){Jiang}, {Liu}, {Zhang}, {Wang}, {Zhang}, {Li},
  {Gao}, \& {Gu}}]{Jiang15}
{Jiang}, X.-J., {Liu}, H.~B., {Zhang}, Q., {et~al.} 2015, \apj, 808, 114

\bibitem[{{Jim{\'e}nez-Serra} {et~al.}(2008){Jim{\'e}nez-Serra}, {Caselli},
  {Mart{\'\i}n-Pintado}, \& {Hartquist}}]{JS08}
{Jim{\'e}nez-Serra}, I., {Caselli}, P., {Mart{\'\i}n-Pintado}, J., \&
  {Hartquist}, T.~W. 2008, \aap, 482, 549

\bibitem[{{Jim{\'e}nez-Serra} {et~al.}(2010){Jim{\'e}nez-Serra}, {Caselli},
  {Tan}, {Hernandez}, {Fontani}, {Butler}, \& {van Loo}}]{JS10}
{Jim{\'e}nez-Serra}, I., {Caselli}, P., {Tan}, J.~C., {et~al.} 2010, \mnras,
  406, 187

\bibitem[{{Jim{\'e}nez-Serra} {et~al.}(2004){Jim{\'e}nez-Serra},
  {Mart{\'\i}n-Pintado}, {Rodr{\'\i}guez-Franco}, \& {Marcelino}}]{JS04}
{Jim{\'e}nez-Serra}, I., {Mart{\'\i}n-Pintado}, J., {Rodr{\'\i}guez-Franco},
  A., \& {Marcelino}, N. 2004, \apjl, 603, L49

\bibitem[{{Kohno} {et~al.}(2021){Kohno}, {Tachihara}, {Torii}, {Fujita},
  {Nishimura}, {Kuno}, {Umemoto}, {Minamidani}, {Matsuo}, {Kiridoshi},
  {Tokuda}, {Hanaoka}, {Tsuda}, {Kuriki}, {Ohama}, {Sano}, {Hasegawa}, {Sofue},
  {Habe}, {Onishi}, \& {Fukui}}]{Kohno21}
{Kohno}, M., {Tachihara}, K., {Torii}, K., {et~al.} 2021, PASJ, 73, S129

\bibitem[{{Kumar} {et~al.}(2020){Kumar}, {Palmeirim}, {Arzoumanian}, \&
  {Inutsuka}}]{Kumar20}
{Kumar}, M.~S.~N., {Palmeirim}, P., {Arzoumanian}, D., \& {Inutsuka}, S.~I.
  2020, \aap, 642, A87

\bibitem[{{Leurini} {et~al.}(2004){Leurini}, {Schilke}, {Menten}, {Flower},
  {Pottage}, \& {Xu}}]{Leurini04}
{Leurini}, S., {Schilke}, P., {Menten}, K.~M., {et~al.} 2004, \aap, 422, 573

\bibitem[{{Leurini} {et~al.}(2007){Leurini}, {Schilke}, {Wyrowski}, \&
  {Menten}}]{Leurini07}
{Leurini}, S., {Schilke}, P., {Wyrowski}, F., \& {Menten}, K.~M. 2007, \aap,
  466, 215

\bibitem[{{Li} {et~al.}(2021){Li}, {Lu}, {Zhang}, {Lee}, {Sanhueza}, {Beuther},
  {Jimenez-Serra}, {Qiu}, {Palau}, {Feng}, {Pillai}, {Kim}, {Liu}, {Girart},
  {Liu}, {Wang}, {Wang}, {Liu}, {Smith}, {Li}, {Lee}, {Li}, {Li}, {Kim}, {Yue},
  \& {Strom}}]{Li21}
{Li}, S., {Lu}, X., {Zhang}, Q., {et~al.} 2021, \apjl, 912, L7

\bibitem[{{Li} {et~al.}(2023){Li}, {Wang}, {Li}, {Liu}, {Yang}, {Zheng}, \&
  {Lu}}]{Li23}
{Li}, Y., {Wang}, J., {Li}, J., {et~al.} 2023, \mnras

\bibitem[{{Lin} {et~al.}(2016){Lin}, {Liu}, {Li}, {Zhang}, {Ginsburg},
  {Pineda}, {Qian}, {Galv{\'a}n-Madrid}, {McLeod}, {Rosolowsky}, {Dale},
  {Immer}, {Koch}, {Longmore}, {Walker}, \& {Testi}}]{Lin16}
{Lin}, Y., {Liu}, H.~B., {Li}, D., {et~al.} 2016, \apj, 828, 32

\bibitem[{{Lin} {et~al.}(2022){Lin}, {Wyrowski}, {Liu}, {Izquierdo},
  {Csengeri}, {Leurini}, \& {Menten}}]{Lin22a}
{Lin}, Y., {Wyrowski}, F., {Liu}, H.~B., {et~al.} 2022, \aap, 658, A128

\bibitem[{{Liow} \& {Dobbs}(2020)}]{Liow20}
{Liow}, K.~Y. \& {Dobbs}, C.~L. 2020, \mnras, 499, 1099

\bibitem[{{Louvet} {et~al.}(2016){Louvet}, {Motte}, {Gusdorf}, {Nguy{\^e}n
  Luong}, {Lesaffre}, {Duarte-Cabral}, {Maury}, {Schneider}, {Hill}, {Schilke},
  \& {Gueth}}]{Louvet16}
{Louvet}, F., {Motte}, F., {Gusdorf}, A., {et~al.} 2016, \aap, 595, A122

\bibitem[{{Mac Low} {et~al.}(2007){Mac Low}, {Toraskar}, {Oishi}, \&
  {Abel}}]{Mclow07}
{Mac Low}, M.-M., {Toraskar}, J., {Oishi}, J.~S., \& {Abel}, T. 2007, \apj,
  668, 980

\bibitem[{{Mezger} \& {Henderson}(1967)}]{Mezger67}
{Mezger}, P.~G. \& {Henderson}, A.~P. 1967, \apj, 147, 471

\bibitem[{{Mocz} \& {Burkhart}(2018)}]{Mocz18}
{Mocz}, P. \& {Burkhart}, B. 2018, \mnras, 480, 3916

\bibitem[{{Molinari} {et~al.}(2016{\natexlab{a}}){Molinari}, {Merello}, {Elia},
  {Cesaroni}, {Testi}, \& {Robitaille}}]{Molinari16b}
{Molinari}, S., {Merello}, M., {Elia}, D., {et~al.} 2016{\natexlab{a}}, \apjl,
  826, L8

\bibitem[{{Molinari} {et~al.}(2016{\natexlab{b}}){Molinari}, {Schisano},
  {Elia}, {Pestalozzi}, {Traficante}, {Pezzuto}, {Swinyard}, {Noriega-Crespo},
  {Bally}, {Moore}, {Plume}, {Zavagno}, {di Giorgio}, {Liu}, {Pilbratt},
  {Mottram}, {Russeil}, {Piazzo}, {Veneziani}, {Benedettini}, {Calzoletti},
  {Faustini}, {Natoli}, {Piacentini}, {Merello}, {Palmese}, {Del Grande},
  {Polychroni}, {Rygl}, {Polenta}, {Barlow}, {Bernard}, {Martin}, {Testi},
  {Ali}, {Andr{\'e}}, {Beltr{\'a}n}, {Billot}, {Carey}, {Cesaroni},
  {Compi{\`e}gne}, {Eden}, {Fukui}, {Garcia-Lario}, {Hoare}, {Huang}, {Joncas},
  {Lim}, {Lord}, {Martinavarro-Armengol}, {Motte}, {Paladini}, {Paradis},
  {Peretto}, {Robitaille}, {Schilke}, {Schneider}, {Schulz}, {Sibthorpe},
  {Strafella}, {Thompson}, {Umana}, {Ward-Thompson}, \&
  {Wyrowski}}]{Molinari16a}
{Molinari}, S., {Schisano}, E., {Elia}, D., {et~al.} 2016{\natexlab{b}}, \aap,
  591, A149

\bibitem[{{Molinari} {et~al.}(2010){Molinari}, {Swinyard}, {Bally}, {Barlow},
  {Bernard}, {Martin}, {Moore}, {Noriega-Crespo}, {Plume}, {Testi}, {Zavagno},
  {Abergel}, {Ali}, {Anderson}, {Andr{\'e}}, {Baluteau}, {Battersby},
  {Beltr{\'a}n}, {Benedettini}, {Billot}, {Blommaert}, {Bontemps}, {Boulanger},
  {Brand}, {Brunt}, {Burton}, {Calzoletti}, {Carey}, {Caselli}, {Cesaroni},
  {Cernicharo}, {Chakrabarti}, {Chrysostomou}, {Cohen}, {Compiegne}, {de
  Bernardis}, {de Gasperis}, {di Giorgio}, {Elia}, {Faustini}, {Flagey},
  {Fukui}, {Fuller}, {Ganga}, {Garcia-Lario}, {Glenn}, {Goldsmith}, {Griffin},
  {Hoare}, {Huang}, {Ikhenaode}, {Joblin}, {Joncas}, {Juvela}, {Kirk},
  {Lagache}, {Li}, {Lim}, {Lord}, {Marengo}, {Marshall}, {Masi}, {Massi},
  {Matsuura}, {Minier}, {Miville-Desch{\^e}nes}, {Montier}, {Morgan}, {Motte},
  {Mottram}, {Muller}, {Natoli}, {Neves}, {Olmi}, {Paladini}, {Paradis},
  {Parsons}, {Peretto}, {Pestalozzi}, {Pezzuto}, {Piacentini}, {Piazzo},
  {Polychroni}, {Pomar{\`e}s}, {Popescu}, {Reach}, {Ristorcelli}, {Robitaille},
  {Robitaille}, {Rod{\'o}n}, {Roy}, {Royer}, {Russeil}, {Saraceno}, {Sauvage},
  {Schilke}, {Schisano}, {Schneider}, {Schuller}, {Schulz}, {Sibthorpe},
  {Smith}, {Smith}, {Spinoglio}, {Stamatellos}, {Strafella}, {Stringfellow},
  {Sturm}, {Taylor}, {Thompson}, {Traficante}, {Tuffs}, {Umana}, {Valenziano},
  {Vavrek}, {Veneziani}, {Viti}, {Waelkens}, {Ward-Thompson}, {White},
  {Wilcock}, {Wyrowski}, {Yorke}, \& {Zhang}}]{Molinari10}
{Molinari}, S., {Swinyard}, B., {Bally}, J., {et~al.} 2010, \aap, 518, L100

\bibitem[{{M{\"o}ller} {et~al.}(2017){M{\"o}ller}, {Endres}, \&
  {Schilke}}]{Moeller17}
{M{\"o}ller}, T., {Endres}, C., \& {Schilke}, P. 2017, \aap, 598, A7

\bibitem[{{Motte} {et~al.}(2022){Motte}, {Bontemps}, {Csengeri}, {Pouteau},
  {Louvet}, {Stutz}, {Cunningham}, {L{\'o}pez-Sepulcre}, {Brouillet},
  {Galv{\'a}n-Madrid}, {Ginsburg}, {Maud}, {Men'shchikov}, {Nakamura}, {Nony},
  {Sanhueza}, {{\'A}lvarez-Guti{\'e}rrez}, {Armante}, {Baug}, {Bonfand},
  {Busquet}, {Chapillon}, {D{\'\i}az-Gonz{\'a}lez}, {Fern{\'a}ndez-L{\'o}pez},
  {Guzm{\'a}n}, {Herpin}, {Liu}, {Olguin}, {Towner}, {Bally}, {Battersby},
  {Braine}, {Bronfman}, {Chen}, {Dell'Ova}, {Di Francesco}, {Gonz{\'a}lez},
  {Gusdorf}, {Hennebelle}, {Izumi}, {Joncour}, {Lee}, {Lefloch}, {Lesaffre},
  {Lu}, {Menten}, {Mignon-Risse}, {Molet}, {Moraux}, {Mundy}, {Nguyen Luong},
  {Reyes}, {Reyes Reyes}, {Robitaille}, {Rosolowsky}, {Sandoval-Garrido},
  {Schuller}, {Svoboda}, {Tatematsu}, {Thomasson}, {Walker}, {Wu}, {Whitworth},
  \& {Wyrowski}}]{Motte22a}
{Motte}, F., {Bontemps}, S., {Csengeri}, T., {et~al.} 2022, \aap, 662, A8

\bibitem[{{Motte} {et~al.}(2018){Motte}, {Bontemps}, \& {Louvet}}]{Motte18}
{Motte}, F., {Bontemps}, S., \& {Louvet}, F. 2018, ARA\&A, 56, 41

\bibitem[{{Motte} {et~al.}(2014){Motte}, {Nguy{\^e}n Luong}, {Schneider},
  {Heitsch}, {Glover}, {Carlhoff}, {Hill}, {Bontemps}, {Schilke}, {Louvet},
  {Hennemann}, {Didelon}, \& {Beuther}}]{Motte14}
{Motte}, F., {Nguy{\^e}n Luong}, Q., {Schneider}, N., {et~al.} 2014, \aap, 571,
  A32

\bibitem[{{Motte} {et~al.}(2003){Motte}, {Schilke}, \& {Lis}}]{Motte03}
{Motte}, F., {Schilke}, P., \& {Lis}, D.~C. 2003, \apj, 582, 277

\bibitem[{{Mottram} {et~al.}(2011){Mottram}, {Hoare}, {Davies}, {Lumsden},
  {Oudmaijer}, {Urquhart}, {Moore}, {Cooper}, \& {Stead}}]{Mottram11}
{Mottram}, J.~C., {Hoare}, M.~G., {Davies}, B., {et~al.} 2011, \apjl, 730, L33

\bibitem[{{Nguyen-Luong} {et~al.}(2013){Nguyen-Luong}, {Motte}, {Carlhoff},
  {Louvet}, {Lesaffre}, {Schilke}, {Hill}, {Hennemann}, {Gusdorf}, {Didelon},
  {Schneider}, {Bontemps}, {Duarte-Cabral}, {Menten}, {Martin}, {Wyrowski},
  {Bendo}, {Roussel}, {Bernard}, {Bronfman}, {Henning}, {Kramer}, \&
  {Heitsch}}]{NL13}
{Nguyen-Luong}, Q., {Motte}, F., {Carlhoff}, P., {et~al.} 2013, \apj, 775, 88

\bibitem[{{Nguyen Luong} {et~al.}(2011){Nguyen Luong}, {Motte}, {Schuller},
  {Schneider}, {Bontemps}, {Schilke}, {Menten}, {Heitsch}, {Wyrowski},
  {Carlhoff}, {Bronfman}, \& {Henning}}]{NL11}
{Nguyen Luong}, Q., {Motte}, F., {Schuller}, F., {et~al.} 2011, \aap, 529, A41

\bibitem[{{Nony} {et~al.}(2023){Nony}, {Galvan-Madrid}, {Motte}, {Pouteau},
  {Cunningham}, {Louvet}, {Stutz}, {Lefloch}, {Bontemps}, {Brouillet},
  {Ginsburg}, {Joncour}, {Herpin}, {Sanhueza}, {Csengeri}, {Towner}, {Bonfand},
  {Fern{\'a}ndez-L{\'o}pez}, {Baug}, {Bronfman}, {Busquet}, {Di Francesco},
  {Gusdorf}, {Lu}, {Olguin}, {Valeille-Manet}, \& {Whitworth}}]{Nony23}
{Nony}, T., {Galvan-Madrid}, R., {Motte}, F., {et~al.} 2023, arXiv e-prints,
  arXiv:2301.07238

\bibitem[{{Padoan} {et~al.}(2020){Padoan}, {Pan}, {Juvela}, {Haugb{\o}lle}, \&
  {Nordlund}}]{Padoan20}
{Padoan}, P., {Pan}, L., {Juvela}, M., {Haugb{\o}lle}, T., \& {Nordlund},
  {\r{A}}. 2020, \apj, 900, 82

\bibitem[{{Panagia}(1973)}]{Panagia}
{Panagia}, N. 1973, \aj, 78, 929

\bibitem[{{Pillai} {et~al.}(2011){Pillai}, {Kauffmann}, {Wyrowski}, {Hatchell},
  {Gibb}, \& {Thompson}}]{Pillai11}
{Pillai}, T., {Kauffmann}, J., {Wyrowski}, F., {et~al.} 2011, \aap, 530, A118

\bibitem[{{Pouteau} {et~al.}(2022){Pouteau}, {Motte}, {Nony}, {Gonzalez},
  {Joncour}, {Robitaille}, {Busquet}, {Galvan-Madrid}, {Gusdorf}, {Hennebelle},
  {Ginsburg}, {Csengeri}, {Sanhueza}, {Dell'Ova}, {Stutz}, {Towner},
  {Cunningham}, {Louvet}, {Men'shchikov}, {Fernandez-Lopez}, {Schneider},
  {Armante}, {Bally}, {Baug}, {Bonfand}, {Bontemps}, {Bronfman}, {Brouillet},
  {Diaz-Gonzalez}, {Herpin}, {Lefloch}, {Liu}, {Lu}, {Nakamura}, {Nguyen
  Luong}, {Olguin}, {Tatematsu}, \& {Valeille-Manet}}]{Pouteau22}
{Pouteau}, Y., {Motte}, F., {Nony}, T., {et~al.} 2022, arXiv e-prints,
  arXiv:2212.09307

\bibitem[{{Priestley} \& {Whitworth}(2021)}]{Priestley21}
{Priestley}, F.~D. \& {Whitworth}, A.~P. 2021, \mnras, 506, 775

\bibitem[{{Redaelli} {et~al.}(2021){Redaelli}, {Bovino}, {Giannetti},
  {Sabatini}, {Caselli}, {Wyrowski}, {Schleicher}, \& {Colombo}}]{Redaelli21}
{Redaelli}, E., {Bovino}, S., {Giannetti}, A., {et~al.} 2021, arXiv e-prints,
  arXiv:2104.06431

\bibitem[{{Rosolowsky} {et~al.}(2008){Rosolowsky}, {Pineda}, {Kauffmann}, \&
  {Goodman}}]{Rosolowsky08}
{Rosolowsky}, E.~W., {Pineda}, J.~E., {Kauffmann}, J., \& {Goodman}, A.~A.
  2008, \apj, 679, 1338

\bibitem[{{Rubin}(1968)}]{Rubin68}
{Rubin}, R.~H. 1968, \apj, 154, 391

\bibitem[{{Sakre} {et~al.}(2021){Sakre}, {Habe}, {Pettitt}, \&
  {Okamoto}}]{Sakre21}
{Sakre}, N., {Habe}, A., {Pettitt}, A.~R., \& {Okamoto}, T. 2021, \pasj, 73,
  S385

\bibitem[{{Schilke} {et~al.}(1997){Schilke}, {Walmsley}, {Pineau des Forets},
  \& {Flower}}]{Schilke97a}
{Schilke}, P., {Walmsley}, C.~M., {Pineau des Forets}, G., \& {Flower}, D.~R.
  1997, \aap, 321, 293

\bibitem[{{Sipil{\"a}} {et~al.}(2013){Sipil{\"a}}, {Caselli}, \&
  {Harju}}]{Sipila13}
{Sipil{\"a}}, O., {Caselli}, P., \& {Harju}, J. 2013, \aap, 554, A92

\bibitem[{{Sipil{\"a}} {et~al.}(2019{\natexlab{a}}){Sipil{\"a}}, {Caselli}, \&
  {Harju}}]{Sipila19a}
{Sipil{\"a}}, O., {Caselli}, P., \& {Harju}, J. 2019{\natexlab{a}}, \aap, 631,
  A63

\bibitem[{{Sipil{\"a}} {et~al.}(2019{\natexlab{b}}){Sipil{\"a}}, {Caselli},
  {Redaelli}, {Juvela}, \& {Bizzocchi}}]{Sipila19b}
{Sipil{\"a}}, O., {Caselli}, P., {Redaelli}, E., {Juvela}, M., \& {Bizzocchi},
  L. 2019{\natexlab{b}}, \mnras, 487, 1269

\bibitem[{{Sipil{\"a}} {et~al.}(2015){Sipil{\"a}}, {Harju}, {Caselli}, \&
  {Schlemmer}}]{Sipila15}
{Sipil{\"a}}, O., {Harju}, J., {Caselli}, P., \& {Schlemmer}, S. 2015, \aap,
  581, A122

\bibitem[{{Snow} \& {Witt}(1996)}]{Snow96}
{Snow}, T.~P. \& {Witt}, A.~N. 1996, \apjl, 468, L65

\bibitem[{{Spitzer}(1978)}]{Spitzer78}
{Spitzer}, L. 1978, {Physical processes in the interstellar medium}

\bibitem[{{Tafalla} {et~al.}(2002){Tafalla}, {Myers}, {Caselli}, {Walmsley}, \&
  {Comito}}]{Tafalla02}
{Tafalla}, M., {Myers}, P.~C., {Caselli}, P., {Walmsley}, C.~M., \& {Comito},
  C. 2002, \apj, 569, 815

\bibitem[{{Takahira} {et~al.}(2018){Takahira}, {Shima}, {Habe}, \&
  {Tasker}}]{Takahira18}
{Takahira}, K., {Shima}, K., {Habe}, A., \& {Tasker}, E.~J. 2018, PASJ, 70, S58

\bibitem[{{Takahira} {et~al.}(2014){Takahira}, {Tasker}, \&
  {Habe}}]{Takahira14}
{Takahira}, K., {Tasker}, E.~J., \& {Habe}, A. 2014, \apj, 792, 63

\bibitem[{{Tremblin} {et~al.}(2014){Tremblin}, {Anderson}, {Didelon}, {Raga},
  {Minier}, {Ntormousi}, {Pettitt}, {Pinto}, {Samal}, {Schneider}, \&
  {Zavagno}}]{Tremblin14}
{Tremblin}, P., {Anderson}, L.~D., {Didelon}, P., {et~al.} 2014, \aap, 568, A4

\bibitem[{{Urquhart} {et~al.}(2018){Urquhart}, {K{\"o}nig}, {Giannetti},
  {Leurini}, {Moore}, {Eden}, {Pillai}, {Thompson}, {Braiding}, {Burton},
  {Csengeri}, {Dempsey}, {Figura}, {Froebrich}, {Menten}, {Schuller}, {Smith},
  \& {Wyrowski}}]{Urquhart18}
{Urquhart}, J.~S., {K{\"o}nig}, C., {Giannetti}, A., {et~al.} 2018, \mnras,
  473, 1059

\bibitem[{{van der Tak} {et~al.}(2007){van der Tak}, {Black}, {Sch{\"o}ier},
  {Jansen}, \& {van Dishoeck}}]{vdt07}
{van der Tak}, F.~F.~S., {Black}, J.~H., {Sch{\"o}ier}, F.~L., {Jansen}, D.~J.,
  \& {van Dishoeck}, E.~F. 2007, \aap, 468, 627

\bibitem[{{V{\'a}zquez-Semadeni} {et~al.}(2019){V{\'a}zquez-Semadeni}, {Palau},
  {Ballesteros-Paredes}, {G{\'o}mez}, \& {Zamora-Avil{\'e}s}}]{VS19}
{V{\'a}zquez-Semadeni}, E., {Palau}, A., {Ballesteros-Paredes}, J.,
  {G{\'o}mez}, G.~C., \& {Zamora-Avil{\'e}s}, M. 2019, \mnras, 490, 3061

\bibitem[{{Whitworth} {et~al.}(1994{\natexlab{a}}){Whitworth}, {Bhattal},
  {Chapman}, {Disney}, \& {Turner}}]{Whitworth94b}
{Whitworth}, A.~P., {Bhattal}, A.~S., {Chapman}, S.~J., {Disney}, M.~J., \&
  {Turner}, J.~A. 1994{\natexlab{a}}, \aap, 290, 421

\bibitem[{{Whitworth} {et~al.}(1994{\natexlab{b}}){Whitworth}, {Bhattal},
  {Chapman}, {Disney}, \& {Turner}}]{Whitworth94a}
{Whitworth}, A.~P., {Bhattal}, A.~S., {Chapman}, S.~J., {Disney}, M.~J., \&
  {Turner}, J.~A. 1994{\natexlab{b}}, \mnras, 268, 291

\bibitem[{{Wiesenfeld} \& {Faure}(2013)}]{Wiesenfeld13}
{Wiesenfeld}, L. \& {Faure}, A. 2013, \mnras, 432, 2573

\bibitem[{{Wu} {et~al.}(2017){Wu}, {Tan}, {Nakamura}, {Van Loo}, {Christie}, \&
  {Collins}}]{Wu17}
{Wu}, B., {Tan}, J.~C., {Nakamura}, F., {et~al.} 2017, \apj, 835, 137

\bibitem[{{Zhang} {et~al.}(2014){Zhang}, {Moscadelli}, {Sato}, {Reid},
  {Menten}, {Zheng}, {Brunthaler}, {Dame}, {Xu}, \& {Immer}}]{Zhang14}
{Zhang}, B., {Moscadelli}, L., {Sato}, M., {et~al.} 2014, \apj, 781, 89

\bibitem[{{Zhang} {et~al.}(2020){Zhang}, {Li}, {Pillai}, {Csengeri},
  {Wyrowski}, {Menten}, \& {Pestalozzi}}]{Zhang20}
{Zhang}, C.-P., {Li}, G.-X., {Pillai}, T., {et~al.} 2020, \aap, 638, A105

\bibitem[{{Zinnecker} \& {Yorke}(2007)}]{ZY07}
{Zinnecker}, H. \& {Yorke}, H.~W. 2007, \araa, 45, 481

\end{thebibliography}

\begin{appendix}
\section{Distribution of dense gas: multiple velocity components}\label{sec:m0s}

\begin{figure*}
\begin{tabular}{p{0.95\linewidth}}
\hspace{.5cm}\includegraphics[scale=0.325]{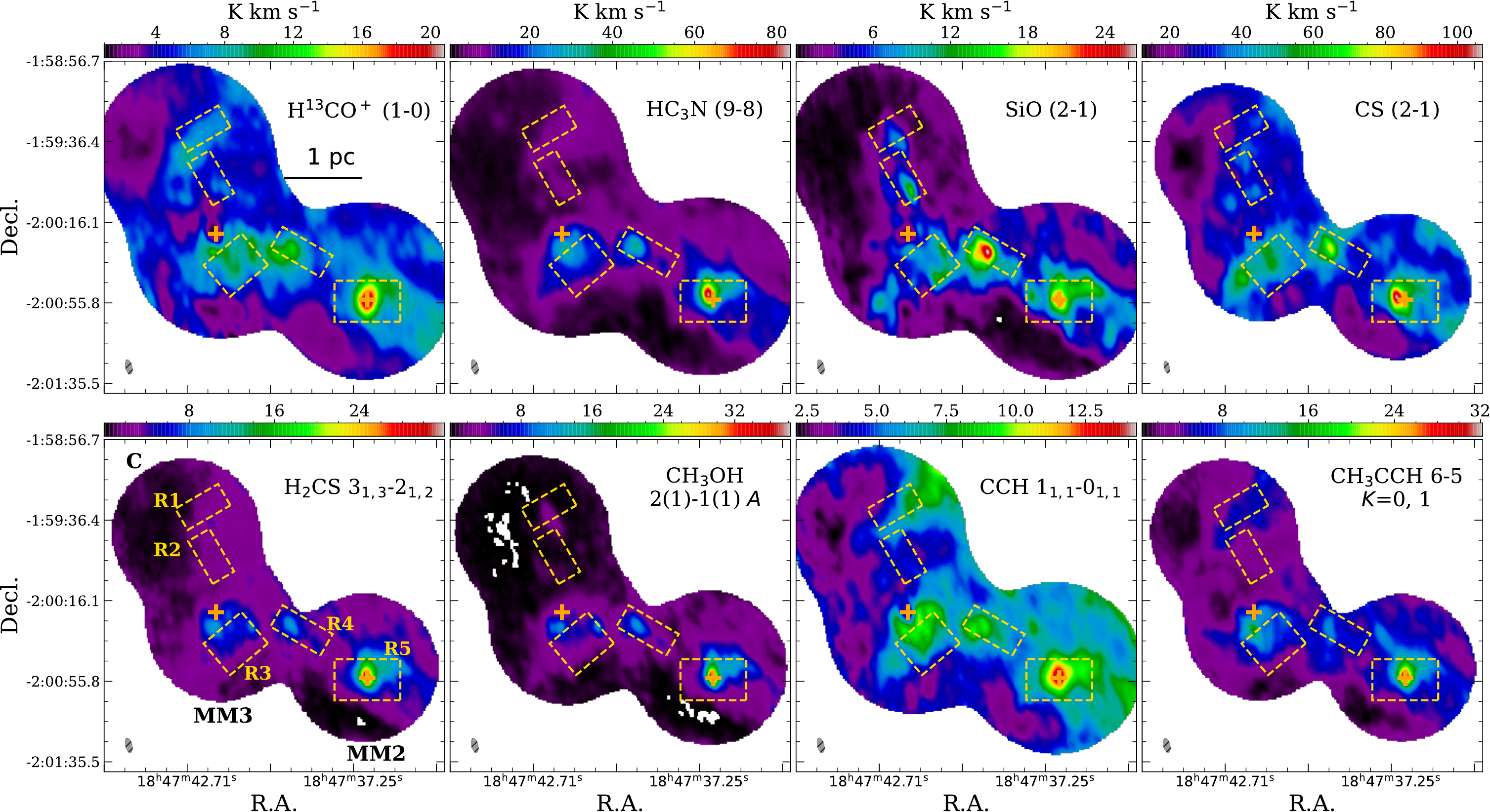}\\
\hspace{.5cm}\includegraphics[scale=0.325]{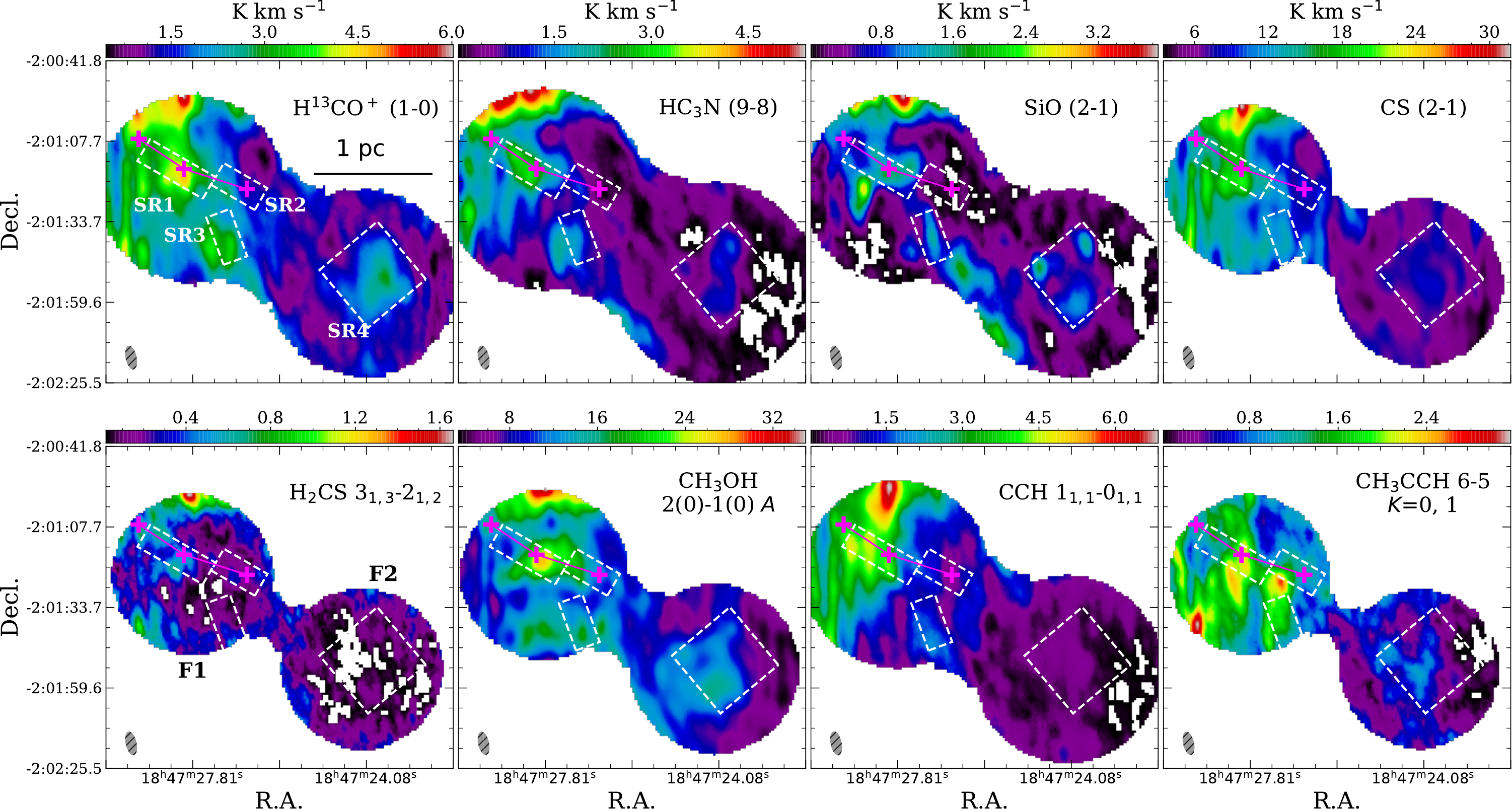}\\
\end{tabular}
\caption{{\emph{Upper panel:}} Integrated intensity maps of multiple molecular lines towards clumps MM2, MM3 and C. {\emph{Lower panel}}:Integrated intensity maps of multiple molecular lines towards clump F1 and F2. For both panels, the velocity range for integration for both regions is [80, 100] km s$^{-1}$. Before generating the integrated intensity maps, all channels with intensity $<$0.1 K (2$\sigma$) are masked.}
\label{fig:m0_maps}
\end{figure*}

\begin{figure*}
\begin{tabular}{p{0.975\linewidth}}
\hspace{-.5cm}\includegraphics[scale=0.3]{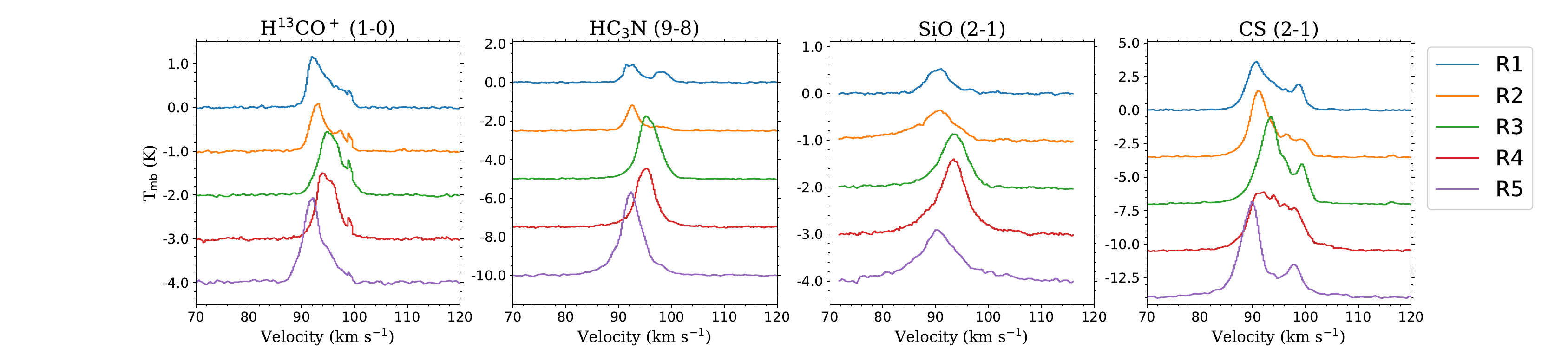}\\
\hspace{-.5cm}\includegraphics[scale=0.3]{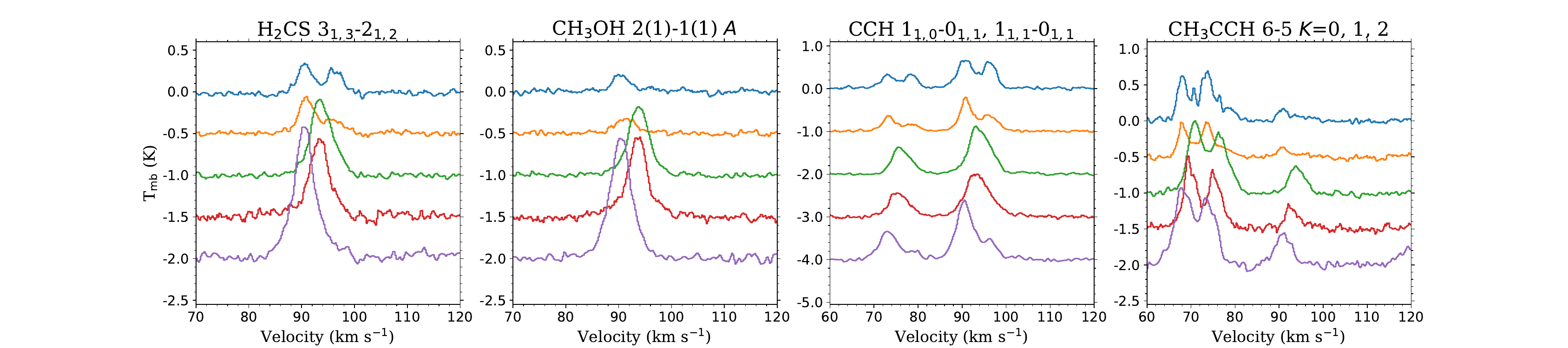}\\
\end{tabular}
\caption{Average spectra in sub-regions R1-R5 associated with clumps MM2, MM3 and C. The sub-regions R1-R5 are shown in Fig. \ref{fig:m0_maps}.}
\label{fig:g307_r5_sp}
\end{figure*}

\begin{figure*}
\begin{tabular}{p{0.975\linewidth}}
\hspace{-.5cm}\includegraphics[scale=0.3]{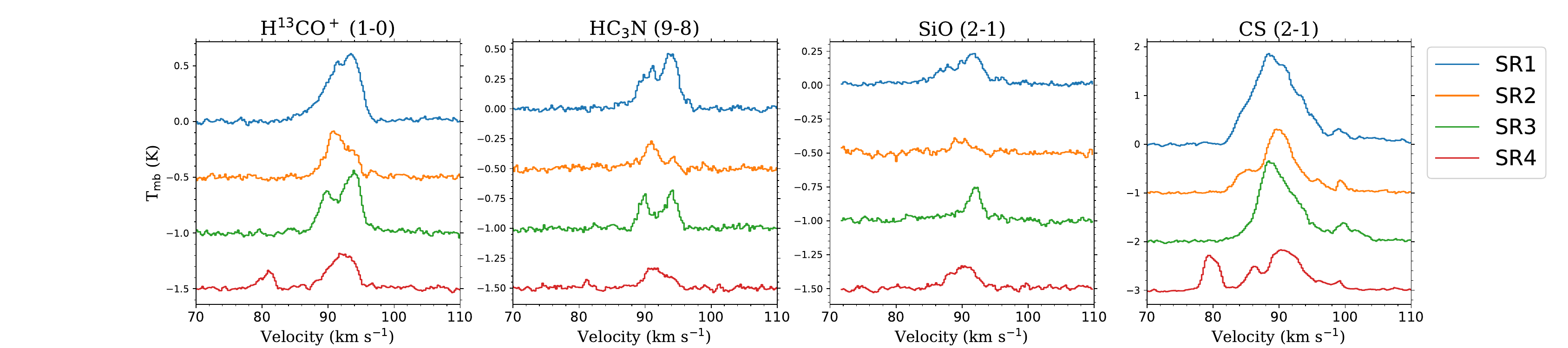}\\
\hspace{-.5cm}\includegraphics[scale=0.3]{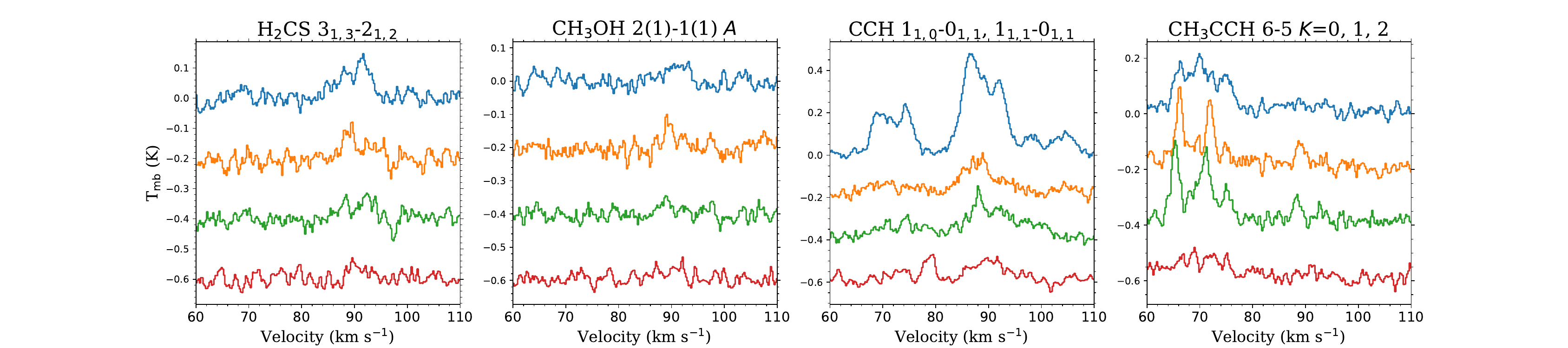}\\
\end{tabular}
\caption{Average spectra in sub-regions SR1-SR4 associated with clump F1 and F2. The sub-regions SR1-SR4 are shown in Fig. \ref{fig:m0_maps}.}
\label{fig:g306_r4_sp}
\end{figure*}

We present in Figure \ref{fig:m0_maps} the integrated intensity maps of molecular lines H$^{13}$CO$^+$ (1-0), HC$_{3}$N (9-8), SiO (2-1), CS (2-1) and H$_{2}$CS 3$_{1,3}$-2$_{1,2}$, CH$_{3}$OH 2(1)-1(1) $A$, CCH 1$_{1,1}$-0$_{1,1}$, CH$_{3}$CCH (6-5), for the three clumps MM2, MM3, C in the north and two clumps F1, F2 in the south (Figure \ref{fig:w43_npts}). For all maps, the velocity ranges for integration are 80-100 km s$^{-1}$. For the northern and southern region, we select 5 and 4 representative sub-regions (R1-R5, SR1-SR4), marked as rectangles in Figure \ref{fig:m0_maps} and the average spectra of these regions are shown in Figure \ref{fig:g307_r5_sp} and \ref{fig:g306_r4_sp}. 

For the northern region encompassing clumps MM2, MM3 and C, the strongest emission of all lines except SiO (2-1) coincides with sub-region R5, the central region of clump MM2, while emission of SiO (2-1) peaks at sub-region R4, the connecting region of clumps MM2 and MM3. Concerning the morphology of the line emission, SiO (2-1) exhibits the most elongated structure along the northeast-southwest direction, concentrating in the central region of the primary beam. 
Comparing the molecular gas distribution to the 3 mm continuum, source MM2 has peak line emission in good correlation with the continuum emission (sub-region R5), while source MM3 has line emission peaking offset (sub-region R3) from the continuum, indicating strong absorption towards the densest UCH\textsc{ii} region. Especially, the H$^{13}$CO$^+$ (1-0) line shows negative absorption dip at 97-101 km s$^{-1}$. 

The southern region composed of F1 and F2 exhibits a less concentrated distribution of line emission, having multiple localised emission peaks.   
In clump F1, except for the core F1-c2, the filament of the 3 mm continuum appears mostly offset from the line emission peak. Clump F2 shows extended and rather uniform line emission, except SiO (2-1), which shows clumpy structures, as can be seen in sub-region SR4. 

\begin{figure*}
\hspace{1.cm}\includegraphics[scale=0.15]{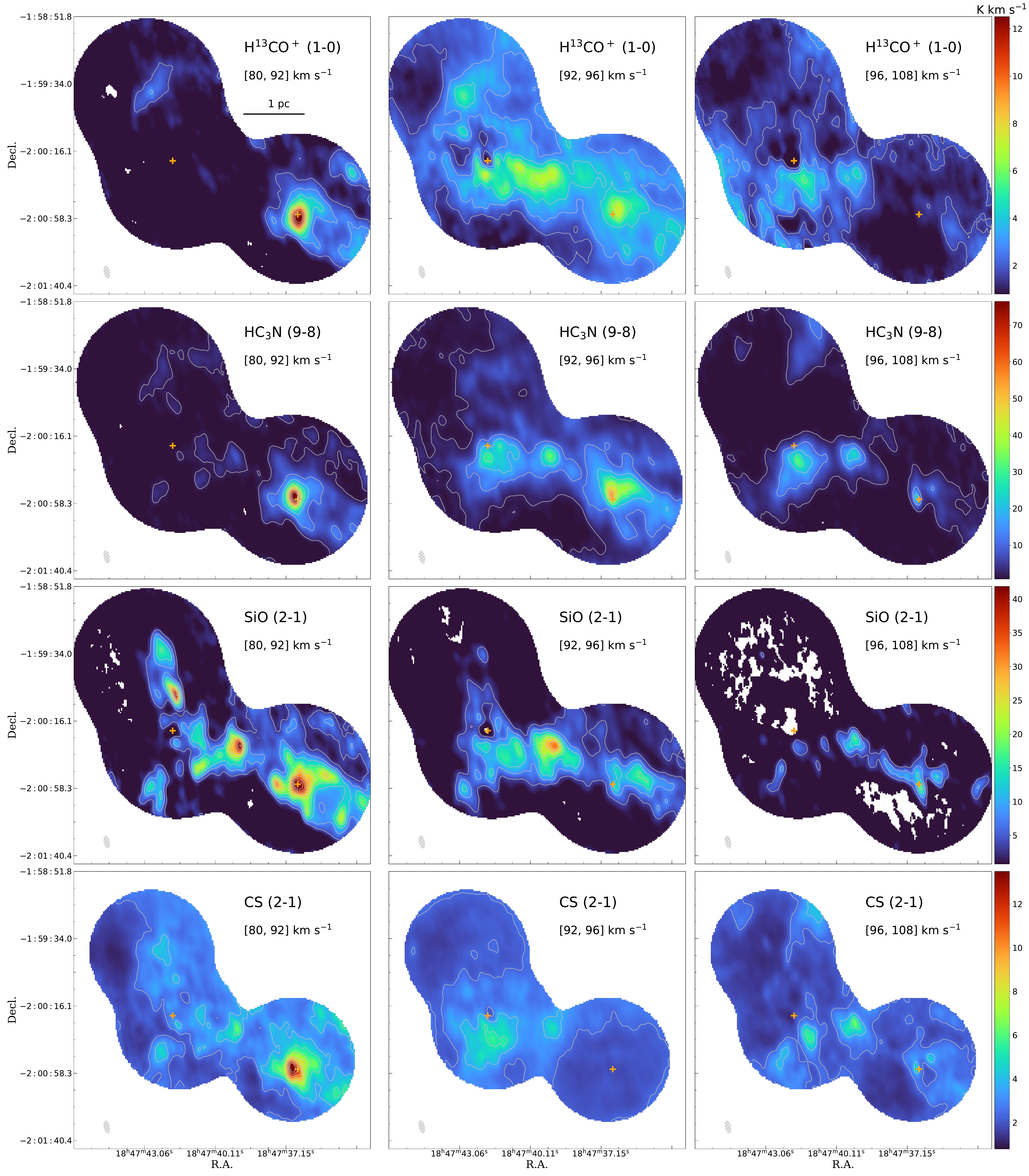}
\caption{Integrated intensity maps of CS (2-1), SO (2$_{3}$-1$_{2}$), H$^{13}$CO$^+$ (1-0), and HC$_{3}$N (9-8). The velocity ranges for integration are indicated in each subplot, which are selected based on the present velocity components of the overall average spectrum of CS (2-1) line. The 3 mm continuum peaks of clump MM3 and MM2 are indicated as orange crosses. Contour levels start from 1.6 K km s$^{-1}$ to the peak integrated intensities of the [80, 92] km s$^{-1}$ range with 8 uniform intervals. The peak integrated intensities are 76.4, 29.0, 12.5, 41.9 K km s$^{-1}$ for CS (2-1), SO (2$_{3}$-1$_{2}$), H$^{13}$CO$^+$ (1-0), and HC$_{3}$N (9-8), respectively.} 
\label{fig:channs_g307}
\end{figure*}

\begin{figure*}
\hspace{1.cm}\includegraphics[scale=0.15]{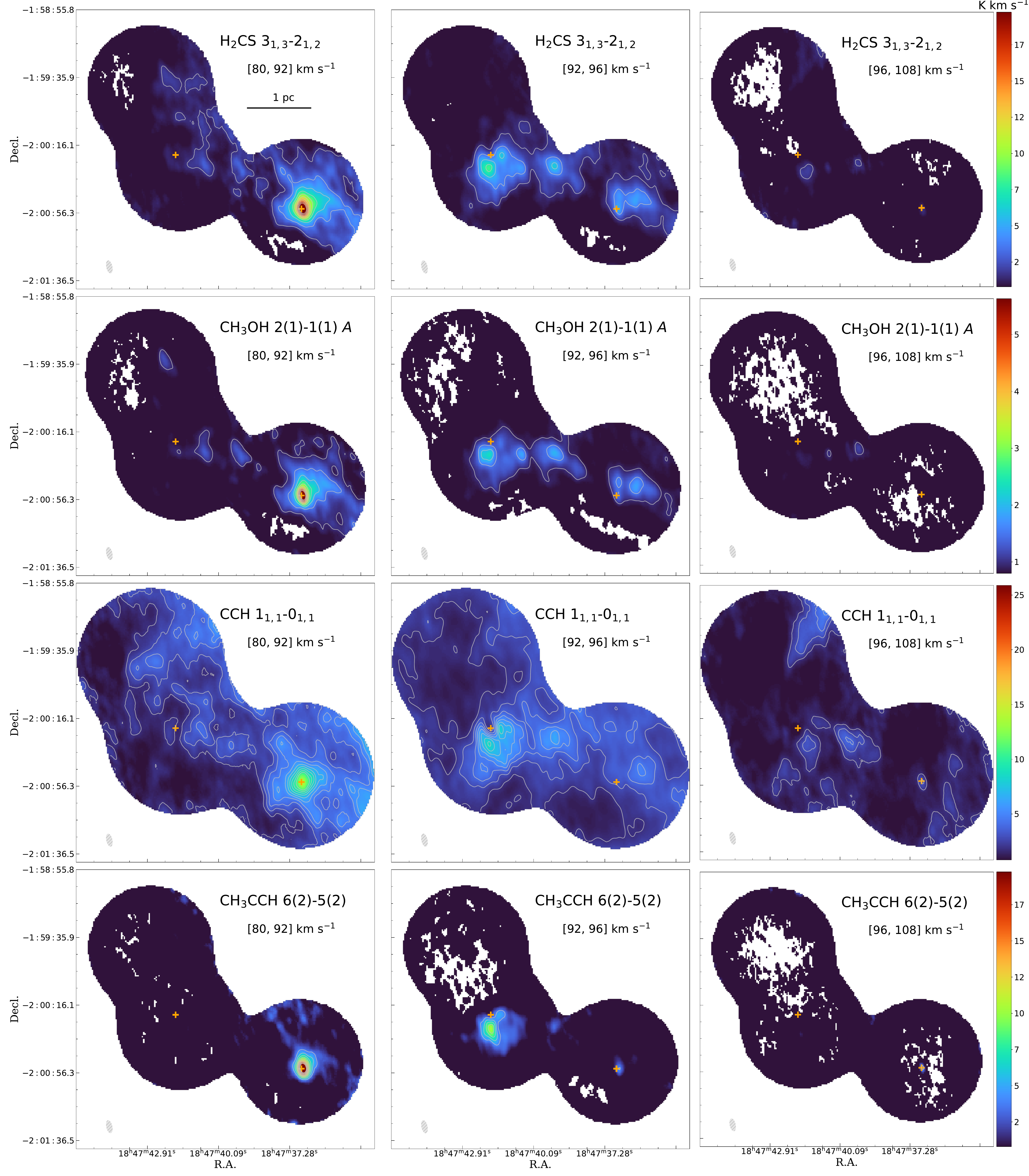}
\caption{Integrated intensity maps of CS (2-1), SO (2$_{3}$-1$_{2}$), H$^{13}$CO$^+$ (1-0), and HC$_{3}$N (9-8). The velocity ranges for integration are indicated in each subplot, which are selected based on the present velocity components of the overall average spectrum of CS (2-1) line. The 3 mm continuum peaks of clump MM3 and MM2 are indicated as orange crosses. Contour levels start from 1.6 K km s$^{-1}$ to the peak integrated intensities of the [80, 92] km s$^{-1}$ range with 8 uniform intervals. The peak integrated intensities are 76.4, 29.0, 12.5, 41.9 K km s$^{-1}$ for CS (2-1), SO (2$_{3}$-1$_{2}$), H$^{13}$CO$^+$ (1-0), and HC$_{3}$N (9-8), respectively. } 
\label{fig:channs_g307_more}
\end{figure*}

\begin{figure*}
\centering
\hspace{-1.cm}\includegraphics[scale=0.12]{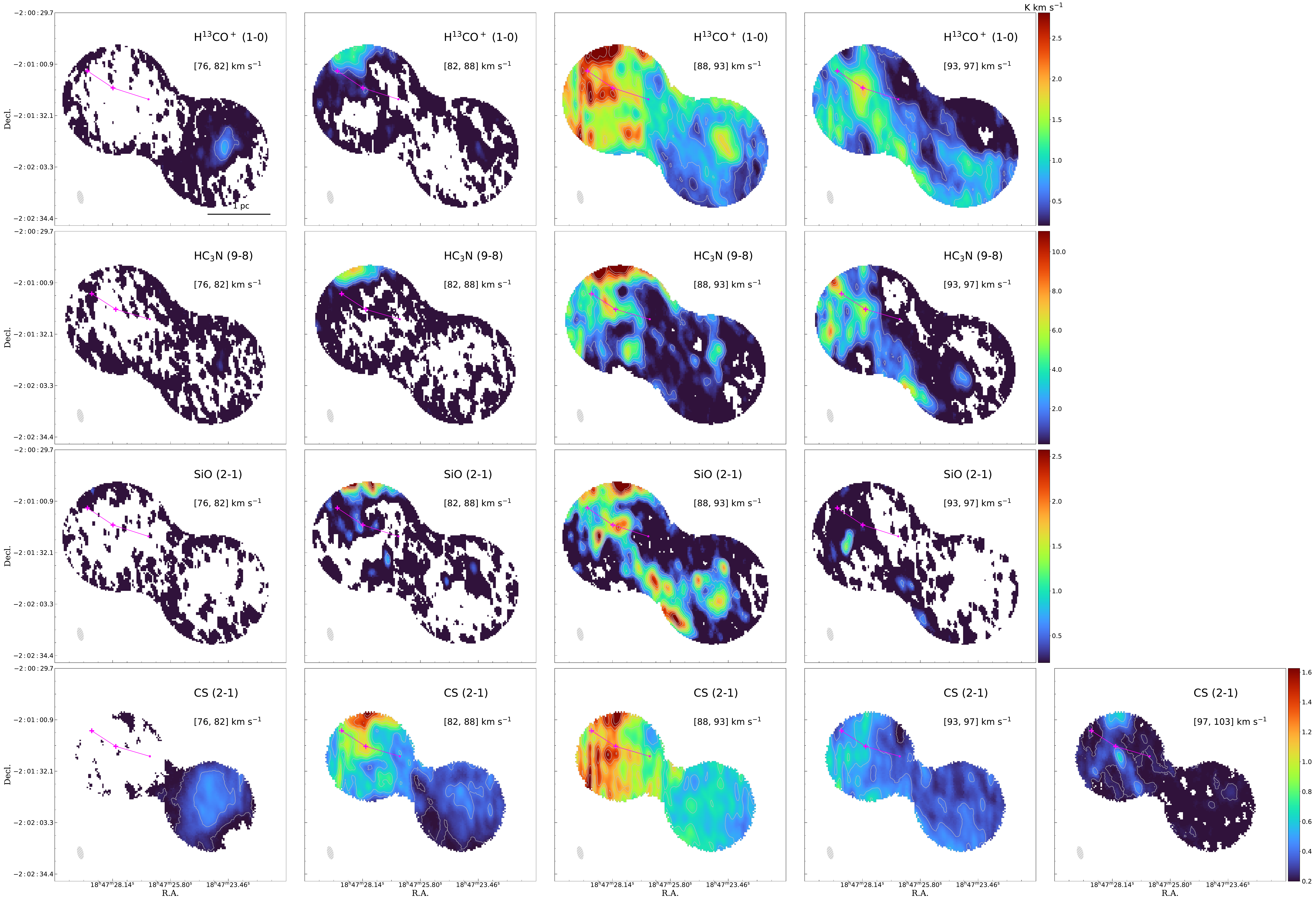}
\caption{Same as Figure \ref{fig:channs_g307}, but for the southern region containing clump F1 and F2. The 3 mm continuum peaks of clump F2 are indicated as orange crosses and linked with lines to indicate the filament ridge. Contour levels start from 0.4 K km s$^{-1}$ to the peak integrated intensities of the [88, 93] km s$^{-1}$ range with 8 uniform intervals. The peak integrated intensities are 11.3, 4.8, 2.8, 2.6 K km s$^{-1}$ for CS (2-1), SO (2$_{3}$-1$_{2}$), H$^{13}$CO$^+$ (1-0), and HC$_{3}$N (9-8), respectively. Integrated intensity maps of certain velocity range for some lines are trimmed if the emission is not significant. }
\label{fig:channs_g306}
\end{figure*}

The individual integrated intensity maps (channel maps) of the same set of lines are shown in Figures \ref{fig:channs_g307}-\ref{fig:channs_g306}. 
The corresponding velocity intervals are chosen according to the line profile of the averaged spectrum of CS (2-1) in the two regions. 

For the northern region, there are at least 3 velocity components peaking in velocity ranges of 80-92 km s$^{-1}$, 92-96 km s$^{-1}$ and 96-108 km s$^{-1}$. 
Compared to emission of H$^{13}$CO$^+$(1-0) and HC$_{3}$N (9-8), emission of CS(2-1) and SiO (2-1) show more extended structures in the 80-92 km s$^{-1}$ (most blue-shifted) velocity range, prominent in the whole northwestern region of the map. 
In this velocity range, towards clump C, the emission morphology of these two set of lines is also different, with CS(2-1) and SO 2$_{3}$-1$_{2}$ showing elongated structures roughly along the north-south direction, and emission of H$^{13}$CO$^+$(1-0) and HC$_{3}$N (9-8) is directed in the northwest-southeast. Clump MM3 hosts many sub-regions of clumpy emission in the south of the 3 mm continuum, which are present in all lines in the velocity range 92-96 km s$^{-1}$ and 96-108 km s$^{-1}$. Towards clump MM2 in the 92-96 km s$^{-1}$ velocity range, the emission of CS (2-1) line is very weak compared to all the other lines, but appears again in the red-shifted 96-108 km s$^{-1}$ range, which is likely due to strong self-absorption. 
Comparing the emission morphology of CS (2-1) and SiO (2-1), it seems the emission region of the red-shifted velocity component and the blue-shifted one mostly intersect with each other at clump MM3 and C, and overlap with each other at clump MM2. We discuss this feature further in Sect.~\ref{sec:vo_vw}, which points to a cloud-cloud collision process.

The southern region containing clumps F1 and F2 (Figure \ref{fig:channs_g306}) has a velocity field of a higher complexity, showing at least five velocity components, in ranges of 76-82, 82-88, 88-93, 93-97.5 and 97.5-103.5 km s$^{-1}$. Clump F2 has dominant emission separated in two velocity ranges 76-82 km s$^{-1}$ and 88-93 km s$^{-1}$, and appears rather compact in the H$^{13}$CO$^+$(1-0) and HC$_{3}$N (9-8) emission, while Clump F1 shows emission in four velocity ranges except 76-82 km s$^{-1}$. The morphology of the line emission of clump F1 is quite complex, which seems to be mainly composed of three parallel filaments in the east-west direction in the velocity range of 88-93 km s$^{-1}$. Interestingly, from the integrated intensity maps of H$^{13}$CO$^+$(1-0) and HC$_{3}$N (9-8) of the velocity range 93-97.5 km s$^{-1}$, the filament associated with clump F1 is directed toward the northeast-southwest. 

In Figure \ref{fig:channs_g307_more} we also show the channel maps of H$_{2}$CS 3$_{1,3}$-2$_{1,2}$, CH$_{3}$OH 2(1)-1(1) $A$, CCH 1$_{1,1}$-0$_{1,1}$, CH$_{3}$CCH (6-5) $K$=2 lines of the northern region, of same velocity ranges as Figure \ref{fig:channs_g307}.

\section{Other maps}
\begin{figure*}[htb]
\begin{tabular}{p{0.95\linewidth}}
\hspace{1cm}\includegraphics[scale=0.4]{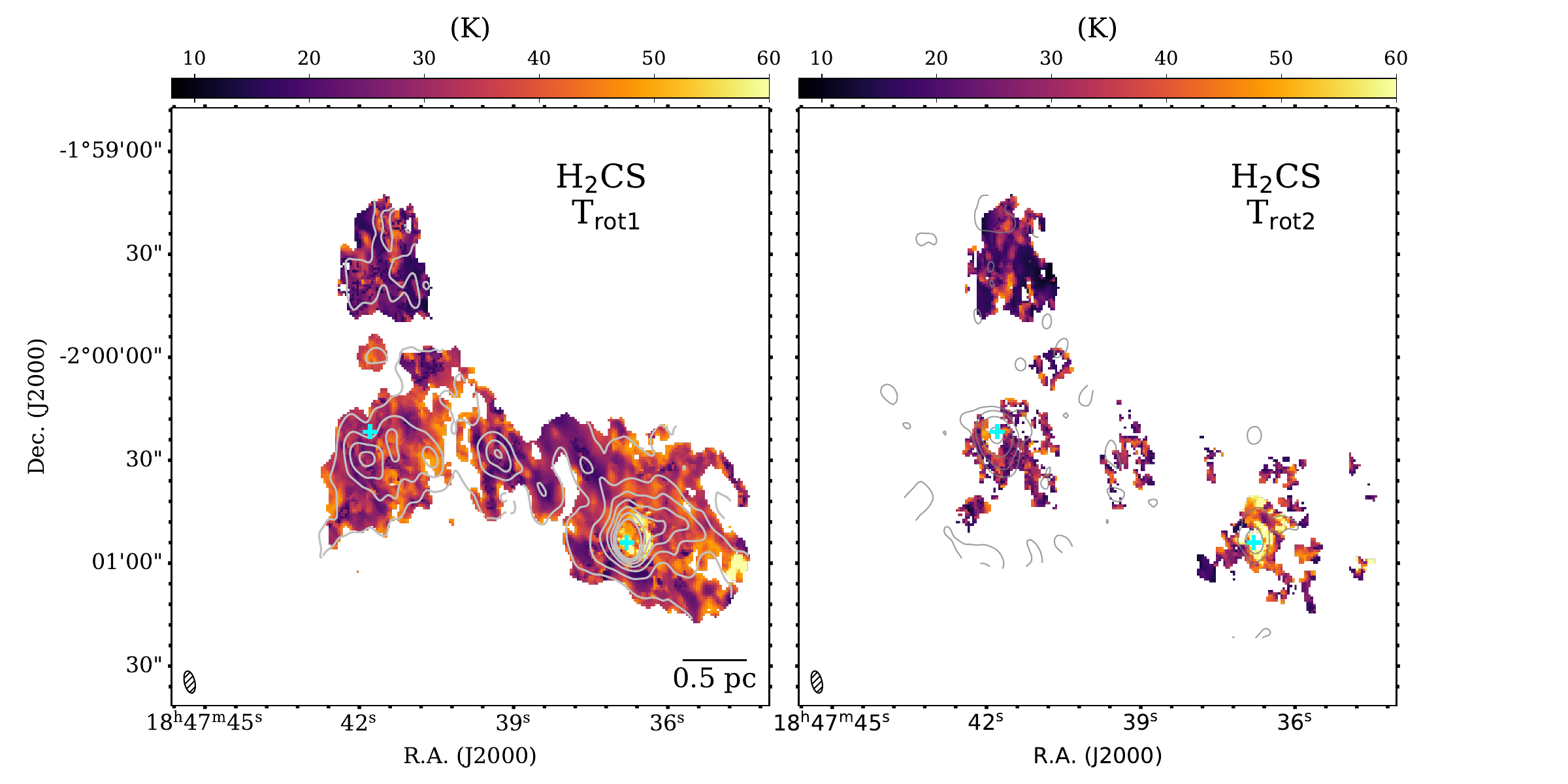}\\
\end{tabular}
\caption{Rotational temperature maps derived by H$_{2}$CS (3-2) lines for clumps MM2, MM3 and C. The left and right panels show the temperatures from the two component LTE models separately. In the left panel the contours represent integrated intensity map of H$_{2}$CS 3(1,3)-2(1,2) line. The velocity range for integration is 80-100 km s$^{-1}$, and the contour levels are from 2.7 K km s$^{-1}$ (6 $\sigma$) to 16.1 K km s$^{-1}$ (0.5 times peak emission value of MM2) with 7 uniform intervals. In the right panel, the contours represent the 3 mm continuum emission, and are logrithmic-spaced from 2$\sigma_{\mathrm{v}}$ ($\sigma_{\mathrm{v}}$ = 1.6 mJy~beam$^{-1}$) to 225.0 mJy~beam$^{-1}$ with 5 intervals (same as Figure \ref{fig:rgb_cont}). In both plots, the cyan crosses indicate the position of the peak intensity of 3 mm emission.}
\label{fig:h2cs_2comp_trot}
\end{figure*}

We present the derived rotational temperature maps from H$_{2}$CS (3-2) lines in Fig. \ref{fig:h2cs_2comp_trot}. Compared to rotational temperature maps derived by CH$_{3}$CCH lines (Fig. \ref{fig:ch3cch_2comp_trot}) the peaking is not as significant, and the overall temperature distribution show mostly localised variations without a clear pattern. 


\section{Two-component {\emph{hfs}} fitting results of the CCH line}\label{app:cch_maps}
\begin{figure*}[htb]
\hspace{-1.5cm}\includegraphics[scale=0.32]{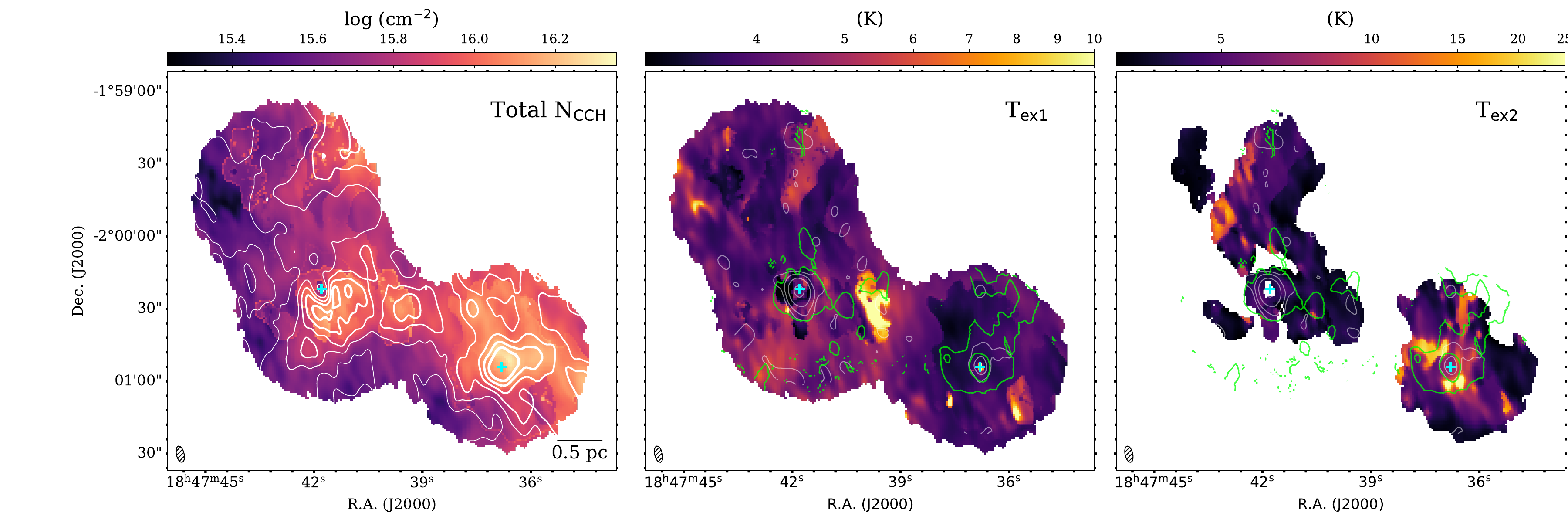}
    \caption{CCH column density and excitation temperatures from the two-component {\emph{hfs}} fitting for clumps MM2, MM3 and C. White contours in the left panels represent the integrated intensity map of CCH 1$_{1,1}$-0$_{1,1}$ line (85-100 km s$^{-1}$), from 3.3 K km s$^{-1}$ (7 $\sigma$) to 9.8 K km s$^{-1}$ (0.9 times the peak emission of MM2) with 6 uniform intervals. The gray and green contours, and crosses follow the same definitions as in Figure \ref{fig:h2cs_2comp_trot}. }
\label{fig:cch_2comp_nmol_tex}
\end{figure*}

\begin{figure*}[htb]
\hspace{-1.5cm} \includegraphics[scale=0.32]{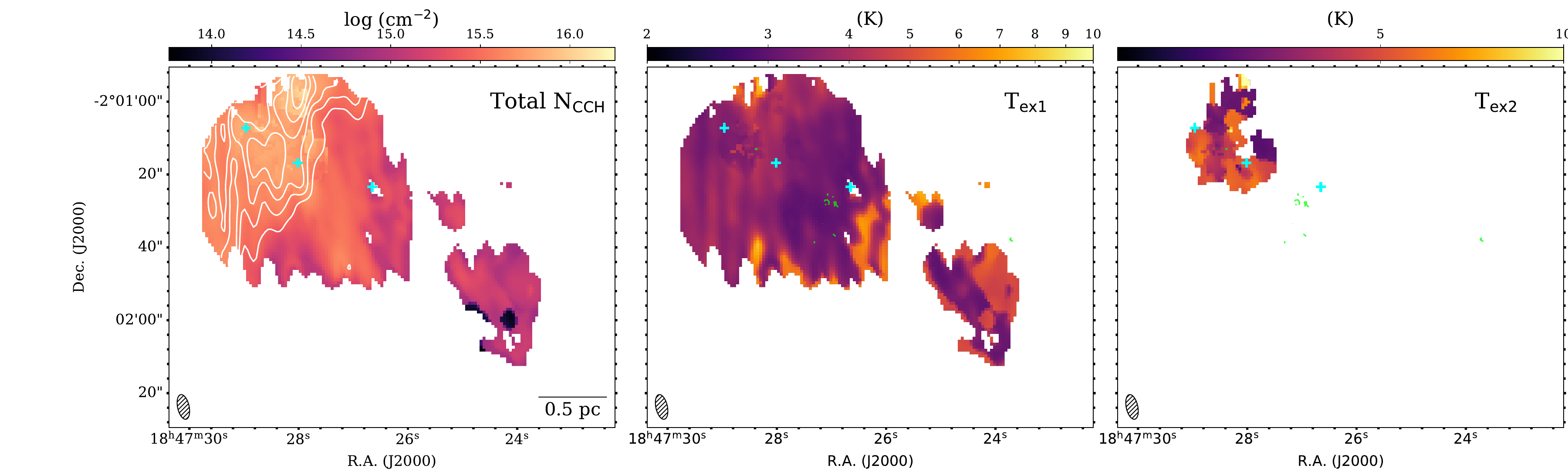}
    \caption{Same as Figure \ref{fig:cch_2comp_nmol_tex}, but for clump F1 and F2. White contours in the left panels represent the integrated intensity map of CCH 1$_{1,1}$-0$_{1,1}$ line (80-95 km s$^{-1}$), from 2.2 K km s$^{-1}$ (7 $\sigma$) to 6.0 K km s$^{-1}$ (0.9 times the peak emission of F1) with 6 uniform intervals.}
\label{fig:cch_2comp_nmol_texs}
\end{figure*}

\begin{figure*}[htb]
\begin{tabular}{p{0.95\linewidth}}
\includegraphics[scale=0.4]{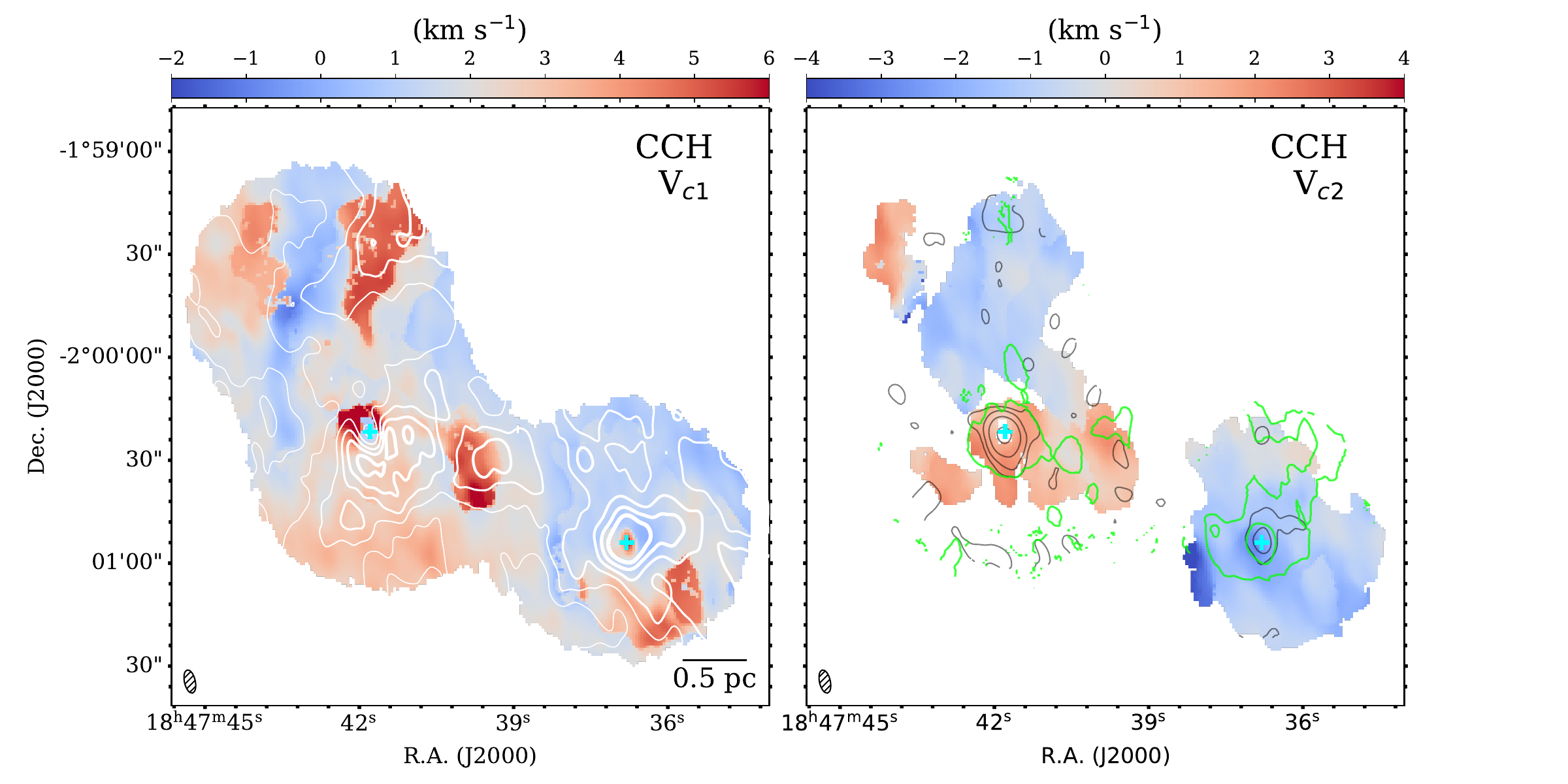}\\
\includegraphics[scale=0.4]{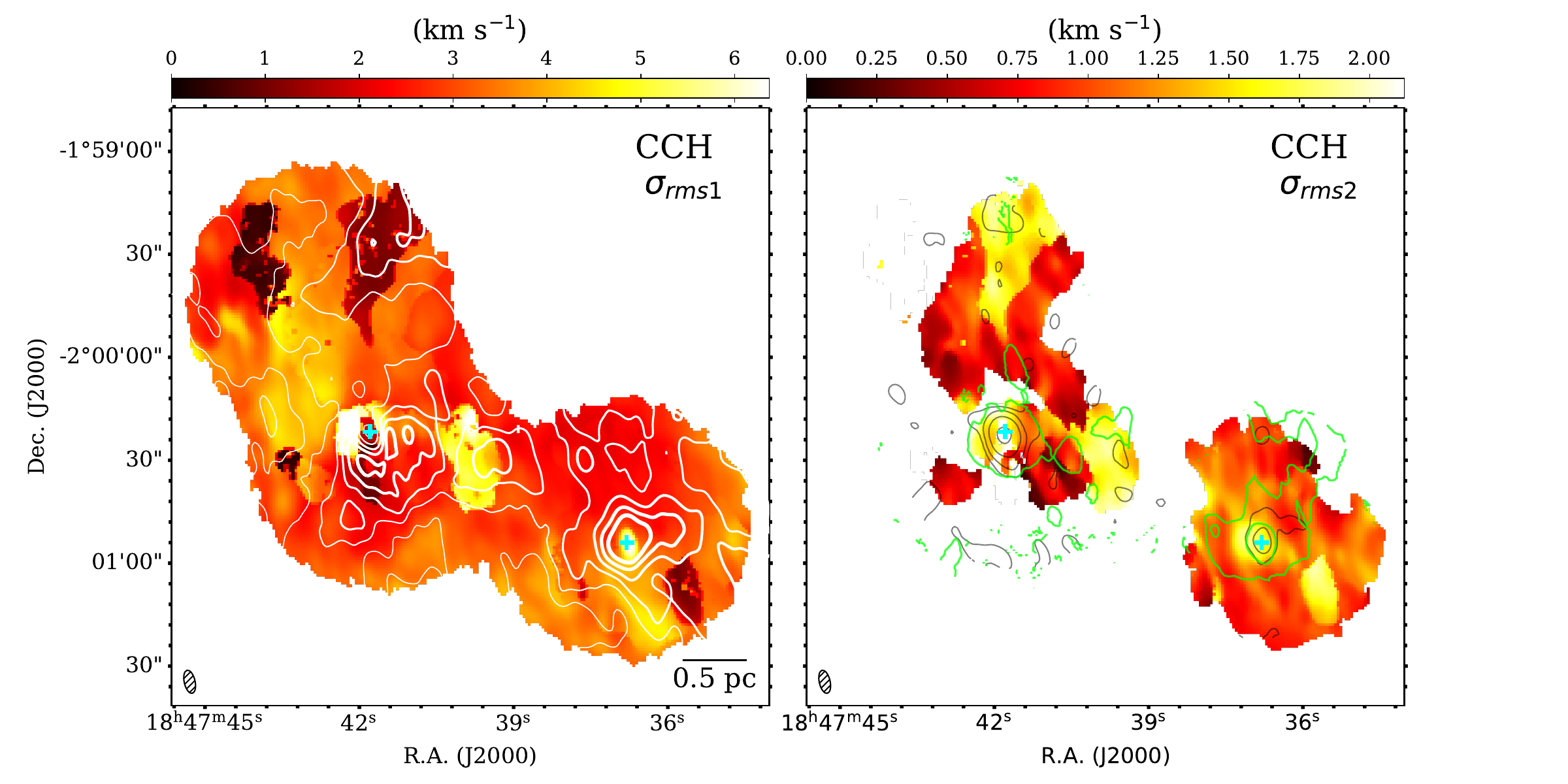}\\
\end{tabular}
\caption{Centroid velocity and velocity dispersion distribution of the two-component Gaussian fits of CCH (1-0) {\emph{hfs}} lines ($\nu$$\sim$87.3 GHz) for clumps MM2, MM3 and C. The upper panel and lower panel shows the two components individually. White contours in left panels follow same definitions as in Figure \ref{fig:cch_2comp_nmol_tex}, representing the integrated intensity map of CCH.
In both upper and lower panel, the gray and green contours, and crosses in left and right plot follow the same definitions as in Figure \ref{fig:h2cs_2comp_trot}. }
\label{fig:cch_2comp_vwvo}
\end{figure*}

\begin{figure*}[htb]
\begin{tabular}{p{0.95\linewidth}}
\includegraphics[scale=0.4]{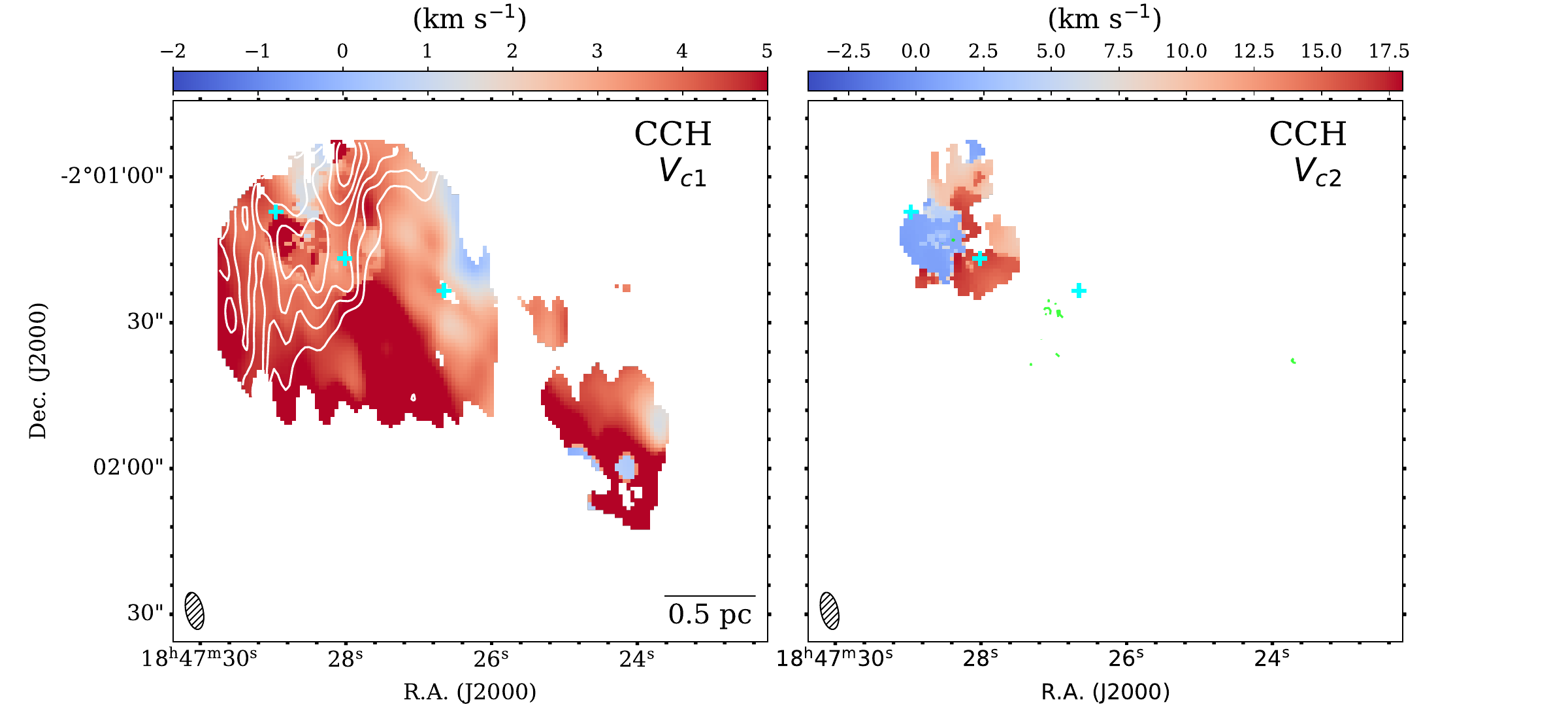}\\
\includegraphics[scale=0.4]{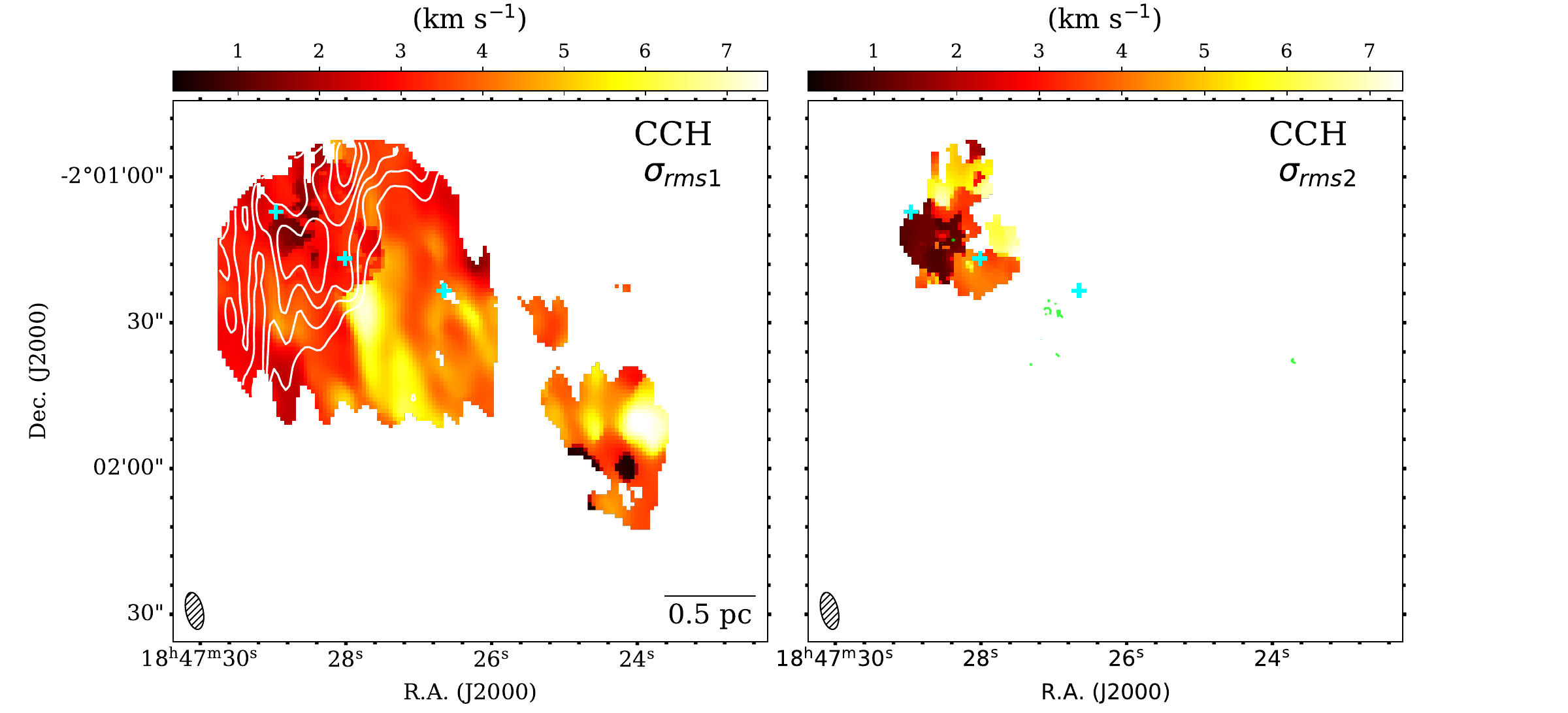}\\
\end{tabular}
\caption{Same as Figure \ref{fig:cch_2comp_vwvo}, but for the clumps F1 and F2. White contours in left panels follow same definitions as in Figure \ref{fig:cch_2comp_nmol_texs}, representing the integrated intensity map of CCH.}
\label{fig:cch_2comp_vwvos}
\end{figure*}

We present the derived parameter maps from LTE modelling of C$_{2}$H (1-0) hyperfine lines in Fig. \ref{fig:cch_2comp_nmol_tex}-\ref{fig:cch_2comp_vwvos}. 

\section{NH$_{2}$D spectra of the identified cores}\label{app:nh2d}
\begin{figure*}
\hspace{2cm}\includegraphics[scale=0.35]{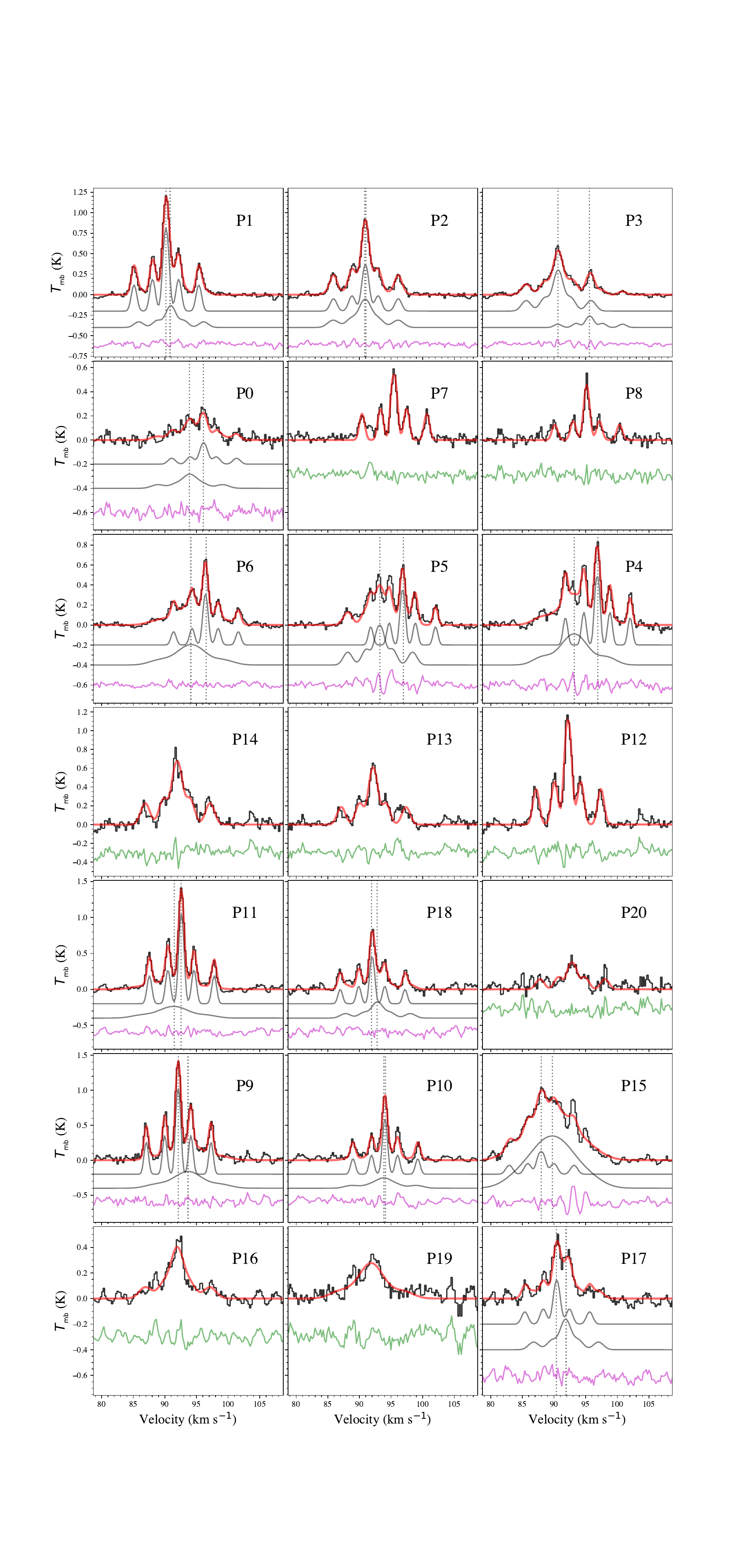}
\caption{NH$_{2}$D spectra of the {\tt{dendrogram}} identified cores in clumps MM2, MM3 and C. For cores that a single {\emph{hfs}} component fit suffices to reproduce the line profile, the model is shown as red line, whereas two-component {\emph{hfs}} fits are shown as blue (sum) line and gray lines (manually adding offset from zero level) for the two components, separately. Vertical dotted lines indicate the peak velocity of the two-component models. The core names are denoted as Pn with n of integers between [0, 20] (Fig. \ref{fig:nh2d_core_map}, upper panels). The subplots are arranged such that the corresponding cores are located from north-east to south-west along the mosaic region.}
\label{fig:nh2d_sps}
\end{figure*}

\begin{figure*}
\hspace{2cm}\includegraphics[scale=0.35]{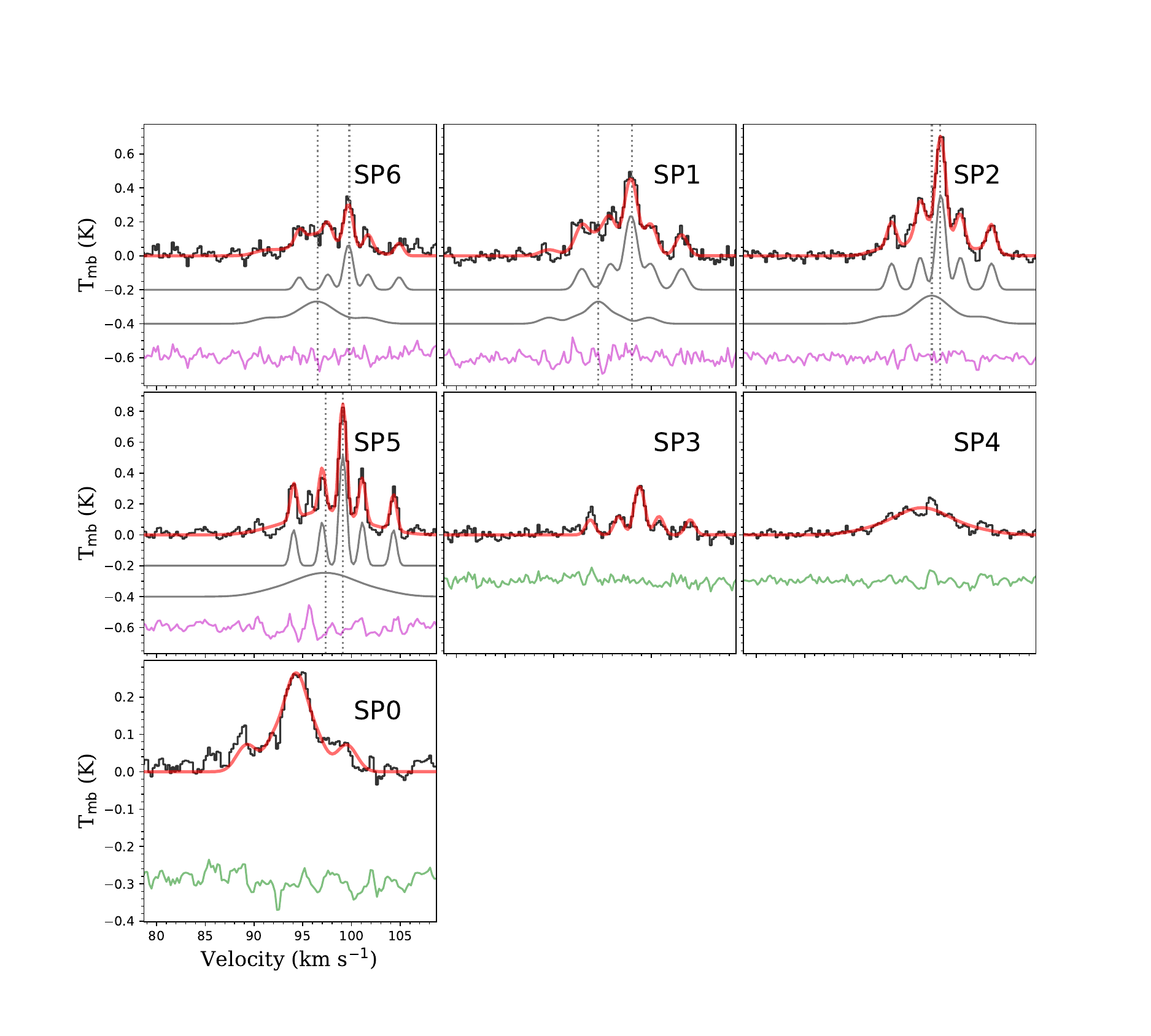}
\caption{Same as Figure \ref{fig:nh2d_sps}, but for NH$_{2}$D spectra of cores residing in clumps F1 and F2. The core names are denoted as SPn with n of integers between [0, 6] (Figure. \ref{fig:nh2d_core_map}, lower panels).}
\label{fig:nh2d_spss}
\end{figure*}

\begin{figure*}
\hspace{0cm}\includegraphics[scale=0.34]{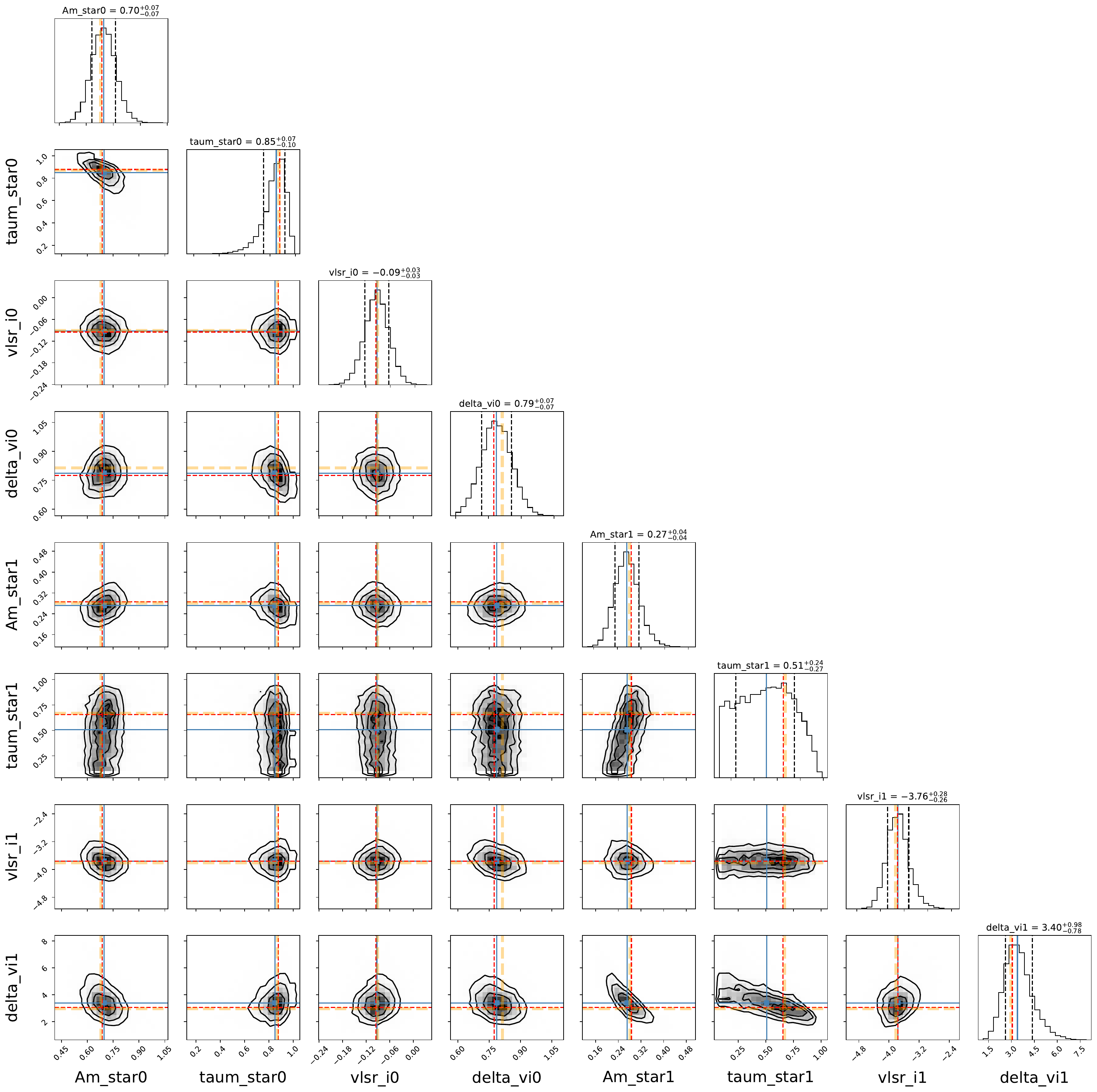}
\caption{Posterior distribution of the parameters from two-component {\emph{hfs}} fitting of core P4. Black vertical dotted lines indicate the 1$\sigma$ quantiles of the distribution. The other colored vertical lines indicate the parameter set corresponding to the least-square fit (red), the median (blue) and maximum-likelihood (orange) values drawn from the posterior distribution.}
\label{fig:nh2d_corner}
\end{figure*}

We plot the core averaged NH$_{2}$D spectra and the fitted spectra in Fig. \ref{fig:nh2d_sps} and \ref{fig:nh2d_spss}. In Fig. \ref{fig:nh2d_corner} an example of posterior distribution of parameters of the two-component fit is shown.

\end{appendix}

\end{document}